\documentclass[UTF8,a4paper]{article}
\usepackage{amsmath}
\usepackage{graphicx}
\usepackage{subfigure}
\usepackage{epstopdf}
\usepackage{geometry}
\usepackage{ulem}
\usepackage{indentfirst}
\usepackage{amsfonts,amssymb,dsfont}
\usepackage{setspace}
\usepackage{threeparttable}
\usepackage{amsmath,mathtools,amsthm}
\usepackage{array}
\usepackage{extarrows}
\usepackage{upgreek}

\makeatletter
\renewcommand*\env@matrix[1][\arraystretch]{%
  \edef\arraystretch{#1}%
  \hskip -\arraycolsep
  \let\@ifnextchar\new@ifnextchar
  \array{*\c@MaxMatrixCols c}}
\makeatother
\begin{document}

%QED 3+1: Appelquist T W, Bowick M, Karabali D and Wijewardhana L C R 1986 Phys. Rev. D 33 3704
%Optical evidence for a Weyl semimetal state in pyrochlore Eu 2 Ir 2 O 7
%Collective Modes of the Massless Dirac Plasma
%relaxation time:Dc and ac transport in silicene
%Femtosecond carrier dynamics and saturable absorption in graphene

\title{\bf Electronic properties and polaronic dynamics of semi-Dirac system within ladder approximation}
\author{Chen-Huan Wu
\thanks{chenhuanwu1@gmail.com}
%\\Key Laboratory of Atomic $\&$ Molecular Physics and Functional Materials of Gansu Province,
\\College of Physics and Electronic Engineering, Northwest Normal University, Lanzhou 730070, China}

\maketitle
\vspace{-30pt}
\begin{abstract}
\begin{large}

%We investigate the properties of the attractive polaron formed by a single impurity dressed with %the particle-hole excitations
%in a Dirac/Weyl system
%at zero-temperature limit.
%Base on the single particle-hole variational ansatz,
 %{Repulsive Fermi Polarons in a Resonant Mixture of Ultracold 6Li Atoms}
%we deduce the expressions of the pair propagator, self-energy (negative), and the non-self%-%consistent medium $T$-matrix.
%Different to the self-consistent $T$-matrix which contains two channels
%due to the many-body effect,
%the non-self-consistent $T$-matrix discussed in this paper 
%contains only the closed channel (i.e., the bare one)
%since we consider only a finite number of the majority particles.
%The chiral factor is also discussed within the pair propagator due to the scattering feature of %the solid state
%(with a certain scattering angle $a$),
%and we found that the fluctuation of the pair propagator is proportional to the ${\rm cos}\ a$.
%Finally, we also discuss the particle spectral function of the polaron in the presence of quantum %many-body effect.\\
%
%
We investigate the electronic properties of the semi-Dirac system and its polaronic dynamics when coupled with a fermi bath
with quadratic dispersion.
The electronic anisotropic transport properties and the semiclassical dynamics of the semi-Dirac system are studied,
including the density-of-states, conductivity, transport relaxation rate, specific heat, electrical current denity, 
  %{Anomalous equilibrium currents for massive Dirac electrons}
  %{Screening-induced temperature-dependent transport in two-dimensional graphene}
and free energy.
The attractive polaron formed as
the semi-Dirac impurity dressed with the particle-hole excitations 
in a two-dimensional system are studied both analytically and numerically.
  %{dark continuum in the spectral function of the resonant fermi polaron}
  %which to our knowledge not being studied before.
The pair propagator, self-energy, spectral function are being detailly calculated and discussed.
The method of medium $T$-matrix approximation (non-self-consistent),
which equivalent to the partially dressed interaction vertex by summing over all ladder diagrams,
is applied,
and compared with some other methods (for many-body problem),
like the leading-order $1/N$ expansion (GW approximation), Hartree-Fock theory, 
and the Nozieres-Schmitt-Rink theory.
Since we foucs on the weak-coupling region, the mean-field approximation is also applicable.
The polaron properties is related to the anisotropic effective masses of the semi-Dirac system.
% i.e., related to the properties of the semi-Dirac system, including the gap in $y$-direction $D$.
That's in contrast to the polarons formed in the surface of normal Dirac systems which has an isotropic dispersion,
since the anisotropic dispersion of the semi-Dirac systems results in anisotropic effective mass and anisotropic charge carrier transport.
Besides, the symmetry between electron and hole is also broken since the effective masses of the electron and hole are different,
  %{polaronic effects in monolayer black phosphorus on polar substrate}
  %{fermi polaron-polaritons in charge-tunable atomically thin semiconductors}
which are affected by the polaronic effect when the semi-Dirac material is deposited on a polar substrate, like hBN.
The self-localization, short-range potential, and experimental methods as well as the 
possible formation of the bose polaron on the surface of semi-Dirac system are also discussed in the end.
Our results are useful also for the investigation of two/three-dimensional bosonic polaron as well as the polarons in other solid state systems, 
like the magnetic matter or the topological systems.\\
\\
% $PACS\ number(s)$: 71.10.Hf, 71.10.Li, 71.36.+c\\
%$ Keywords$: Fermi polaron;
%Medium T-matrix;
%Self-energy;
%Spectral function;
%Pair propagator;
%Ladder approximation\\

\end{large}

\end{abstract}
\begin{large}

\section{Introduction}

We investigate the electronic properties of semi-Dirac system as well as the related polaronic dynamics as a semi-Dirac quasiparticle
(impurity) immersed into a medium (two-dimensional (2D) electron gas).
Semi-Dirac 2D material,
which exhibits a relativistic dispersion in
one direction and nonrelativistic in another, has surge a great research interest\cite{Pyatkovskiy P K,Carbotte J P}.
Besides, due to the existence of nonadiabatic feature in the nonrelativistic direction,
the polaronic effect would be foud as the minority semi-Dirac quasiparticles interact with the quadratic majority particles 
(particle-hole excitations) in 2D electron gas.
%{Fermi-Bose Quantum Degenerate 40K-87Rb Mixture with Attractive Interaction}
%The polaronic effect in dilute Fermi gases as well as the BEC are widely studied, 
For semi-Dirac system,
the electronic transport properties, semiclassical dynamics, and the polaronic dynamics are affected by the anisotropic dispersion,
and becomes direction-dependent.
To deal with the anisotropicity, we transform the momentum coordinate to the polar coordinate, and apply it into the study of polaron.

The fermionic polaron formed by a mobile light bosonic impurity dressed by the polarized fermions 
is widely studied in the fermi gases as well as the superfluid quantum state,
  %{stability condition of a strongly interacting boson-fermion mixture across an interspecies feshbach rwsonance}
and the bosonic polaron is also widely explored in Bose-Einstein condensate (BEC)\cite{Park J W,Onofrio R} using the method of momentum-resolved radio-frequency (rf) spectroscopy\cite{Hu M G}, 
  %{Fermi polarons in two dimensions}
or in a bosonic bath with the induced lattice vibration (distortion) in a solid state system.
  %{Diagrammatic Monte Carlo study of the acoustic and the Bose–Einstein condensate polaron}
The repulsive polaron has also been realized by experiments\cite{Hu M G,J?rgensen N B,Mistakidis S I2} in a BEC.
Besides, the scenarios that a fermionic (bosonic) impurity embedded in a fermi (boson) 
bath (i.e., the fermi-fermi (boson-boson) mixture) are also arouse great interest, like the ${}^{6}{\rm Li}-{}^{40}{\rm K}$
\cite{Kohstall C,Mistakidis S I} 
and the ${}^{133}{\rm Cs}-{}^{87}{\rm Rb}$\cite{Spethmann N} mixures.
The formation of the many-body bound state for bose polaron, like the bipolaron and the tripolaron have also been investigated\cite{Alexandrov A S,Camacho-Guardian A,Bonca J,Moszkowski S,Wellein G}.
  %{Bose Polarons in the Strongly Interacting Regimek}
  %{impurity-induced multibody resonances in a bose gas}
However, the fermi polaron in the solid state is inverstigated less, 
but it is also meaningful to study, e.g., in the battery application.
%like a scenario that some single bosonic impurity adsorbed to a solid state material as happen  
%in the battery application.
%{shengyuan yang}
%In this article, 
%we investigate the properties of the attractive polaron formed by a single impurity dressed with the particle-hole excitations 
%in a two-dimensional semi-Dirac system.
We note that, recently, 
the plasmon-polaron mode formed by a two-dimensional electron gas occupying the surface of topological matter has also been reported 
in a recent work\cite{Shvonski A}.

Polaron as a dressed quasiparticle
is also related to the strength of the interspecies and intraspecies coupling as well as the mean-field energy. 
In fact, the mean-field approximation overestimates the interaction effect in the strong interacting regime\cite{Li W}, 
thus the mean-field results also overestimate the critical value of the attractive interaction strength 
that a stable system begins to collapse, as a phenomenon induced by the quantum pressure. 
It's also found that the manipulation of the band dispersion 
(formed by the free electrons away from the Dirac cone) can used to stabilize the system and change the stability criterial. 
The spin-orbit coupling is important in stabilizing the collapsed system.
  %{The phase diagram and stability oftrapped D-dimensional spin-orbit coupled Bose-Einstein condensate} 
Besides, 
the strength of spin-orbit coupling can be used to manipulating the polaronic effect (like the polaron self-energy) 
since it's related directly to the Dirac mass of the Dirac system, 
expecially in the topological insulators with the intrinsic spin-momentum locking\cite{1,2,3,4}.
It's also found that the modified band dispersion 
(away from the simple Bogoliubov spectrum by the spin-orbit coupling),
can stabilizes the system by counteracting the attractive interactions.

The variational wave function of the Fermi gas system as firstly reported in Ref.\cite{Chevy F} 
is much simpler than the one of the solid state system (see Appendix.A for details).
The differences are mainly come from the properties (like interaction effect) 
of the bath (reservoir of the majority particles) but ont the center mobile quantum impurity.
By taking the intraspecies coupling into consideration, we can obtain a more accurate result for the polaron in the solid state system.
That also directly revealed in the formula of the bare coupling parameter (see text)
which contains the bandwidth factor in the denominator (see text)
(while in the vacuum limit with $k_{F}\rightarrow 0$, the bandwidth is nearly zero,
e.g., the molecule whose binding energy equals to the center-of-mass kinetic energy\cite{Parish M M}).
  %{Highly polarized Fermi gases in two dimensions}
The formation of the polaron (or even the trion) and the polariton
have been realized in semiconductor 
heterostructure by using the technique of cavity coupling in the presence of exciton-electron interactions\cite{Sidler M,Ganchev B}.
For semi-Dirac system, which has very interesting electronic and topological properties under irradiation\cite{Islam S K F,Saha K,Huang S},
although its anisotropic dispersion,
we can still approximate it as the one similar to the two-dimensional electron gas when the carrier density is just slightly larger 
than the band gap ($D\gg \mu_{\uparrow}-D>0$) and with small momentum.
%{Dynamical polarization and plasmons in a two-dimensional system with merging Dirac points}
    %{Semiclassical Boltzmann transport theory of few-layer black phosphorous in various phases}
In the mean time, the chemical potential of the impurity can be treated as a negative unphysical parameter $\mu_{\downarrow}<-\mu_{\uparrow}$
during the calculation of the polaron energy (i.e., the pole of the dressed Green's function).
  %{Diagrammatic Monte Carlo study of the Fermi polaron in two dimensions}
As a quasiparticle,
the attractive polaron is metastable as directly seen from the spectral function as well as the effective mass and quasiparticle residue,
that also revealed by its complex energy as obtained by the diagrammatic method within ladder approximation,
and such result is also in agreement with the one in Ref.\cite{Sidler M}.
  %{Highly polarized Fermi gases in two dimensions}

In Sec.2, we present the model and the related electronic quantities of the semi-Dirac system.
In Sec.3, we discuss the semiclassical dynamics.
In Sec.4, we study the polaronic effects in terms of the anisotropic and isotropic treatments.
In Sec.5, we discuss the negative gap case.
In Sec.6, we discuss the remained Coulomb interaction effect in the prsence of low carrier-density approximation.
In the end of this article,
we also present a detail discussion of the theoretical framework about the 
Chevy-type variational wave function within ladder approximation (Appendox.A),
and the possible self-localization as well as the short-range (but not zero) potential (Appendix.B).
In Appendix.C, we present the experimental method which is possible to demonstrate our theories,
and discuss the possible realization of the bose polaron for semi-Dirac system deposited on a polar substrate.

We note that, in this paper, we mainly focus on the zero temperature case (or the low-temperature limit), 
and using the non-self-consistent $T$-matrix theory
which provides a more accurate polaron energy estimation than the self-consistent one\cite{Hu H} due to the cancelling effect of the
 higher order multi-particle-hole excitations\cite{Combescot R2}.
Besides, during the calculation of polaronic dynamics-related quantities,
we mainly consider the motion of impuirty (or polaron) in the nonadiabatic direction ($p_{x}$-direction),
no matter in anisotropic or isotropic treatments.

\section{Model}

  %{Plasmon-polaron of the topological metallic surface states}
We consider a model that the semi-Dirac impurity interact attractively with the excitations
    %(like the phonon excited by a polar substrate or the cavity photon, or simply, the hole in valene band\cite{Castella H}) embedded
    %immersed
in a fermion bath (free 2D electron gas).
    %{Dropping an impurity into a Chern insulator A polaron view on topological matter}
    %{Plasmon-polaron of the topological metallic surface states}
    %partially polarized 2D semi-Dirac system at zero-temperature limit.
    %{Discriminative generation and hydrogen modulation of the Dirac-Fermi polarons at graphene edges and atomic vacancies}
Unlike the point fermi surface aspect in zero gap intrinsic materials,
  %{Phenomenology of a semi-Dirac semi-Weyl semimetal}
we set the finite fermi energy (chemical potential in noninteracting case)
  %{Damping of long wavelength collective modes in spinor Bose-Fermi mixtures}
  %{Phenomenology of a semi-Dirac semi-Weyl semimetal}
for the quasiparticles in semi-Dirac material during the evolution of energy spectrum.
  %{Dynamical polarization and plasmons in a two-dimensional system with merging Dirac points}
Thus the Landau damping (or Beliaev damping), due to the pairing interaction effect (or fermi surface effect),
   %{Absence of damping of low-energy excitations in a quasi-two-dimensional dipolar Bose gas}
   %{Damping of long wavelength collective modes in spinor Bose-Fermi mixtures}
emerges,
and it exists as long as the interaction between impurity particle and the majority one is nonzero.
Also, the existence of chemical potential leads to the screening to the long-range Coulomb potential,
thus we can assume the contact potential (interspecies) is dominating in our model,
and due to the short-range nature, the anisotropy of interaction can
be ignored.
  %{hase diagram of Rydberg-dressed Fermi gas in two dimensions}
We note that, while in the limit of zero doping,
the ratio of Coulomb energy to the kinetic energy is largely increased\cite{Jafari S A}, and 
the orbital susceptibility and fermi surface orbiting frequency will diverge\cite{Banerjee S2}. 

At zero-temperature limit, the summation over Matsubara frequencies can be replaced
by the integrals over the continuous frequencies,
which can be measured from the quasiparticle dispersion as well as the self-energy\cite{Combescot R}.
  %{mass imbalance effect in resonant bose-fermi mixtures}
And in single impurity case, the intervalley scattering does not have to be taken into account.
  %{fermi polaron-polaritons in charge-tunable atomically thin semiconductors}
For the Fermionic reservoir, we apply the semi-Dirac model with the merging Dirac points
in order to containing both the linear Dirac dispersion and the effective mass (hopping).
The phases of the semi-Dirac system is dependent on the Dirac-mass $D$:
for positive $D$, the system is in insulator phase, while for negative $D$, the system is in semimetal phase, which is in contrast to the usual phase transition to a metal due to the disorder (like the Coulomb interaction) \cite{Roy B}. 
  %{interaction corrections to the polarization function of graphene}
The evolution of the dispersion with phase is shown in Fig.1.

We assume the number-density of the semi-Dirac quasiparticle is much lower than that from fermi bath, thus we
regard the former as the impurity, while the latter as the majority component.
   %{Mass imbalance effect in resonant Bose-Fermi mixtures}
Then Hamiltonian of such mixture reads
\begin{equation} 
\begin{aligned}
H=H_{i}+H_{m}+g_{b},
\end{aligned}
\end{equation}
where $g_{b}$ is the bare (unrenormalized) contact interaction.
The Hamiltonian of the quasiparticle in doped 2D semi-Dirac system reads
\begin{equation} 
\begin{aligned}
H_{i}=(\frac{p_{x}^{2}}{2m_{x}}+D)\sigma_{x}+v_{y}p_{y}\sigma_{y}-\mu_{i}\sigma_{0},
%\hbar v_{x}k_{x}\sigma_{x}+(D+ak_{y}^{2})\sigma_{y}+g_{\psi\phi}\sigma_{0}-\mu_{\uparrow}\sigma_{0},
\end{aligned}
\end{equation}
  %{thermodynamics of spin-orbit-coupled bose-einstein condensates}
and the Hamiltonian of isotropic majority component (electron-hole pair excited by the semi-Dirac quasiparticle) reads
\begin{equation} 
\begin{aligned}
H_{m}=\frac{(k-q)^{2}}{2m_{\uparrow}}-2\mu_{m}.
\end{aligned}
\end{equation}
The Dirac mass\cite{Ludwig A W W,Ezawa M} term here reads $D=M\sigma_{z}$ where $M$ is the band gap.
For zero momentum and zero magnetic field, $\sigma_{z}=1/2$.
The matrix form of $H_{i}$ reads
\begin{equation} 
\begin{aligned}
H_{i}=\begin{pmatrix}
-\mu_{i}  &  \frac{p_{x}^{2}}{2m_{x}}-iv_{y}p_{y}+D\\
\frac{p_{x}^{2}}{2m_{x}}+iv_{y}p_{y}+D   &   -\mu_{i}
\end{pmatrix},
\end{aligned}
\end{equation}
with the eigenenergy 
\begin{equation} 
\begin{aligned}
\varepsilon=&\pm\sqrt{D^{2}+\frac{Dp_{x}^{2}}{m_{x}}+\frac{p_{x}^{4}}{4m_{x}^{2}}+v_{y}^{2}p_{y}^{2}}-\mu_{i}\\
=&\pm\sqrt{(\frac{p_{x}^{2}}{2m_{x}}+D)^{2}+v_{y}^{2}p_{y}^{2}}-\mu_{i},
\end{aligned}
\end{equation}
To dealing with the anisotropic dispersion of the semi-Dirac quasiparticle,
we use the following substitution,
\begin{equation} 
\begin{aligned}
p_{x}=&p_{0}\sqrt{(r{\rm cos}\theta-\frac{D}{\varepsilon_{0y}})\frac{\varepsilon_{0y}}{\varepsilon_{0x}}}
=p_{0}\sqrt{(r{\rm cos}\theta-\frac{D}{\varepsilon_{0y}})\frac{2m_{x}v_{y}}{p_{0}}},\\
p_{y}=&p_{0}r{\rm sin}\theta,\\
\varepsilon_{0x}=&\frac{p_{0}^{2}}{2m_{x}}\approx v_{x}p_{0}=\sqrt{\frac{2|D|}{m_{x}}}p_{0},\\
\varepsilon_{0y}=&v_{y}p_{0}\approx \frac{p_{0}^{2}}{2m_{y}}=\frac{p_{0}^{2}v_{y}^{2}}{2D}.
\end{aligned}
\end{equation}
where $p_{0}$ is the unit momentum which of the order of $a^{-1}$ ($a$ is the lattice constant).
$\theta=\frac{p_{y}v_{y}}{\frac{p_{x}^{2}}{2m_{x}}+D}$.
During the following integrating process, the maximum value of $\theta$ is $\pi/2$ for $D=0$
and ${\rm arccos}(\frac{D}{\varepsilon_{0y}r})$ for $D>0$ and $|D/\varepsilon_{0y}|<r$. 
   %{Semiclassical Boltzmann transport theory of few-layer black phosphorus in various phases}
$v_{y}\sim at_{y}/\hbar$ is the velocity along the $y$-direction with $t_{y}$ the hopping in $y$-direction.
  %{Optical conductivity of multi-Weyl semimetals}
  %{Collective modes in multi-Weyl semimetals support}
  %{Semiclassical Boltzmann transport theory of few-layer black phosphorous in various phases}
    %{Signatures of merging Dirac points in optics and transport}
    %{Dynamical polarization and plasmons in a two-dimensional system with merging Dirac points}
    %{Semiclassical Boltzmann transport theory of few-layer black phosphorous in various phases}
    %{Frequency and orientation dependent conductivity of a Semi-Dirac system}
Note that the approximation in third line is valid for negative $D$,
while that
in the last line is valid only at small momentum with positive $D$.
    %{Dynamical polarization and plasmons in a two-dimensional system with merging Dirac points}
Base on the above two equations, we obtain the band velocity as
  %{Semiclassical Boltzmann transport theory of few-layer black phosphorus in various phases}
\begin{equation} 
\begin{aligned}
v_{x}=&\frac{p_{x}(D+\frac{p_{x}^{2}}{2m_{x}})}{m_{x}(\varepsilon+\mu_{i})},\\
v_{y}=&\frac{v_{y}^{2}p_{y}}{\varepsilon+\mu_{i}}.
\end{aligned}
\end{equation}

The above hamiltonian matrix can be transformed to
\begin{equation} 
\begin{aligned}
H_{i}=&\begin{pmatrix}
0  &  \varepsilon_{0x}\frac{p_{x}^{2}}{p_{0}^{2}}-i\varepsilon_{0y}\frac{p_{y}}{p_{0}}+D\\
\varepsilon_{0x}\frac{p_{x}^{2}}{p_{0}^{2}}+i\varepsilon_{0y}\frac{p_{y}}{p_{0}}+D  &   0
\end{pmatrix}-\mu_{i}\\
=&\varepsilon_{0y}r\begin{pmatrix}
0  &  e^{-i\theta}\\
e^{i\theta}  &   0
\end{pmatrix}-\mu_{i},
\end{aligned}
\end{equation}
whose eigenenergy reads $\varepsilon=\pm\varepsilon_{0y}r-\mu_{i}$ and the corresponding
spinor part of eigenvector can be written as
   %{Semiclassical Boltzmann transport theory of few-layer black phosphorus in various phases}
   %{Collective excitations on a surface of topological insulator}
   %{Plasmon-polaron of the topological metallic surface states}
\begin{equation} 
\begin{aligned}
|\psi\rangle=\frac{1}{\sqrt{2}}\begin{pmatrix}
1\\
\pm e^{\pm i\theta}
\end{pmatrix}.
\end{aligned}
\end{equation}
Then through the Jacobian transformation, we have
\begin{equation} 
\begin{aligned}
dp_{x}dp_{y}=&
\begin{vmatrix}
\frac{\partial p_{x}}{\partial r} & \frac{\partial p_{x}}{\partial \theta}\\
\frac{\partial p_{y}}{\partial r} & \frac{\partial p_{y}}{\partial \theta}
\end{vmatrix}drd\theta\\
=&\begin{vmatrix}
p_{0}\frac{1}{2}\frac{{\rm cos}\theta    \frac{\varepsilon_{0y}}{\varepsilon_{0x}}       }
{\sqrt{({\rm cos}\theta-\frac{D}{\varepsilon_{0y}})\frac{\varepsilon_{0y}}{\varepsilon_{0x}}   }}
&
p_{0}\frac{1}{2}\frac{-{\rm sin}\theta r \frac{\varepsilon_{0y}}{\varepsilon_{0x}}         }
{\sqrt{({\rm cos}\theta-\frac{D}{\varepsilon_{0y}})\frac{\varepsilon_{0y}}{\varepsilon_{0x}}  }}
\\
p_{0}{\rm sin}\theta & p_{0}r{\rm cos}\theta
\end{vmatrix}drd\theta\\
=&\begin{vmatrix}
\xi {\rm cos}\theta & \xi r(-{\rm sin}\theta)\\
p_{0}{\rm sin}\theta & p_{0}r{\rm cos}\theta
\end{vmatrix}drd\theta\\
=&\xi p_{0}rdrd\theta,
\end{aligned}
\end{equation}
    %{Semiclassical Boltzmann transport theory of few-layer black phosphorous in various phases}
    %{Frequency and orientation dependent conductivity of a Semi-Dirac system}
where we define $\xi=p_{0}\frac{1}{2}\frac{    \frac{\varepsilon_{0y}}{\varepsilon_{0x}}       }
{\sqrt{(r{\rm cos}\theta-\frac{D}{\varepsilon_{0y}})\frac{\varepsilon_{0y}}{\varepsilon_{0x}}   }}$.
Then base on the bare Green's function $G^{0}_{p}(\omega)=1/(\omega+i0-\varepsilon)$,
the electronic denisty of states (DOS),
which is an essential quantity for nanosystems,
  %{Tunable spin transport and quantum phase transitions in silicene materials and superlattices}
 can be obtained as
  %{Emergent Anisotropic Non-Fermi Liquid at a Topological Phase Transition in Three Dimensions support}
\begin{equation} 
\begin{aligned}
\rho(\omega)
=&-\frac{1}{(2\pi)^{2}}\int dp_{x}\int dp_{y}\frac{1}{\pi}{\rm Im}{\rm Tr}G^{0}_{p}(\omega)\\
=&\frac{1}{(2\pi)^{2}}\int^{\infty}_{0} dr\int^{\pi/2}_{0} d\theta \xi p_{0}r \delta[\omega-(\pm\varepsilon_{0y}r-\mu_{i})]\\
=&\frac{1}{(2\pi)^{2}}\frac{1.85407p^{2}_{0}
\sqrt{\frac{m_{x}v_{y}}{p_{0}}}
\sqrt{\frac{\pm \mu_{i}\pm \omega}{p_{0}v_{y}}}
\Theta(\frac{\pm\mu_{i}\pm\omega}{p_{0}v_{y}})}
{|p_{0}v_{y}|}\\
=&\frac{1}{(2\pi)^{2}}1.85407
\frac{\sqrt{m_{x}(\pm\mu_{i}\pm \omega)}}{|v_{y}|}
\Theta(\frac{\pm \mu_{i}\pm\omega}{p_{0}v_{y}}),
\end{aligned}
\end{equation}
where $\Theta(x)$ is the Heaviside step function.
Note that the upper (lower) sign corresponds to the positive (negative) impurity energy.
The factor $1.85407$ is related to the complete elliptic integrals of the first kind $K(1/2)$.
   %{Semiclassical Boltzmann transport theory of few-layer black phosphorus in various phases}
   %{Semiclassical Boltzmann transport theory of few-layer black phosphorus in various phases}
   %{A universal Hamiltonian for motion and merging of Dirac points in a two-dimensional crystal}
For positive gap $D>0$ (insulating phase),
\begin{equation} 
\begin{aligned}
\rho(\omega)
=&\frac{1}{(2\pi)^{2}}K(\frac{\omega+\mu_{i}-D}{2(\omega+\mu_{i})})
\frac{\sqrt{m_{x}(\mu_{i}+ \omega)}}{|v_{y}|}
\Theta(\frac{ \mu_{i}+\omega}{p_{0}v_{y}}).
\end{aligned}
\end{equation}
For negative gap (energy offset) with $\omega<-D$,
\begin{equation} 
\begin{aligned}
\rho(\omega)
=&\frac{1}{(2\pi)^{2}}K(\frac{2(\omega+\mu_{i})}{\omega+\mu_{i}-D})
\frac{\sqrt{2}(\mu_{i}+ \omega)}{\sqrt{\mu_{i}+ \omega-D}}
\frac{\sqrt{m_{x}}}{|v_{y}|}
\Theta(\frac{ \mu_{i}+\omega}{p_{0}v_{y}}).
\end{aligned}
\end{equation}

 While for the case that the initial momentum of impurity is along the $x$-direction,
i.e., $\theta=0$,
the DOS can be obained as
\begin{equation} 
\begin{aligned}
\rho(\omega)\equiv \rho_{x}
=&\frac{1}{(2\pi)^{2}}\frac{0.707107p^{2}_{0}
\sqrt{\frac{m_{x}v_{y}}{p_{0}}}
\sqrt{\frac{\pm \mu_{i}\pm \omega}{p_{0}v_{y}}}
\Theta(\frac{\pm\mu_{i}\pm\omega}{p_{0}v_{y}})}
{|p_{0}v_{y}|}\\
=&\frac{1}{(2\pi)^{2}}0.707107
\frac{\sqrt{m_{x}(\pm\mu_{i}\pm \omega)}}{|v_{y}|}
\Theta(\frac{\pm \mu_{i}\pm\omega}{p_{0}v_{y}}),
\end{aligned}
\end{equation}
Through further calculation, we will find that the DOS of impurity whose initial direction (before scattering) is along the $x$-axis
(contributes to the formation of polaron)
is much smaller than the case when the initial direction is closes to the $y$-axis (e.g., $\theta=0.499\pi$).
For different initial polar angle $\theta$ and Dirac mass,
the DOSs are ploted in Fig.2,
and compared to the result obtained by integrating all the possible $\theta$ (blue line; see Eq.(12)).
It is obvious that the DOSs have $\rho(\omega)\propto \frac{\sqrt{m_{x}\omega}}{v_{y}}$,
that is different from the results obtained by low-density approximation as presented below.
  %{Phenomenology of a semi-Dirac semi-Weyl semimetal}
  %{Semiclassical Boltzmann transport theory of few-layer black phosphorus in various phases}
  %{Mean-field quantum phase transition in graphene and in general gapless systems}
  %{New Magnetic Field Dependence of Landau Levels in a Graphenelike Structure}
Also, we found that the DOS is linear with energy in $\omega\ll D$ region.

Since the fermi wave vector $p_{F}$ is the same for both the parabolic and linear dispersions\cite{Hwang E H},
we can savely write the $p_{F}$ of semi-Dirac system as (ignore the degenerate parameter 4 here)
\begin{equation} 
\begin{aligned}
p_{F}=&\sqrt{\pi n_{\downarrow}}\\
=&\sqrt{\pi \int^{\mu_{i}}_{0}d\omega \rho(\omega)}\\
=&\sqrt{\pi 
\left[
\frac{(1.23605 p_{0}^{2} \sqrt{\frac{m_{x} v_{y}}{p_{0}}} 2\mu_{i} \sqrt{\frac{2\mu_{i}}{p_{0} v_{y}}})}
{| p_{0} v_{y}|}
-
\frac{(1.23605 p_{0}^{2} \sqrt{\frac{m_{x} v_{y}}{p_{0}}} \mu_{i} \sqrt{\frac{\mu_{i}}{p_{0} v_{y}}})}
{| p_{0}v_{y}|}
\right]
}\\
=&\sqrt{\pi 
\left[
\frac{(1.23605 p_{0}^{2} \sqrt{\frac{m_{x} v_{y}}{p_{0}}} \mu_{i} \sqrt{\frac{2\mu_{i}}{p_{0} v_{y}}})}
{| p_{0} v_{y}|}
\right]
}\\
\end{aligned}
\end{equation}
Thus the fermi momentum has $p_{F}\propto \frac{\sqrt{m_{x}}}{v_{y}}$.
  %{}
In low carrier density limit, 
$p_{F}a_{B}\propto \sqrt{\pi\frac{\sqrt{m_{x}}}{v_{y}}}\sqrt{\epsilon}{m_{x}e^{2}}=\sqrt{\frac{\pi}{v_{y}}}\frac{\epsilon}{e^{2}}m_{x}^{-3/4}
\ll 1$, where $a_{B}$ is the effective Bohr radius in $k_{x}$ direction and $\epsilon$ is the dielectric constant,
  %{Fermi polaron-polaritons in charge-tunable atomically thin semiconductors support}
  %{Exchange intervalley scattering and magnetic phase diagram of transition metal dichalcogenide monolayers}
and that can be realized by turning up the value of $v_{y}$.
We note that, in perspective of polaron in weak coupling regime,
  %{Normal State of Highly Polarized Fermi Gases: Simple Many-Body Approaches}
the fermi wave vector can also be written as
$p_{F}=\sqrt{2\pi m_{r}\mu_{i}/a_{\psi\phi}}=\sqrt{2\pi m_{r}\Sigma(0,0)/a_{\psi\phi}}$.

At low-temperature, the specific heat of semi-Dirac system can be obtained by the DOS,
which reads (averaging over energy)
  %{Screening-induced temperature-dependent transport in two-dimensional graphene}
  %{A universal Hamiltonian for motion and merging of Dirac points in a two-dimensional crystal}
\begin{equation} 
\begin{aligned}
C_{v}=&\frac{1}{2T^{2}}\int^{\infty}_{0}\frac{d\omega}{2\pi}\frac{\omega^{2}\rho(\omega)}{{\rm cosh}^{2}\frac{\omega}{2T}}\\
=&\frac{0.927037p_{0}^{2}\sqrt{\frac{m_{x}v_{y}}{p_{0}}}\int d\omega \omega^{2}\sqrt{\frac{\mu_{i}+\omega}{p_{0}v_{y}}}
{\rm sech}^{2}(\frac{\omega}{2T})}{T^{2}|p_{0}v_{y}|},
\end{aligned}
\end{equation}
and the specific heat of polaron can be obtained just by replacing the DOS $\rho(\omega)$ with the polaron one.
In low carrier density approximation where the DOS is independent of the quasiparticle energy (see below) in the region
of $\omega>D-\mu_{i}\lesssim 1$,
the above equation becomes
\begin{equation} 
\begin{aligned}
C_{v}=&\frac{1}{2T^{2}}\int^{\infty}_{0}\frac{d\omega}{2\pi}\frac{\omega^{2}\rho }{{\rm cosh}^{2}\frac{\omega}{2T}}\\
%=&\frac{0.927037p_{0}^{2}\sqrt{\frac{m_{x}v_{y}}{p_{0}}}\int d\omega \omega^{2}\sqrt{\frac{\mu_{i}+\omega}{p_{0}v_{y}}}
%{\rm sech}(\frac{\omega}{2T})^{2}}{T^{2}|p_{0}v_{y}|},
=& \frac{\rho (8 T^{3} {\rm Li}_{2}( -e^{-(\omega/T)}) + 
   2 T \omega (-\omega - 4 T {\rm ln}[1 + e^{-(\omega/T)}] + \omega {\rm tanh}[\frac{\omega}{2 T}]))}
   {2 T^{2}}
\bigg|_{\omega=0}^{\omega_{\Lambda}}.
\end{aligned}
\end{equation}
The result of $C_{v}$ in low carrier density approximation is shown in Fig.3 (black line).

In the perspective of temperature-depenent free energy,
the specific heat of semi-Dirac system can also be written as
  %{Emergent Anisotropic Non-Fermi Liquid at a Topological Phase Transition in Three Dimensions}
\begin{equation} 
\begin{aligned}
C_{v}=-T\frac{\partial^{2}F}{\partial T^{2}},
\end{aligned}
\end{equation}
where the free energy reads
\begin{equation} 
\begin{aligned}
F
=&-\frac{T}{V}{\rm ln}  \mathcal{Z}\\
=&-\frac{T}{V}{\rm ln}  {\rm Det}(\frac{1}{T}\frac{1}{G_{0}})\\
=&-\frac{T}{V}{\rm ln}  {\rm Det}(\frac{1}{T}((\omega+i0)\sigma_{0}-H))\\
=&-\frac{T}{V}{\rm ln}  {\rm Det}(\frac{1}{T}((\omega+i0)\sigma_{0}-H))\\
=&-\frac{T}{V}{\rm ln}  {\rm Det}
\begin{pmatrix}
\frac{1}{T}(\omega+i0) & \frac{1}{T}(-\varepsilon_{0y}re^{-i\theta})\\
\frac{1}{T}(-\varepsilon_{0y}re^{i\theta}) & \frac{1}{T}(\omega+i0)\\
\end{pmatrix}
\\
=&-\frac{T}{V}{\rm ln} 
\prod_{\omega}\prod_{r}
[-((\varepsilon_{0y}^{2} r^{2})/T^{2}) + \omega^{2}/T^{2} + (2 i \omega z)/T^{2} - z^{2}/T^{2}]
\\
=&-\frac{T}{V}V\int \frac{\xi p_{0}rdrd\theta}{(2\pi)^{2}}
\sum_{\omega} {\rm ln} 
[
-((\varepsilon_{0y}^{2} r^{2})/T^{2}) + \omega^{2}/T^{2} + (2 i \omega z)/T^{2} - z^{2}/T^{2}
]\\
\approx &-\frac{T}{V}V\int \frac{\xi p_{0}rdrd\theta}{(2\pi)^{2}}
\sum_{\omega} {\rm ln} 
[
\frac{\omega^{2}}{T}-\frac{\varepsilon_{0y}^{2}r^{2}}{T^{2}}
]\\
= &-T\int \frac{\xi p_{0}rdrd\theta}{(2\pi)^{2}}
{\rm ln} 
[-\varepsilon_{0y}^{2}r^{2}T^{-2\omega_{\Lambda}}
\frac{\Gamma(1-\varepsilon_{0y}r-1+\omega_{\Lambda})}{\Gamma(1-\varepsilon_{0y}r)}
\frac{\Gamma(1+\varepsilon_{0y}r-1+\omega_{\Lambda})}{\Gamma(1+\varepsilon_{0y}r)}
].
\end{aligned}
\end{equation}
where $\omega_{\Lambda}$ denotes the energy-cutoff.
Since it is necessary to guarantees the term within the square bracket of above expression is positive,
we apply the eigenvalue of
valence band which with negative energe,
then the free energy can be obtained as
\begin{equation} 
\begin{aligned}
F=&-\frac{T}{V}V\int \frac{\xi p_{0}rdrd\theta}{(2\pi)^{2}}
\sum_{\omega} {\rm ln} 
[
\frac{\omega^{2}}{T^{2}}+\frac{\varepsilon_{0y}^{2}r^{2}}{T^{2}}
]\\
=&-T \int \frac{\xi p_{0}rdrd\theta}{(2\pi)^{2}}
 [\frac{\varepsilon_{0y}r}{T}+2{\rm ln}(1+e^{-\varepsilon_{0y}r/T})+{\rm const.}].
\end{aligned}
\end{equation}
According to the low carry-density approximation as will presented below,
the free energy can be obtained as, for $\theta=0$,
\begin{equation} 
\begin{aligned}
F
=&-\frac{T}{V}V\int \frac{\frac{\sqrt{m_{x}m_{y}}}{m}RdRd\Phi}{(2\pi)^{2}}
\sum_{\omega} {\rm ln} 
\left[
\frac{\omega^{2}}{T^{2}}+\frac{\varepsilon_{0y}^{2}(\frac{1}{\sqrt{2}}R^{2})^{2}}{T^{2}}
\right]\\
=&-\frac{T}{V}V\int \frac{\frac{\sqrt{m_{x}m_{y}}}{m}RdRd\Phi}{(2\pi)^{2}}
\sum_{\omega} {\rm ln} 
\left[
\frac{\omega^{2}}{T^{2}}+\frac{\varepsilon_{0y}^{2}(\frac{1}{\sqrt{2}}R^{2})^{2}}{T^{2}}
\right]\\
=&-\frac{T}{V}V\int \frac{\frac{\sqrt{m_{x}m_{y}}}{m}RdRd\Phi}{(2\pi)^{2}}
[\frac{\varepsilon_{0y}r}{T}+2{\rm ln}(1+e^{-\varepsilon_{0y}r/T})+{\rm const.}].
\end{aligned}
\end{equation}
After subtracting the temperature-independent part of free energy
\begin{equation} 
\begin{aligned}
F
=&-\frac{T}{V}V\int \frac{\frac{\sqrt{m_{x}m_{y}}}{m}RdRd\Phi}{(2\pi)^{2}}
[2{\rm ln}(1+e^{-\varepsilon_{0y}r/T})]\\
=&-T\frac{1}{(2\pi)^{2}}\frac{2\sqrt{m_{x}m_{y}}}{m}
[\frac{R^{2}( \varepsilon_{0y}R^{2}+4T{\rm ln}(1+e^{\frac{-\varepsilon_{0y}R^{2}}{2T}})
                                  -4T{\rm ln}(1+e^{\frac{\varepsilon_{0y}R^{2}}{2T}})
)}{8T}\\
&-
\frac{T {\rm Li}_{2}(-e^{\frac{\varepsilon_{0y}R^{2}}{2T}})}{ \varepsilon_{0y}}]\bigg|_{R=0}^{2^{1/4}\Lambda}.
\end{aligned}
\end{equation}
Then the specific heat is easy to obtained through Eq.(19).
As presented in Fig.3 (blue line),
unlike the specific heat calculated above,
we found that below a certain threshold temperature there is no notable value of specific heat.
Note that when the polaronic effects are taken into account,
the bare Green's function within above expression should be replaced by the dressed one 
$G^{-1}=\omega+i0-\varepsilon-\Sigma(p,\omega)$.
The result of $C_{v}$ in perspective of temperature-dependent free energy is shown in Fig. (blue line).
By comparison,
we see that the $C_{v}$ in perspective of temperature-dependent free energy is zero at low-enough temperature,
while that in the low carier density approximation is linear with temperature at the begining and last for a finite range of $T$,
and both of these two curves converge to a finite nonzero constant in the high-temperature limit.

At high temperature, the specific heat of semi-Dirac system at critical point $D=0$ is the same as that of the normal 3D electron gas, i.e., 
$\frac{3}{2}k_{B}-\mu_{i}$, according to equipartition theorem.
   %{Phenomenology of a semi-Dirac semi-Weyl semimetal}
Unlike both the high-temperature and low-temperature limit,
the low temperature heat capacity\cite{Banerjee S2} per semi-Dirac particle is linear with the 
$C_{v}=\frac{2}{3}\alpha m_{x}k_{B}^{2}T\sqrt{\frac{\omega}{\varepsilon_{0}}}
\approx 0.218504m_{x}k_{B}^{2}T\sqrt{\frac{\omega}{\varepsilon_{0}}}$,
where $\alpha\approx 0.327756$ for integral over the $\theta'$ from 0 to $\pi/2$ as shown in the formula of DOS in above,
and $\varepsilon_{0}=2m_{x}v_{y}^{2}$ is the scaling constant of energy.
  %{Phenomenology of a semi-Dirac semi-Weyl semimetal}
  %{Frequency and orientation dependent conductivity of a Semi-Dirac system}

\section{semiclassical dynamics}

Then the semiclassical conductivity of semi-Dirac quasiparticle can be obtained through the Einstein relation
$\sigma_{xy}=e^{2}\rho(\omega)\xi_{xy}$.
$\xi_{xy}$ is the diffusion coefficient which can be obtained through the autocorrelation function
%\cite{Kinaci A}
  %{On calculation of thermal conductivity from Einstein relation in equilibrium molecular dynamics}
  %{Phenomenology of a semi-Dirac semi-Weyl semimetal}
  %{Semiclassical Boltzmann transport theory of few-layer black phosphorus in various phases}
  %{Signatures of merging Dirac points in optics and transport}
\begin{equation} 
\begin{aligned}
\xi_{xy}=\frac{1}{N}\int^{\tau_{p}}_{0}      (\sum_{i}^{N}\sqrt{v^{i2}_{x}(t)+v^{i2}_{y}(t)}      \sqrt{v^{i2}_{x}(0)+v^{i2}_{y}(0)})dt,
\end{aligned}
\end{equation}
with $\tau_{p}$ the relaxation time and $N$ the number of impurity paritlce.
  %{Absence of damping of low-energy excitations in a quasi-two-dimensional dipolar Bose gas}
  %{Single-particle relaxation time versus transport scattering time in a two-dimensional graphene layer}
  %{Damping of long wavelength collective modes in spinor Bose-Fermi mixtures.}
For short-enough relaxation time,
and with $v_{x}(0)=v_{y}(0)=0$,
  %{Electronic transport in graphene: A semiclassical approach including midgap states}
  %{Acoustic phonon scattering limited carrier mobility in two-dimensional extrinsic graphene}
%the velocity can be treated as time-independent,
we can approximately obtain
\begin{equation} 
\begin{aligned}
\xi_{xy}=\frac{1}{2}(v_{x}^{2}+v_{y}^{2})\tau_{p}=\frac{1}{2}v^{2}\tau_{p},
\end{aligned}
\end{equation}
where $v$ is the anisotropic in-plane velocity.
  %{Phenomenology of a semi-Dirac semi-Weyl semimetal}
Since we focus on the $x$-direction nonrelativistic momentum of the semi-Dirac particle,
the longitudinal conductivity can be obtained as
\begin{equation} 
\begin{aligned}
\sigma_{xx}=e^{2}\rho(\omega)\frac{1}{2}v_{x}^{2}\tau_{p}.
\end{aligned}
\end{equation}
Note that while for conductivity at fermi energy, it reads $\sigma_{xx}=e^{2}\rho(E_{F})\frac{1}{2}v_{x}^{2}\tau_{p_{F}}$.

To obtain the conductivity of polaron, the DOS and relaxation time must be replaced by the polaron ones,
and the band velocity $v_{x}$ and $v_{y}$ must be evaluated from the polaron dispersion (modified by the self-energy effect).
For polaron formation,
the transport scattering time $\tau_{p}$ at zero temperature reads
(we assume both the excited electron and impurity are above the fermi surface, as evidented by Eq.(40))
\begin{small}
\begin{equation} 
\begin{aligned}
\tau_{p}=&\frac{2\pi N}{\hbar S}\sum_{k}(1-{\rm cos}(\theta'-\theta))
\delta[\omega-(\frac{p_{'x}^{2}}{2m_{x}}+D-\mu_{i})
-(\frac{(k-q)^{2}}{2m}-\mu_{m})]
|g_{b}|^{2}|\langle p+q-k|p \rangle|^{2}\\
=&\frac{2\pi N}{\hbar S}\sum_{k}(1-{\rm cos}(\theta'-\theta))
\delta[-2\mu_{i}-4\mu_{m}+\frac{k^{2}p_{0}^{2}}{m}-\frac{2kp_{0}^{2}q}{m}+\frac{p_{0}^{2}q^{2}}{m}-2\omega]
|g_{b}|^{2}{\rm cos}^{2}(\frac{\theta'-\theta}{2})\\
=&\frac{1}{\hbar S}8.88577|g_{b}|^{2}m_{x}Np_{0}v_{y}(1-{\rm cos}(\theta-\theta'))
{\rm cos}^{2}(\frac{\theta'-\theta}{2})\\
&\int^{\Lambda}_{0}dr_{k}\int d\theta'
\frac{r_{k}
\delta[-2\mu_{i}-4\mu_{m}+\frac{r_{k}^{2}p_{0}^{2}}{m}-\frac{2r_{k}p_{0}^{2}q}{m}+\frac{p_{0}^{2}q^{2}}{m}
-2r_{k}p_{0}v_{y}+2pp_{0}v_{y}+2p_{0}qv_{y}
-2\omega]}
{\sqrt{\frac{m_{x}v_{y}(-r_{k}+r_{p}+r_{q}-\frac{D}{p_{0}v_{y}})}{p_{0}}}}\\
=&
-((0.55536 g_{b}^{2} N p_{0}^{3} v_{y} ((p_{0} q + m v_{y})/p_{0} - \sqrt{(
        2 m \mu_{i} + 4 m \mu_{m} - 2 m p p_{0} v_{y} + m^{2} v_{y}^{2} + 2 m \omega)/
        p_{0}^{2}}) \\
		&\sqrt{-((
       m_{x} (D - p p_{0} v_{y} - p_{0} q v_{y} + 
          p_{0} v_{y} ((p_{0} q + m v_{y})/p_{0} - \sqrt{(
             2 m \mu_{i} + 4 m \mu_{m} - 2 m p p_{0} v_{y} + m^{2} v_{y}^{2} + 2 m \omega)/
             p_{0}^{2}})))/p_{0}^{2})}\\
       &{\rm Boolean}[0 < (p_{0} q + m v_{y})/p_{0} - \sqrt{(
          2 m \mu_{i} + 4 m \mu_{m} - 2 m p p_{0} v_{y} + m^{2} v_{y}^{2} + 2 m w)/p_{0}^{2}} < 
         \Lambda \\
&|| \Lambda < (p_{0} q + m v_{y})/p_{0} - \sqrt{(
          2 m \mu_{i} + 4 m \mu_{m} - 2 m p p_{0} v_{y} + m^{2} v_{y}^{2} + 2 m w)/p_{0}^{2}} < 
         0]\\
&		 (2 \theta' + {\rm sin}(2 (\theta - \theta'))))/(\hbar S (D - p p_{0} v_{y} - 
        p_{0} q v_{y} + 
        p_{0} v_{y} ((p_{0} q + m v_{y})/p_{0} \\
&- \sqrt{(
           2 m \mu_{i} + 4 m \mu_{m} - 2 m p p_{0} v_{y} + m^{2} v_{y}^{2} + 2 m \omega)/
           p_{0}^{2}}))\\
&		   {\rm Abs}[(
       p_{0}^{2} q + m p_{0} v_{y} - 
        p_{0}^{2} ((p_{0} q + m v_{y})/p_{0} \\
&- \sqrt{(
           2 m \mu_{i} + 4 m \mu_{m} - 2 m p p_{0} v_{y} + m^{2} v_{y}^{2} + 2 m \omega)/
           p_{0}^{2}}))/m]))\big|^{\theta'_{max}}_{0},
%\frac{2\pi N}{\hbar S}\int^{\Lambda}_{0}dr_{k}\int d\theta'\xi p_{0}r_{k}
% (1-{\rm cos}(\theta'-\theta))
%\delta[-2\mu_{i}-4\mu_{m}+\frac{k^{2}p_{0}^{2}}{m}-\frac{2kp_{0}^{2}q}{m}+\frac{p_{0}^{2}q^{2}}{m}-2\omega]
%|g_{b}|^{2}{\rm cos}^{2}(\frac{\theta'-\theta}{2})\\
\end{aligned}
\end{equation}
\end{small}
where $\delta[\cdot\cdot\cdot]$ is the Dirac-$\delta$ function,
and ${\rm Abs}[\cdot\cdot\cdot]$ denotes the absolute value.
Note that the scattering time here is unrelated to the Landau damping due to the zero-temperature limit.
Also, we note that the particle-particle scattering amplitude is independent of temperature in such limit.
  %{Damping in 2D and 3D dilute Bose gases}
In fact, the formula in above is different with both the Beliaev damping and Laudau damping\cite{Chung M C,Liu W V}
  %{Damping in 2D and 3D dilute Bose gases}
  %{Theoretical Study of the Damping of Collective Excitations in a Bose-Einstein Condensate}
due to the existence of impurity frequency term (before scattering) $\omega$ within the $\delta$-function.
But when the thermal excitations are considered
(like the positive band gap in insulator phase), 
  %{Semiclassical Boltzmann transport theory of few-layer black phosphorus in various phases}
the Laudau damping is possible as the momentum of polaron damping back to $p$.
  %{Absence of damping of low-energy excitations in a quasi-two-dimensional dipolar Bose gas}
  %{Damping of Long-Wavelength Collective Modes in Spinor Bose-Fermi Mixtures}
While the Beliaev damping requires the energy conservation and mainly happen
for $|{\bf p}|\approx |{\bf p'}|\approx |{\bf (q-k)}|$.
  %{Damping in 2D and 3D dilute Bose gases}
  %{Fate of the Bose polaron at finite temperature}

From this Dirca-$\delta$ function, we can easily know that the scattering here is inelastic since
the energy of impurity changed by the scattering.
  %{Vertex corrections to the dc conductivity in anisotropic multiband systems}
$\theta'_{max}$ denotes the maximum value of the possible $\theta'$
which depends on the value of $r_{k}$.
The above expression, in the case of large $\Lambda$,
is valid under the condition
$(p_{0} (m v_{y} + 
      p_{0} (q - \sqrt{\frac{(m (2 \mu_{i} + 4 \mu_{m} - 2 p p_{0} v_{y} + m v_{y}^{2} + 2 w))}
         {p_{0}^{2}}})) > 0$
and
$  q + (m v_{y})/p_{0} < 
   \Lambda + \sqrt{\frac{(m (2 \mu_{i} + 4 \mu_{m} - 2 p p_{0} v_{y} + m v_{y}^{2} + 2 w))}{p_{0}^{2}}})$.
Note that here we assume the hole is static with $q=0$, and
we ignore approximately apply $r_{p'}=r_{p}-r_{k}$,
which ignores the effect of angle $\theta''$ between $p$ and scattering wave vector $q-k$:
$\theta''$ can be obatined by the relation $\theta'={\rm arccos}\frac{p+(q-k){\rm cos}\theta''}{\sqrt{p^{2}+(q-k)^{2}+2p(q-k){\rm cos}\theta''}}$.
  %{Frequency-dependent polarizability, plasmons, and screening in the two-dimensional pseudospin-1 dice lattice}
  %{Collective modes in multi-Weyl semimetals support}
  %{Dynamic screening and low-energy collective modes in bilayer graphene}
 %{Dynamical Screening and Excitonic Instability in Bilayer Graphene}
 %{Dynamical screening in bilayer graphene}
This transport scattering time is different from the single particle relaxation time because it is related to the pairing scattering as well as
the scattering wave vector.
When consider the effect of finite temperature,
the transport scattering time 
reads
\begin{equation} 
\begin{aligned}
\tau_{p}=&\frac{2\pi N}{\hbar S}\sum_{k}(1-{\rm cos}(\theta'-\theta))(1-N_{F}(\varepsilon_{p})-N_{F}(\varepsilon_{p-k}))\\
&\delta[\omega-(\frac{p_{x}^{2}}{2m_{x}}+D-\mu_{i})
-(\frac{(k-q)^{2}}{2m}-\mu_{m})]
|g_{b}|^{2}|\langle p+q-k|p \rangle|^{2}\\
=&\frac{2\pi N}{\hbar S}\sum_{k}(1-{\rm cos}(\theta'-\theta))(1-N_{F}(\varepsilon_{p}))\frac{1-N_{F}(\varepsilon_{p-k})}{1-N_{F}(\varepsilon_{p})}\\
&\delta[-2\mu_{i}-4\mu_{m}+\frac{k^{2}p_{0}^{2}}{m}-\frac{2kp_{0}^{2}q}{m}+\frac{p_{0}^{2}q^{2}}{m}-2\omega]
|g_{b}|^{2}{\rm cos}^{2}(\frac{\theta'-\theta}{2}),
\end{aligned}
\end{equation}
where we still assume $q=0$, and in second line we use the relation
$1-N_{F}(\varepsilon_{k})-N_{F}(\varepsilon_{p-k})=(1-N_{F}(\varepsilon_{k}))\frac{1-N_{F}(\varepsilon_{p-k})}{1-N_{F}(\varepsilon_{p})}$
  %{Vertex corrections to the dc conductivity in anisotropic multiband systems}
which is correct in large $k$ and $p$ region as well as the low-temperature case.
$N$ and $S$ denotes the total number of impurity particles and the number of unit cells, respectively.
Note that the $N_{F}$ here is equilibrium Fermi-Dirac distribution function, i.e., without affected by the effective electric field
and thus $\frac{\partial N_{F}'}{\partial t}=e{\bf E}\cdot {\bf v}(-\frac{\partial N_{F}}{\partial \varepsilon})=0$\cite{Zhang X G,Stauber T}.
  %{Vertex corrections to the dc conductivity in anisotropic multiband systems}
  %{Semiclassical Boltzmann transport theory of few-layer black phosphorus in various phases}
  %{First-principles method for electron-phonon coupling and electron mobility Applications to two-dimensional materials}
  %{Conductivity of metallic films and multilayers}

The above transport scattering time satisfy the linearized Boltzmann transport equation
\begin{equation} 
\begin{aligned}
1=&\sum_{k}\left(
w_{pp'}\frac{1-N_{F}(\varepsilon_{p'})}{1-N_{F}(\varepsilon_{p})}\tau_{p}
-w_{p'p}\frac{1-N_{F}(\varepsilon_{p})}{1-N_{F}(\varepsilon_{p'})}\tau_{p'}{\rm cos}\theta_{pp'}
\right)\\
=&\sum_{k}\left(w_{pp'}\frac{1-N_{F}(\varepsilon_{p'})}{1-N_{F}(\varepsilon_{p})}\tau_{p}
-w_{pp'}\frac{N_{F}(\varepsilon_{p})}{N_{F}(\varepsilon_{p'})}
\frac{1-N_{F}(\varepsilon_{p'})}{1-N_{F}(\varepsilon_{p})}
\frac{1-N_{F}(\varepsilon_{p})}{1-N_{F}(\varepsilon_{p'})}\tau_{p'}\frac{p\cdot p'}{|p||p'|}
\right)\\
=&\sum_{k}w_{pp'}\frac{1-N_{F}(\varepsilon_{p'})}{1-N_{F}(\varepsilon_{p})}
\left[\tau_{p}
-\frac{N_{F}(\varepsilon_{p})}{N_{F}(\varepsilon_{p'})}
\frac{1-N_{F}(\varepsilon_{p})}{1-N_{F}(\varepsilon_{p'})}\tau_{p'}\frac{p\cdot p'}{|p||p'|}\right]\\
=&\sum_{k}w_{pp'}\frac{1-N_{F}(\varepsilon_{p'})}{1-N_{F}(\varepsilon_{p})}
\left[\tau_{p}
-\frac{N_{F}(\varepsilon_{p})}{N_{F}(\varepsilon_{p'})}
\frac{1-N_{F}(\varepsilon_{p})}{1-N_{F}(\varepsilon_{p'})}\tau_{p'}
\frac{v_{p'}\cdot v_{p}}{|v_{p}|^{2}}
\right]\\
=&\sum_{k}w_{pp'}\frac{1-N_{F}(\varepsilon_{p'})}{1-N_{F}(\varepsilon_{p})}
\left[\tau_{p}
-\frac{N_{F}(\varepsilon_{p})}{N_{F}(\varepsilon_{p'})}
\frac{1-N_{F}(\varepsilon_{p})}{1-N_{F}(\varepsilon_{p'})}\tau_{p'}
\frac{|v_{p'}|{\rm cos}\theta_{pp'}}{|v_{p}|}
\right]\\
\approx &\sum_{k}w_{pp'}\frac{1-N_{F}(\varepsilon_{p'})}{1-N_{F}(\varepsilon_{p})}
\left[\tau_{p}
-\frac{N_{F}(\varepsilon_{p})}{N_{F}(\varepsilon_{p'})}
\frac{1-N_{F}(\varepsilon_{p})}{1-N_{F}(\varepsilon_{p'})}\tau_{p}
\frac{v_{p'}\cdot v_{p}}{|v_{p'}||v_{p}|}
\right]\\
= &\sum_{k}w_{pp'}\frac{1-N_{F}(\varepsilon_{p'})}{1-N_{F}(\varepsilon_{p})}
\left[\tau_{p}
-\frac{N_{F}(\varepsilon_{p})}{N_{F}(\varepsilon_{p'})}
\frac{1-N_{F}(\varepsilon_{p})}{1-N_{F}(\varepsilon_{p'})}\tau_{p}
{\rm cos}\theta_{pp'}
\right],
\end{aligned}
\end{equation}
where we define the group velocity (band velocity) vector as $v_{p}=v_{x}+v_{y}$ and the approximation in last two lines 
we apply the relation $\tau_{p}|v_{p}|=\tau_{p'}|v_{p'}|$.
We write the scattering rate in Born approximation as
\begin{equation} 
\begin{aligned}
w_{pp'}=\frac{2\pi}{\hbar}|g_{b}|^{2}(1-N_{F}(\varepsilon_{p}))\delta\left[\omega-(\frac{p_{x}^{2}}{2m_{x}}+D-\mu_{i})
-(\frac{(k-q)^{2}}{2m}-\mu_{m})\right],
\end{aligned}
\end{equation}
and for inelastic scattering the detailed balance\cite{Kawamura T,Gunst T}
$w_{pp'}N_{F}(\varepsilon_{p})(1-N_{F}(\varepsilon_{p'}))=
w_{p'p}N_{F}(\varepsilon_{p'})(1-N_{F}(\varepsilon_{p}))$ should be obeyed.
  %{Phonon-scattering-limited electron mobilities in Al Gal As GaAs heterojnnctions}
  %{Vertex corrections to the dc conductivity in anisotropic multiband systems}
  %{First-principles method for electron-phonon coupling and electron mobility:Applications to two-dimensional materials}
  %{Semiclassical Boltzmann transport theory of few-layer black phosphorus in various phases}
  %{Acoustic phonon scattering limited carrier mobility in two-dimensional extrinsic graphene}
  %{Boltzmann transport and residual conductivity in bilayer graphene}
  %{Screening-induced temperature-dependent transport in two-dimensional graphene}
Through the above expression, we can also obtain 
\begin{equation} 
\begin{aligned}
{\rm cos}\theta_{pp'}
\frac{N_{F}(\varepsilon_{p})}{N_{F}(\varepsilon_{p'})}
\frac{1-N_{F}(\varepsilon_{p})}{1-N_{F}(\varepsilon_{p'})}=\frac{|v_{p'}|}{|v_{p}|}=\frac{|\tau_{p}|}{|\tau_{p'}|}.
\end{aligned}
\end{equation}
  %{First-principles method for electron-phonon coupling and electron mobility:Applications to two-dimensional materials}
  %{Phonon-scattering-limited electron mobilities in Al Gal As/GaAs heterojnnctions}
  %{Vertex corrections to the dc conductivity in anisotropic multiband systems}
In the presence of external potential which can be expressed in a electric field form $(-e){\bf E}=-\nabla U$ ($e>0$),
  %{Anomalous equilibrium currents for massive Dirac electrons}
  %{Conductivity of metallic films and multilayers}
  %{Screening-induced temperature-dependent transport in two-dimensional graphene}
  %{density and current response functions in strongly disordered electron systems: diffusion, electrical conductivity and einstein relation}
the time derivative of the nonequilibrium distribution function reads
\begin{equation} 
\begin{aligned}
\frac{dN'_{F}}{dt}&=\frac{d[N_{F}+\tau_{p}(-e)({\bf E}\cdot{\bf v})\frac{\partial N_{F}}{\partial \varepsilon}]}{dt}\\
&=(-e){\bf E}\cdot\frac{\partial N_{F}}{\partial {\bf p}}-e({\bf E}\cdot{\bf v})\frac{\partial N_{F}}{\partial \varepsilon}\\
&=-2e({\bf E}\cdot{\bf v})\frac{\partial N_{F}}{\partial \varepsilon},
%\int\frac{d^{2}p}{(2\pi)^{2}}e^{2}v^{2}|{\bf E}|{\rm cos}\phi(\frac{\partial N_{F}(\varepsilon_{p})}{\partial \varepsilon})\tau_{p},
\end{aligned}
\end{equation}
and the dc transport of electrical currents can be obtained as
\begin{equation} 
\begin{aligned}
J
&=\int\frac{d^{2}p}{(2\pi)^{2}}e^{2}v{\bf E}\cdot{\bf v}(\frac{\partial N_{F}(\varepsilon_{p})}{\partial \varepsilon})\tau_{p}\\
&=\int\frac{d^{2}p}{(2\pi)^{2}}e^{2}v^{2}|{\bf E}|{\rm cos}\beta(\frac{\partial N_{F}(\varepsilon_{p})}{\partial \varepsilon})\tau_{p}.
\end{aligned}
\end{equation}
where $\beta$ here is the angle between directions of electrin field and electrical current.
  %{Anomalous equilibrium currents for massive Dirac electrons}
The electrical currents can be related to the dc conductivity as 
\begin{equation} 
\begin{aligned}
J
&={\pmb \sigma}\cdot{\bf E}e\\
&=\frac{e^{2}v^{2}}{2}\rho(\omega)\tau_{p}|{\bf E}|e{\rm cos}\beta\\
&=\frac{e^{2}v^{2}}{2}\rho(\omega)\tau_{p}ipU{\rm cos}\beta,
\end{aligned}
\end{equation}
where we also have ${\bf E}=\frac{1}{e}\nabla U=\frac{1}{e}i{\bf k}U$,
$|{\pmb \sigma}|=\frac{e^{2}v^{2}}{2}\rho(\omega)\tau_{p}{\rm cos}\beta=ev^{2}
    \frac{\partial N_{F}}{\partial \varepsilon_{p}}\tau_{p}{\rm cos}\beta$,
and $\frac{e}{2}\rho(\omega)=\frac{\partial N_{F}}{\partial \varepsilon_{p}}$.
The dc conductivity presented here is also consistent wih the result\cite{Burkov A A} $\sigma=e^{2}p_{F}^{2}\ell$
where $\ell=v\tau_{p}\gg 1$ is the mean-free path in semiclassical limit and $p_{F}\propto\sqrt{\rho(\omega)\frac{v}{2}}$
is in agree with the Eq.(16).

To study the Hall current, we can apply the static electric field along $x$-direction,
  %{Suppression of the persistent spin Hall current by defect scattering}
which will leads to Hall current $J_{y}(x)=(-e)v\psi^{\dag}(x)\sigma_{y}\psi(x)$
with $\psi(x)=\Psi(x,y)e^{-ip_{y}y}$ the plane wave solution of Hamiltonian Eq.(3) ($p_{y}$ can be treated invariant here),
  %{Anomalous equilibrium currents for massive Dirac electrons}
  %{Electron collimation at van der Waals domain walls in bilayer graphene}
and that also related to the Hall transition conductivity (in real space) by the Wick's theorem
%{Anomalous Hall effect Naoto Nagaosa}Ohm's law
  %{Integer quantum Hall transition An alternative approach and exact results}
  %{Observation of three-dimensional massless Kane fermions in a zinc-blende crystal}
\begin{equation} 
\begin{aligned}
\sigma_{xy}=\frac{J_{y}(x)}{eE_{x}}
=\frac{i}{e\Omega_{x}}\int\frac{d\omega}{2\pi}\int\frac{d^{2}p}{(2\pi)^{2}}
{\rm Tr}[\sigma_{y}G_{0}(p,\omega)G_{0}(p,\omega+\Omega_{x})],
\end{aligned}
\end{equation}
where $\Omega_{x}$ corresponds to the energy induced by the electric field,
  %{Integer quantum Hall transition An alternative approach and exact results}
  %{Signatures of merging Dirac points in optics and transport}
  %{Charge Transport in Weyl Semimetals}
  %{Dynamical conductivity of AA-stacked bilayer graphene}
  %{Suppression of the persistent spin Hall current by defect scattering}
  %{Optical conductivity of black phosphorus with a tunable electronic structure}
and the effect of

The Hall conductivity at finite temperature can also be obtained by the Kubo formula
\begin{equation} 
\begin{aligned}
\sigma_{xy}(\Omega_{x})
&=\frac{e^{2}}{\Omega_{x}}\int\frac{d\omega}{2\pi}[N_{F}(\omega)-N_{F}(\omega+\Omega_{x})]
\int\frac{d^{2}p}{(2\pi)^{2}}{\rm Tr}[v_{x}\hat{A}(p,\omega)v_{y}\hat{A}(p,\omega+\Omega_{x})]\\
&=\frac{e^{2}}{\Omega_{x}}\int\frac{d\omega}{2\pi}[N_{F}(\omega)-N_{F}(\omega+\Omega_{x})]
\int\frac{d^{2}p}{(2\pi)^{2}}\\
&\frac{(\delta(\omega-\mu_{i}-\varepsilon_{p})+\delta(\omega-\mu_{i}+\varepsilon_{p}))
      (\delta(\omega+\Omega_{x}-\mu_{i}-\varepsilon_{p})+\delta(\omega+\Omega_{x}-\mu_{i}+\varepsilon_{p}))}{4\varepsilon_{p}^{2}}\\
&[
 (\frac{p_{x}^{2}}{2m_{x}}+D+iv_{y})(\frac{p_{x}^{2}}{2m_{x}}+D-iv_{y}p_{y})^{2}(\frac{p_{x}}{m_{x}}+iv_{y}p_{y}+D)\\
&+(\frac{p_{x}^{2}}{2m_{x}}+D-iv_{y})(\frac{p_{x}}{m_{x}}+D-iv_{y}p_{y})(\frac{p_{x}^{2}}{2m_{x}}+iv_{y}p_{y}+D)^{2}\\
&+(\frac{p_{x}^{2}}{2m_{x}}+D+iv_{y})(\frac{p_{x}}{m_{x}}+D-iv_{y}p_{y})(-\mu_{i}+\omega)(-\mu_{i}+\omega+\Omega_{x})\\
&+(\frac{p_{x}^{2}}{2m_{x}}+D-iv_{y})(\frac{p_{x}}{m_{x}}+D+iv_{y}p_{y})(-\mu_{i}+\omega)(-\mu_{i}+\omega+\Omega_{x})
],
\end{aligned}
\end{equation}
where the velocity matrices reads $v_{x/y}=\frac{\partial H}{\hbar\partial p_{x/y}}=\pm v\sigma_{x/y}$
  %{Integer and half-integer quantum Hall effect in silicene: Influence of an external electric field and impurities}
  %{Valley polarized Quantum Hall effect and topological insulator phase transitions in silicene}
Here $\hat{A}(p,\omega)$ is the spectral density matrix
which is related to the matrix of Green's function by 
$G(p,\omega+i\eta)=\int^{\infty}_{-\infty}\frac{d\omega}{2\pi}\lim_{\eta\rightarrow 0}\frac{A(p,\omega)}{i\eta}$.
  %{Dynamical conductivity of AA-stacked bilayer graphene}
  %{Optical conductivity of bilayer graphene with and without an asymmetry gap}
  %{Transport of Dirac quasiparticles in graphene: Hall and optical conductivities}
Note that we assume the effect of electric field provides a photon-like energy and leads to a vertical transition
in the dispersion (with zero relativa momentum).
  %{Dynamical conductivity of AA-stacked bilayer graphene}
  %{Simulations of vibronic profiles in two-photon absorption}
  %{Infrared and Raman spectra of AA-stacking bilayer graphene}

Note that the spectral function reads
\begin{equation} 
\begin{aligned}
A(p,\omega)=&-\frac{1}{\pi}{\rm Tr}{\rm Im}G_{0}(p,\omega)\\
=&-\frac{1}{\pi}{\rm Tr}{\rm Im}\frac{1}{\omega+i0-H}\\
=&-\frac{1}{\pi}{\rm Tr}{\rm Im}
\begin{pmatrix}
-\mu_{i}+\omega+i0 & \frac{p_{x}^{2}}{2m_{x}}-iv_{y}p_{y}+D\\
\frac{p_{x}^{2}}{2m_{x}}+iv_{y}p_{y}+D  & -\mu_{i}+\omega+i0
\end{pmatrix}^{-1}\\
=&\frac{-1}{\pi}{\rm Tr}\frac{-\pi{\rm sgn}[\omega]}{2\varepsilon_{p}}
(\omega+H)[\delta(\omega-\mu_{i}-\varepsilon_{p})+\delta(\omega-\mu_{i}+\varepsilon_{p})]\\
=&\frac{1}{\pi}
\frac{\pi{\rm sgn}[\omega](\delta(\omega-\mu_{i}-\varepsilon_{p})+\delta(\omega-\mu_{i}+\varepsilon_{p}))(\omega-\mu_{i})}{\varepsilon_{p}},
\end{aligned}
\end{equation}
and for nonzero spectral function, the above formula can be reduced to
\begin{equation} 
\begin{aligned}
A(p,\omega)
=&\frac{1}{\pi}
\frac{\pi{\rm sgn}[\omega](\delta(\omega-\mu_{i}-\varepsilon_{p})+\delta(\omega-\mu_{i}+\varepsilon_{p}))(\omega-\mu_{i})}{\varepsilon_{p}}\\
=&\delta(\omega-\mu_{i}-\varepsilon_{p})+\delta(\omega-\mu_{i}+\varepsilon_{p}).
\end{aligned}
\end{equation}
%&=\frac{e^{2}}{\Omega_{x}}\int\frac{d\omega}{2\pi}[N_{F}(\omega)-N_{F}(\omega+\Omega_{y})]
%\int\frac{d^{2}p}{(2\pi)^{2}}{\rm Tr}[v_{x}{\rm Im}G(p,\omega+i0)v_{y}{\rm Im}G(p,\omega+\Omega_{y}+i0)]\\
%&=\frac{e^{2}}{\Omega_{x}}\int\frac{d\omega}{2\pi}[N_{F}(\omega)-N_{F}(\omega+\Omega_{y})]
%\int\frac{d^{2}p}{(2\pi)^{2}}
%\sum_{\alpha,\alpha'=v,c}\langle \alpha|v_{x}|\alpha'\rangle \langle \alpha'|v_{y}|\alpha\rangle
%\frac{1}{2}
%[\delta(\omega+\frac{\Omega_{y}}{2})\delta(\omega+\frac{\Omega_{y}}{2})
%+\delta(\omega-\frac{\Omega_{y}}{2})\delta(\omega+\frac{\Omega_{y}}{2})]
%\end{aligned}
%\end{equation}
where
  %{Electron collimation at van der Waals domain walls in bilayer graphene}
  %{Anomalous equilibrium currents for massive Dirac electrons}
  %{Conductivity of metallic films and multilayers}
 %{Signatures of merging Dirac points in optics and transport}
  %{density and current response functions in strongly disordered electron systems: diffusion, electrical conductivity and einstein relation}
  %{Integer quantum Hall transition An alternative approach and exact results}
  %{Emergent Anisotropic Non-Fermi Liquid at a Topological Phase Transition in Three Dimensions}
and $\omega$ is the Matsubara frequency.
In terms of the sum over scattering states, the imaginary part of Green's function can also be expressed as
 ${\rm Im}G(\omega+i0)=-\pi\sum_{\alpha=v,c}|\alpha\rangle\langle\alpha|\delta(\omega-\varepsilon_{p\alpha})$,
  %{Conductivity of metallic films and multilayers}
where $v,c$ denote the quantum states in valence band and conduction band, respectively.
Through this relation, the conductivity can also be expressed by the response of current to the vector potential:
\cite{wums18,Ahn S,Orlita M}
  %{Optical conductivity of multi-Weyl semimetals}
  %{Observation of three-dimensional massless Kane fermions in a zinc-blende crystal}
\begin{equation} 
\begin{aligned}
\sigma_{xy}=\frac{-ie^{2}}{\hbar}\int\frac{d^{2}p}{(2\pi)^{2}}\sum_{\alpha,\alpha'=v,c}
\frac{N_{F}(\omega)-N_{F}(\omega+\Omega_{y})}{\varepsilon_{p\alpha}-\varepsilon_{p\alpha'}}
\frac{\langle \alpha|v_{x}|\alpha'\rangle \langle \alpha'|v_{y}|\alpha\rangle}
{\omega+i0+\varepsilon_{p\alpha}-\varepsilon_{p\alpha'}}.
\end{aligned}
\end{equation}
Note that, since we consider the finite chemical potential,
the interband transition has a gap as large as $2\mu_{i}$ (for monolayer) due to the Pauli blocking,
and thus the vertical transition requires $\Omega_{x}>2\mu_{i}$,
while the intraband transition with a Drude peak in zero limit of $\Omega_{x}$.
  %{Optical properties of a semi-Dirac material}
  %{Optical conductivity of multi-Weyl semimetals}
  %{Dynamical conductivity of AA-stacked bilayer graphene}

The broadened spectral function due to the quasiparticle scattering rate can be expressed by the Lorentzian representation
 $\delta(\omega-\mu_{i}-\varepsilon_{p})=\frac{\frac{1}{\pi}\frac{1}{2\tau_{p}}}{(\omega-\mu_{i}-\varepsilon_{p})^{2}+\frac{1}{2\tau_{p}}}$
where ${\rm Re}\Sigma=0$ and ${\rm Im}\Sigma=\Gamma=2\eta=\frac{1}{\tau_{p}}$ where $\Gamma$ is the transport scattering rate
amd $\eta$ is the quasiparticle scattering rate.
  %{Dynamical conductivity of AA-stacked bilayer graphene}
  %{Signatures of merging Dirac points in optics and transport}

\section{Polaronic effects: pair propagator and self-energy}

\subsection{Potential and the polaronic interspecies coupling}

We consider the contact interaction (zero-range potential) 
within the environment with low fermi-energy and fermi wave vector\cite{Wu1182}, like the $\delta$-type impurity field, 
  %{fermi polaron-polaritons in charge-tunable atomically thin semiconductors}
which can also be replaced by the Gaussian broadening 
(due to the scattering effect at finite temperature, like the cases with thermal de Broglie wavelength).
  %{Bose–Einstein condensation of exciton polaritons}
%{spin-dependent polaron formation in pristine graphene}
%{evidence for majorana bound states in an iron-based superconductor}
This broadening is more observable in the solid state like the Dirac/Weyl systems, 
as evidened by the measured conductivity as well as the resonance spectrum.
Due to the presence of short-range interaction (as widely observed in the low-density regime of the fermi gases), the long-range Coulomb potential as well as the frozen ripple scattering is reduced, 
%{Resonant Scattering by Realistic Impurities in Graphene}
while the resonance scattering is more dominating as in presence of, e.g., the ion irradiation\cite{Chen J H} (resonance impurity). 
%{Diffusion and Criticality in Undoped Graphene with Resonant Scatterers}
 Even in the absence of artificial irradiation, 
the scattering mechanism analogous to the Fano-Feshbach resonance is possible in the solid state systems 
base on the virtual transition as discussed in the graphene\cite{Gaul C}.
The impurity effect and the related electronic transport can be detected by the resonance scattering, 
which means that the $s$-wave scattering length $a$ can also be used in our calculation for the solid state (Dirac) systems. 
And since we ignore the resonance scattering effect, the interspecies scattering length can be referred to the background scattering length.

For zero-range model in mean-field approximation, 
the inversed interspecies coupling $g_{\psi\phi}^{-1}$ and the pair propagator $\Pi(p+q,\omega+\Omega)$ are ultraviolet divergent
(restricted by a momentum cutoff scale $\Lambda$),
in the absence of the renormalization-term---$\frac{2m_{r}}{k^{2}}$.
%the unusual bare scalar potential propagator $D_{0}=
  %{variational study of polarons in Bose-Einstein condensates}
  %{polaron and dressed molecules near narrow feshbach resonances}
The resonance scattering is detectable in the vacancies of the lattice system, 
where the on-site potential is infinitely strong and with the divergent scattering length $a\rightarrow\infty$\cite{Ostrovsky P M},
%{Resonant Scattering by Realistic Impurities in Graphene}
that means the scattering length is of the order of inversed potential range $R^{-1}$,
similar to the momentum cutoff $\Lambda$.
That can also easily be verified numberically 
that, the propagation of the traveling plane wave state (in momentum space) will destroyed by the strong on-site potential of a single vacancy,
%{Diffusion and Criticality in Undoped Graphene with Resonant Scatterers}
%{Transport of spin-orbit coupled Bose–Einstein condensates in lattice with defects}
%{normal state of highly polarized fermi gases: simple many-body approximation}
except when there is a strong enough interspecies interaction 
(between the particles with opposite spins) which with a longer scattering length than the on-site impurity one.
That will be particularly obvious for the Anderson localization with the disordered media\cite{Asatryan A A}.
These are different to the cases where the long-range interactions are taken into consideration and play a main role\cite{Pocock S R}.
For the bare interspecies coupling constant is bound to $g_{\psi\phi}\rightarrow 0^{-}$ in the zero-range limit
(the negative sign is due to the attractive interaction),
that can be easily obtained from its expression,
\begin{equation} 
\begin{aligned}
%g_{\psi\psi}=&\frac{4\pi \hbar^{2}a_{\psi\psi}}{m_{\psi}},\\
%g_{\psi\psi}=\simeq 
g_{\psi\phi}=&[\frac{m_{r}}{2\pi \hbar^{2}a_{\psi\phi}}-\int\frac{d^{3}k}{(2\pi)^{3}}\frac{2m_{r}}{k^{2}}]^{-1}\\
=&[\frac{m_{r}}{2\pi \hbar^{2}a_{\psi\phi}}-\frac{m_{r}\Lambda}{\pi^{2}}]^{-1},
\end{aligned}
\end{equation}
which becomes $\widetilde{g}_{\psi\phi}=\frac{m_{r}}{2\pi \hbar^{2}a_{\psi\phi}}$ after the renormalization which is also the result obtained by the first-order Born approximation for the low-energy collision\cite{Vlietinck J}.
$g_{\psi\phi}<0$ here due to the attractive interaction and with the
negative $s$-wave scattering length $a_{\psi\phi}$ (like the fermionic ${}^{6}{\rm Li}$)\cite{Bohn J L}.

\subsection{anisotropic treatment}

Similar to the procedure as we carried in Ref.\cite{polaron4}, we consider the chiral effect into the pairing scattering event:
an electron-hole pair (electron with momentum $k$ and hole with momentum $q$)
is excited in the fermi bath by the chiral semi-Dirac quasiparticle which with initial momentum $p$ and final momentum $p+q-k$.
The non-self-consistent $T$-matrix theory takes into account the higher-order pairing fluctuation than the Gaussian fluctuation\cite{Tsuchiya S},
although the propagators of the two quasiparticle excitation are bare.
Then the pair propagator can be written as
  %{Polarons and dressed molecules near narrow Feshbach resonances}
  %{Polarons, dressed molecules and itinerant ferromagnetism in ultracold Fermi gases}
  %{Mass imbalance effect in resonant Bose-Fermi mixtures}
\begin{equation} 
\begin{aligned}
\Pi(p,\omega)=\int\frac{d^{2}k}{(2\pi)^{2}}\frac{1-N_{F}(\varepsilon_{k})-N_{F}(\varepsilon_{p+q-k})}
{\omega+i0-\varepsilon_{p+q-k}-(\varepsilon_{k}-\varepsilon_{q})}F_{\lambda\lambda'}.
\end{aligned}
\end{equation}
Since the semi-Dirac quasiparticle is anisotropic which with relativistic behavior in $y$-direction and 
non-relativistic behavior in $x$-direction.
Obviously, the quadratic dispersion in $x$-direction with nonadiabatic feature (lower velocity compare to the relativistic one)
gives the mainly contribution to the formation of polaron, i.e.,
the interaction between the momentum of electron-hole pair and the $x$-direction momentum of impurity
is much stronger than the $y$-direction one,
thus for the impurity term in above equation, we only integral over the $x$-direction momentum.
Note that while for the excited electron-hole pair, we suppose it as isotropic in momentum space.
We continue to use the substitutions
\begin{equation} 
\begin{aligned}
(p+q-k)_{x}=&p_{0}\sqrt{(r_{p+q-k}{\rm cos}\theta'-\frac{D}{\varepsilon'_{0y}})\frac{\varepsilon'_{0y}}{\varepsilon'_{0x}}},\\
%p_{y}=&p_{0}r{\rm sin}\theta,\\
%\varepsilon_{0x}=&\frac{p_{0}^{2}}{2m_{x}},\\
%\varepsilon_{0y}=&v_{y}p_{0}\approx \frac{p_{0}^{2}}{2m_{y}}=\frac{p_{0}^{2}v_{y}^{2}}{2D}.
k=p_{0}r_{k},\\
q=p_{0}r_{q}
\end{aligned}
\end{equation}
Note that $\theta'=\frac{p'_{y}v_{y}}{\frac{p_{x}^{'2}}{2m_{x}}+D}$
and $r_{p+q-k}=r_{p}+r_{q}-r_{k}\equiv r_{p'}$ when $\theta'=0$.
The chiral factor reads
  %{Dynamical spin susceptibility of silicene}
  %{Kinked plasmon dispersion in borophene-borophene and borophene-graphene double layers}
  %{Frequency-dependent polarizability, plasmons, and screening in the two-dimensional pseudospin-1 dice lattice}
  %{Collective excitations on a surface of topological insulator}
  %{Plasmon-polaron of the topological metallic surface states}
\begin{equation} 
\begin{aligned}
F_{\lambda\lambda'}&=\langle p+q-k|p\rangle \langle k-q|0\rangle\\
&=\frac{1}{2}
\begin{pmatrix}
e^{ i\theta'/2} &
e^{-i\theta'/2}
\end{pmatrix}
\begin{pmatrix}
e^{-i\theta/2} \\
e^{i\theta/2}
\end{pmatrix}\\
&=
{\rm cos}\frac{\theta'-\theta}{2},
\end{aligned}
\end{equation}
where $|0\rangle$ denotes the initial state of electron-hole pair beforce scattering
(there is not initial occupation in mode $k-q$)
but
we assume the excited electron-hole pair is along the initial direction and thus $\langle k-q|0\rangle=1$.
This chiral form factor shows that the interaction vertex during the pairing scattering is dependent on both the initial momentum $p$
and scattering wave vector $p-k$,
similar feature can be found in some other models with electron-phonon interaction\cite{Giustino F,Marchand D J J,Yan J A},
while in the absence of such kind of transition matrix element (between initial state and excited state),
  %{Sharp Transition for Single Polarons in the One-Dimensional Su-Schrieffer-Heeger Model}
  %{Analytical properties of polaron systems or: Do polaronic phase transitions exist or not}
 the interaction term becomes dependent only on the scattering wave vector,
like in Refs.\cite{Sidler M,Christensen R S,Parish M M}.
   %{Quasiparticle Properties of a Mobile Impurity in a Bose-Einstein Condensate}
   %{Highly polarized Fermi gases in two dimensions}
   %{Molecule and Polaron in a Highly Polarized Two-Dimensional Fermi Gas with Spin-Orbit Coup}
As we mentioned above, the Landau damping (as a conjugate process of the  Beliaev coupling) 
here is related to the value of scattering wave vector or $p$\cite{Natu S S,Pixley J H},
  %{Absence of damping of low-energy excitations in a quasi-two-dimensional dipolar Bose gas}
  %{Damping of Long-Wavelength Collective Modes in Spinor Bose-Fermi Mixtures}
i.e., in long-wavelength limit with $q-k\rightarrow 0$ (or in the strong coupling limit),
  %{Bulk Bogoliubov excitations in a Bose-Einstein condensate}
the damping vanishes due to the destructive quantum interference.
  %{Bulk Bogoliubov excitations in a Bose-Einstein condensate}
At finite temperature, the Landau damping of polaron here is easy to recognized as the process that the polaron decays into
two modes: $p+q-k\rightarrow p,(q-k)$.
Note that the 
destructive quantum interference can be destroyed by the adiabatic dynamics,
e.g., the linear dispersion (gapless) in $y$-direction of the semi-Dirac system.
  %{Decoherence of an impurity in a one-dimensional fermionic bath with mass imbalance}
In adiabatic limit, the superposition of impurity state and the majority one will
turns into the statistical mixture.
While in the nonadiabatic case, the quantum interference patterns exist and the superposition between states 
  %{Normal State of Highly Polarized Fermi Gases: Full Many-Body Treatment}
  %{Decoherence of an impurity in a one-dimensional fermionic bath with mass imbalance}
  %{Damping of Long-Wavelength Collective Modes in Spinor Bose-Fermi Mixtures}
  %{Bose-Einstein Condensate in a Honeycomb Optical Lattice: Fingerprint of Superfluidity at the Dirac Point}
can be ensured by the equal masses between impurity and majority particles.
In this case, the behavior of dissipationless flow (in superfluid phase) can be observed in a Bose gas\cite{Natu S S} where the interactions
support the linear dispersion in low-momentum region, unlike the Dirac system which owns linear dispersion in low-momentum region 
in noninteracting case.
Due to the interaction effect,
the transport of polaron in $x$-direction will not be ballistic, and the momentum will not conserved during the transport.
Also, in the non-spin-degenerate case, the spin polarization is possible to produced by the electric
field-driven acceleration of electrons unlike the adiabatic case\cite{Silvestrov P G}.
  %{Tight-Binding Modeling and Low-Energy Behavior of the Semi-Dirac Point}
  %{Mesoscopic Spin-Hall Effect in 2D Electron Systems with Smooth Boundaries}

In this paper, we mainly focus on the nonadiabatic case.
In momentum space, the above chiral factor can also be rewritten as
    %{Dynamical polarization and plasmons in a two-dimensional system with merging Dirac points}
\begin{equation} 
\begin{aligned}
F_{\lambda\lambda'}=
\frac{(\frac{p_{x}^{2}}{2m_{x}}+D)(\frac{p_{x}^{'2}}{2m_{x}}+D)+v_{y}^{2}p_{y}p'_{y}}
{\varepsilon_{p}\varepsilon_{p+q-k}}.
\end{aligned}
\end{equation}
The final expression of the pair propagator at zero temperature reads
\begin{equation} 
\begin{aligned}
\Pi(p,\omega)=\int dr_{k}\int d\theta'
\frac{\xi' p_{0}r_{p'}}{\omega+i0-(\frac{
\left[p_{0}\sqrt{(r_{p'}{\rm cos}\theta'-\frac{D}{\varepsilon'_{0y}})\frac{\varepsilon'_{0y}}{\varepsilon'_{0x}}}\right]
^{2}}{2m_{x}}+D-\mu_{i})-(\frac{p_{0}^{2}(r_{k}-r_{q})^{2}}{2m_{\uparrow}}-2\mu_{m})}{\rm cos}\frac{\theta'-\theta}{2}
\end{aligned}
\end{equation}
where
$\xi'=p_{0}\frac{1}{2}\frac{    \frac{\varepsilon'_{0y}}{\varepsilon'_{0x}}       }
{\sqrt{(r_{p'}{\rm cos}\theta'-\frac{D}{\varepsilon'_{0y}})\frac{\varepsilon'_{0y}}{\varepsilon'_{0x}}   }}$.
$r_{p'}=\sqrt{r_{p}^{2}+r_{q-k}^{2}+2r_{p}r_{q-k}{\rm cos}\theta}$.
Note that unlike the calculation of DOS,
we set a momentum cutoff the upper limit of integral over $r_{k}$
to prevent the problem of inconvergence, as we presented previously in Ref.\cite{polaron4}.
Similar to the above transport scattering time, we here only consider the contribution from $x$-component of momentum of semi-Dirac quasiparticle
but consider all components of momentum of the majority particle,
i.e., the nonrelativistic parts.
The results are presented in Fig.5-7.
 The nonanalyticity are obvious in pair-propagator and self-energy due to the logarithmic divergence.
  %{Nonanalytic paramagnetic response of itinerant fermions away and near a ferromagnetic quantum phase transition}
  %{Nonanalytic corrections to the specific heat of a three-dimensional Fermi liquid}
  %{Valleytronics in merging Dirac cones: All-electric-controlled valley filter, valve, and universal reversible logic gate}
We note that the integral over scattered impurity wave vector can savely be replaced by the integral over the scattering wave vector
in any cases.
  %{Acoustic phonon scattering limited carrier mobility in two-dimensional extrinsic graphene}
  %{Boltzmann transport and residual conductivity in bilayer graphene}
  %{Screening-induced temperature-dependent transport in two-dimensional graphene}
  %{Temperature- and frequency-dependent optical and transport conductivities in doped buckled honeycomb lattices}
  %{Phonon linewidths and electron-phonon coupling in graphite and nanotubes}
  %{First-principles analysis of electron-phonon interactions in graphene}
  %{Phonon linewidths and electron-phonon coupling in graphite and nanotubes}
  %
  %{Intrinsic electrical transport properties of monolayer silicene and MoS 2 from first principles}
  %{Electron-Phonon Interaction and Transport in Semiconducting Carbon Nanotubes}
  %{First-principles method for electron-phonon coupling and electron mobility Applications to two-dimensional materials}
Thus for scattered semi-Dirac quasiparticle (polaron), the DOS (at semi-Dirac phase) can be obtained by the expression
\begin{equation} 
\begin{aligned}
\rho(\omega)=&\int^{\infty}_{0} dr_{k}\int^{\pi/2}_{0} d\theta \xi p_{0}r \delta[\omega-(\pm\varepsilon_{0y}r_{p'}-\mu_{i})].
\end{aligned}
\end{equation}

The above-mentioned polaron band velocity (after pairing scattering) can then be obtained as
\begin{equation} 
\begin{aligned}
v_{x}&=\frac{\partial \varepsilon}{\partial p'_{x}}
=\frac{\partial \varepsilon}{\partial r_{p'}}
 \frac{\partial r_{p'}}{\partial p'_{x}},\\
&=\frac{\partial \varepsilon}{\partial r_{p}}
  \frac{\partial r_{p}}{\partial r_{p'}}
  \frac{\partial r_{p'}}{\partial r_{p}}
  \frac{\partial r_{p}}{\partial p'_{x}}\\
&=\frac{\partial \varepsilon}{\partial r_{p}}
  \frac{\partial r_{p}}{\partial p'_{x}},\\
v_{y}&=\frac{\partial \varepsilon}{\partial p'_{y}}
=\frac{\partial \varepsilon}{\partial r_{p'}}
 \frac{\partial r_{p'}}{\partial p'_{y}}\\
&=\frac{\partial \varepsilon}{\partial r_{p}}
  \frac{\partial r_{p}}{\partial r_{p'}}
  \frac{\partial r_{p'}}{\partial r_{p}}
  \frac{\partial r_{p}}{\partial p'_{y}}\\
&=\frac{\partial \varepsilon}{\partial r_{p}}
  \frac{\partial r_{p}}{\partial p'_{y}},\\
\end{aligned}
\end{equation}
where the polaron energy, unlike the above mentioned bare energy of
semi-Dirac quasiparticle, reads $\varepsilon=\sqrt{(\frac{p^{'2}_{x}}{2m_{x}}+D)^{2}+v_{y}^{2}p^{'2}_{y}}-\mu_{i}+\Sigma(p,\omega)$.
here the polaronic effect induced self-energy can be obtained as
\begin{equation} 
\begin{aligned}
\Sigma(p,\omega)=\frac{1}{g_{b}^{-1}-\Pi(p,\omega)},
\end{aligned}
\end{equation}
where $g_{b}$ is the coupling constant obtained by integrating over the momentum 
(not the relative (transfer) momentum and the center-of-mass momentum).
  %{Fermi polaron-polaritons in charge-tunable atomically thin semiconductors} 
  %{Highly polarized Fermi gases in two dimensions}
Note that at low impurity density limit ($p_{F}a_{B}\ll 1$), the transfer momentum-dependence of the interaction term
can usually be neglected\cite{Miserev D,Sidler M}, i.e., the $g_{b}$ can be treated approximated as $(q-k)$-independent.
While for large transfer momentum, e.g., $q=2p_{F}$, the Kohn anomaly appears only in the presence of spin (or pseudospin)
splitting\cite{Maslov D L},
and the polarization bubble (not the pair-scattering one) in second order diagram (to second order of the nonlocal interaction $g_{q}$)
  %{Nonanalytic paramagnetic response of itinerant fermions away and near a ferromagnetic quantum phase transition}
 behaviors as
$\Pi'(q',\omega')=-\rho(E_{F})+|\omega'|/q'$ where $q'$ is the transferred momentum of the impurity and $\omega'$ is the
transferred energy.
  %{Nonanalytic paramagnetic response of itinerant fermions away and near a ferromagnetic quantum phase transition}
  %{Exchange intervalley scattering and magnetic phase diagram of transition metal dichalcogenide monolayers}
  %{Nonanalytic magnetic response of Fermi and non-Fermi liquids}
Note that here the dynamical polarization is analytic (does not contains the logarithmic term),
which implies that the nonanalyticities from both the small $q'$ region and $q'\sim 2p_{F}$ (with backscattering due to the weak coupling)
Landau damping are neglected.
While the logarithmic divergence bringed by the Landau damping with thermal excitations
has a cutoff at energy $\varepsilon\sim T$,
  %{Nonanalytic magnetic response of Fermi and non-Fermi liquids}
unlike the configuration we discussed above which has a cutoff at energy of the order of impurity bandwidth.
The pair propagator, self-energy, and spectral function are presented in Figs. for insulator phase ($D=0.1$; solid lines)
and semi-Dirac phase ($D=0$; dashed lines).
The spectral function shift rightward with the turning up of Dirac mass term,
and we can also see that the removing a fast impurity will cost larger energy $|-\omega|$.
  %{Observation of Fermi Polarons in a Tunable Fermi Liquid of Ultracold Atoms}
  %{Observation of Fermi Polarons in a Tunable Fermi Liquid of Ultracold Atoms}
For stronger coupling $g_{b}$, this energy cost will descrases.

Through the above-mentioned formulas we have
\begin{equation} 
\begin{aligned}
v_{x}&=\frac{2 p_{0} v_{y} {\rm cos}   \theta' (D + p_{0} v_{y} (-(D/(p_{0} v_{y})) + (-k + p + q) {\rm cos}\theta')) + 
 2 p_{0}^{2} (-k + p + q) v_{y}^{2} ({\rm sin}\theta')^{2}}{2 \sqrt{(D + 
    p_{0} v_{y} (-(D/(p_{0} v_{y})) + (-k + p + q) {\rm cos}\theta'))^{2} + 
  p_{0}^{2} (-k + p + q)^{2} v_{y}^{2} ({\rm sin}\theta')^{2}}}\\
&\times
\frac{\sqrt{2}\sqrt{\frac{m_{x}v_{y}(-\frac{D}{p_{0}v_{y}}+(p+q-k){\rm cos}\theta')}{p_{0}}}}{m_{x}v_{y}{\rm cos}\theta'},\\
v_{y}&=\frac{2 p_{0} v_{y} {\rm cos}   \theta' (D + p_{0} v_{y} (-(D/(p_{0} v_{y})) + (-k + p + q) {\rm cos}\theta')) + 
 2 p_{0}^{2} (-k + p + q) v_{y}^{2} ({\rm sin}\theta')^{2}}{2 \sqrt{(D + 
    p_{0} v_{y} (-(D/(p_{0} v_{y})) + (-k + p + q) {\rm cos}\theta'))^{2} + 
  p_{0}^{2} (-k + p + q)^{2} v_{y}^{2} ({\rm sin}\theta')^{2}}}\\
&\times
\frac{1}{p_{0}{\rm sin}\theta'},
\end{aligned}
\end{equation}
For semiclassical case, as stated above, the product of $v_{x}$ and $v_{y}$ can be used directly to obtain the
semiclassical conductivity.
While for the quantum-like conductivity, 
  %{Frequency and orientation dependent conductivity of a Semi-Dirac system}
the product of $v_{x}$ and $v_{y}$ should be replaced by the product of their respective matrix elements
$\langle \psi|v_{x}|\psi'\rangle$ and $\langle \psi|v_{y}|\psi'\rangle$,
i.e., containing the effect of wave function overlap\cite{wums18}.

It is obvious that the free energy calculated above is the noninteracting result, we now take the polaronic coupling 
into account.
To containing the interaction effect,
the free energy can be rewritten as\cite{wupara}
  %{Anomalous orbital magnetism in Dirac-electron systems Role of pseudospin paramagnetism}
  %{Exotic electronic and transport properties of graphene}
  %{Exchange intervalley scattering and magnetic phase diagram of transition metal dichalcogenide monolayers}
  %{Hall quantization and optical conductivity evolution with variable Berry phase in the α-T3 model}
\begin{equation} 
\begin{aligned}
F=\Omega-\mu_{i}(T)\frac{\partial \Omega}{\partial \mu_{i}(T)},
\end{aligned}
\end{equation}
where $\Omega=\Omega_{0}+\Omega_{1}$ is the grand canonical potential (thermodynamic potential) which reads
  %{Exchange intervalley scattering and magnetic phase diagram of transition metal dichalcogenide monolayers}
\begin{equation} 
\begin{aligned}
\Omega_{0}= & T\int\frac{d\omega}{2\pi}\int\frac{d^{2}p}{(2\pi)^{2}}{\rm ln}\frac{1}{\omega+i0-\varepsilon_{p}}\\
\Omega_{1}= &-T\int\frac{d\omega}{2\pi}\int\frac{d^{2}p}{(2\pi)^{2}}\int\frac{d\omega'}{2\pi}\int\frac{d^{2}p'}{(2\pi)^{2}}
\frac{\pi}{m_{x}}[g_{b}\frac{1}{\omega+i0-\varepsilon_{p}}\frac{1}{\omega'+i0-\varepsilon_{p'}}],
\end{aligned}
\end{equation}
$\Omega_{1}$ is the first-order interaction correction to the grand canonical potential which 
shows that the
interaction effect can be contained even just consider the bare Green's function.
$\mu_{i}(T)$ is the temperature-dependent chemical potential,
which can be obtained by solving 
$n_{\downarrow}(T)=\int^{\mu_{i}}_{0}d\omega\rho(\omega,T)$
where $\rho(\omega,T)\propto N_{F}(\omega)$ is the DOS at finite temperature.
The concentration can also be related to the Green's function at finite temperature by
$n_{\downarrow}=T\sum_{p,\omega}G(p,\omega)\approx T\frac{1}{(2\pi)^{3}}\int\int dRd\Phi G(p,\omega)$,
or related to the chemical potential by $\partial {\rm Re}\Sigma(p,\omega)/\partial (n_{\downarrow}S)=\mu_{i}$
(or ${\rm Re}\Sigma(p,\omega)=\int^{n_{\downarrow}}_{0}\mu_{i}n_{\downarrow}S dn_{\downarrow}$)
where we ignore the residual polaron-polaron interaction.
  %{Attractive Fermi polarons at nonzero temperatures with a finite impurity concentration}
  %{Fate of the Bose polaron at finite temperature}
The calculated results shows that, at finite temperature,
even the noninteracting grand canonical potential
diverges from the simple result\cite{Illes E,Miserev D}
  %{Hall quantization and optical conductivity evolution with variable Berry phase in the α-T 3 model}
  %{Exchange intervalley scattering and magnetic phase diagram of transition metal dichalcogenide monolayers.pdf}
$\Omega_{0}=\mu_{i}\int^{\mu_{i}}_{0}\rho(\omega)d\omega=\mu_{i}n_{\downarrow}$,
and becomes rather complex for the semi-Dirac system especially when the polaronic effect is taken into account.
The polaronic interaction will decreases the grand canonical potential,
although it can be neglected if the coupling is weak enough.
The impurity density here is also temperature-dependent,
  %{Fate of the Bose polaron at finite temperature}
$n_{\downarrow}=-\frac{\partial \Omega}{\partial \mu_{i}}$,
which implies that for small impurity density, it is more easy to adjust the chemical potential by gating.
  %{Engineering a  superconductor Comparison of topological insulator and Rashba spin-orbit-coupled materials}
Note that the relation between the eigenenergy (or the excited quasiparticle energy) and the grand canonical potential
can be obtained by the thermodynamic Bethe Ansatz equation for equilibrium
states\cite{Wang S}.
  %{Wang S, Yin X, Chen Y Y, et al. Emergent ballistic transport of Bose-Fermi mixtures in one dimension[J]. arXiv preprint arXiv:1910.06224, 2019.}

\subsection{Conductivity}

  %{Transport of Dirac quasiparticles in graphene Hall and optical conductivities}
  %{Suppression of the persistent spin Hall current by defect scattering}
  %{Dynamical conductivity of AA-stacked bilayer graphene}
  %{Dc and ac transport in silicene}
  %{Anomalous integer quantum Hall effect in A A-stacked bilayer graphene}
  %{Quantum hall effect in bilayer and multilayer graphene with finite fermi energy}
Base on the Kubo formula, the real part of conductivity in dc-limit can also be written as
\begin{equation} 
\begin{aligned}
{\rm Re}\sigma_{xy}=\lim_{\Omega_{x}\rightarrow 0}\frac{{\rm Im}\Pi(\Omega_{x}+i0)}{\Omega_{x}},
\end{aligned}
\end{equation}
where the current-current correlation reads
\begin{equation} 
\begin{aligned}
\Pi(i\Omega_{x})
=&T\frac{d^{2}p}{(2\pi)^{2}}\frac{d\omega}{2 \pi)}
{\rm Tr}[J_{x}G(\omega+\Omega_{x})J_{y}G(\omega)]\\
=&-e^{2}T\frac{d^{2}p}{(2\pi)^{2}}\frac{d\omega}{2 \pi)}
{\rm Tr}[\hat{v}_{x}G(\omega+\Omega_{x})\hat{v}_{y}G(\omega)]
\end{aligned}
\end{equation}

Temporally ignore the contribution from the polaronic attractive interaction-assisted hopping to the conductivity,
  %{Polaronic Impurity Hopping Conduction}
  %{Damping of Long-Wavelength Collective Modes in Spinor Bose-Fermi Mixtures}
the Hall conductivity at finite temperature can be written by the Kubo formula as presented above
\begin{equation} 
\begin{aligned}
{\rm Re}\sigma_{xy}(\Omega_{x})
&=\frac{e^{2}}{\Omega_{x}}\int\frac{d\omega}{2\pi}[N_{F}(\omega)-N_{F}(\omega+\Omega_{x})]
\int\frac{d^{2}p}{(2\pi)^{2}}{\rm Tr}[\hat{v}_{x}\hat{A}(p,\omega)\hat{v}_{y}\hat{A}(p,\omega+\Omega_{x})].
\end{aligned}
\end{equation}
where the nondiagonal elements of the new velocity matrices in $(r,\theta)$ coordinate reads 
\begin{equation} 
\begin{aligned}
\hat{v}_{x}^{12}=&\frac{i\sqrt{2}e^{-i\theta}p_{0}\sqrt{\frac{m_{x}v_{y}(r{\rm cos}\theta-\frac{D}{v_{y}p_{0}})}{p_{0}}}{\rm csc}\theta}{m_{x}}
+\frac{\sqrt{2}e^{-i\theta}p_{0}\sqrt{\frac{m_{x}v_{y}(r{\rm cos}\theta-\frac{D}{v_{y}p_{0}})}{p_{0}}}{\rm sec}\theta}{m_{x}},\\
\hat{v}_{x}^{21}=&-\frac{i\sqrt{2}e^{i\theta}p_{0}\sqrt{\frac{m_{x}v_{y}(r{\rm cos}\theta-\frac{D}{v_{y}p_{0}})}{p_{0}}}{\rm csc}\theta}{m_{x}}
+\frac{\sqrt{2}e^{i\theta}p_{0}\sqrt{\frac{m_{x}v_{y}(r{\rm cos}\theta-\frac{D}{v_{y}p_{0}})}{p_{0}}}{\rm sec}\theta}{m_{x}},\\
\hat{v}_{y}^{12}=&\frac{e^{-i\theta}v_{y}p_{0}{\rm csc}\theta}{p_{0}}-\frac{ie^{-i\theta}v_{y}p_{0}{\rm sec}\theta}{p_{0}},\\
\hat{v}_{y}^{21}=&\frac{e^{i\theta}v_{y}p_{0}{\rm csc}\theta}{p_{0}}+\frac{ie^{i\theta}v_{y}p_{0}{\rm sec}\theta}{p_{0}}.
\end{aligned}
\end{equation}
%Here $\hat{A}(p,\omega)$ is the spectral density matrix
%which is related to the matrix of Green's function by 
%$G(p,\omega+i\eta)=\int^{\infty}_{-\infty}\frac{d\omega}{2\pi}\lim_{\eta\rightarrow 0}\frac{A(p,\omega)}{i\eta}$.
  %{Dynamical conductivity of AA-stacked bilayer graphene}
  %{Optical conductivity of bilayer graphene with and without an asymmetry gap}
  %{Transport of Dirac quasiparticles in graphene: Hall and optical conductivities}
  %{Optical conductivity of multi-Weyl semimetals}
Then the matrix elements reads
\begin{equation} 
\begin{aligned}
\langle \alpha|\hat{v}_{x}|\alpha'\rangle=&\pm\frac{2i\sqrt{2}e^{i\theta}p_{0}\sqrt{-\frac{m_{x}(D-p_{0}rv_{y}{\rm cos}\theta)}{p_{0}^{2}}}{\rm sec}\theta}
{(-1+e^{2i\theta})m_{x}},\\
\langle \alpha|\hat{v}_{y}|\alpha'\rangle=&
\pm\frac{4ie^{2i\theta}p_{0}v_{y}}{\sqrt{1-2e^{4i\theta}+e^{8i\theta}}p_{0}}.
\end{aligned}
\end{equation}

The spectral density matrix reads
\begin{equation} 
\begin{aligned}
\hat{A}(p,\omega)
=&\frac{[\delta(\omega-\mu_{i}-\varepsilon_{p})-\delta(\omega-\mu_{i}+\varepsilon_{p})]}{2\varepsilon_{p}}{\rm sgn}[\omega]
\begin{pmatrix}
-\mu_{i}+\omega & e^{-i\theta}v_{y}p_{0}r\\
e^{i\theta}v_{y}p_{0}r & -\mu_{i}+\omega
\end{pmatrix}.
\end{aligned}
\end{equation}

In contrast with the diffusion one,
the hopping conductivity which
requires finite temperature reads, for low-temperature limit,
  %{Small polaron conduction in V2O5 P2O5 glasses}
\begin{equation} 
\begin{aligned}
\sigma_{xx}=&\frac{e^{2}}{TV}\sum_{\alpha}
N_{F}(\varepsilon_{p\alpha})(1-N_{F}(\varepsilon_{p\alpha}))\frac{|\langle \alpha|v_{x}|\alpha\rangle|^{2}}{i\Omega_{x}+\tau_{p}^{-1}}\\
=&\frac{e^{2}}{V}\sum_{\alpha}
\delta(\varepsilon_{p\alpha}-\mu_{i})\frac{|\langle \alpha|v_{x}|\alpha\rangle|^{2}}{i\Omega_{x}+\tau_{p}^{-1}}.
\end{aligned}
\end{equation}
  %{Dc and ac transport in silicene}
  %{Linear response theory revisited III: One-body response formulas and generalized Boltzmann equations}
  %{Integer and half-integer quantum Hall effect in silicene: Influence of an external electric field and impurities}
The relation 
$\frac{1}{T}N_{F}(\varepsilon_{p\alpha})(1-N_{F}(\varepsilon_{p\alpha}))=\delta(\varepsilon_{p\alpha}-\mu_{i})$ ($T\rightarrow 0$) is used here.
  %{Dc and ac transport in silicene}
  %{Linear response theory revisited III: One-body response formulas and generalized Boltzmann equations}
We can see that the transfer matrix element (expectation value of velocity operator) only between the states with same energy,
  %{}
but when the polaronic attraction is taken into account,
the hopping between donor impurity states with different energies becomes possible,
as evidented in the systems with slow phonons\cite{Greaves G N,Schnakenberg J},
  %{Damping of Long-Wavelength Collective Modes in Spinor Bose-Fermi Mixtures}
and the conductivity then has (consider the hopping only in $x$-direction)
\begin{equation} 
\begin{aligned}
\sigma_{xx}\propto
\int dx e^{-i p_{x}x}
\frac{1}{TV}\sum_{x}I_{xx'}^{2}R_{xx'}^{2},
\end{aligned}
\end{equation}
where $I_{xx'}$ is the resonance integral which can be proved proportional to the coupling constant here
through the polaron canonical transformation\cite{Bryksin V V}.
  %{Small-polaron theory with allowance for the influence of lattice vibrations on the resonance integral}
$R_{xx'}$ is the distance between hopping sites in real space.

We note that unlike the isotropic system (e.g., with the $C_{4}$ crystal symmetry) where $\sigma_{xx}=\sigma_{yy}$\cite{wums18},
  %{Emergent Anisotropic Non-Fermi Liquid at a Topological Phase Transition in Three Dimensions support}
the longitudianl optical conductivities in semi-Dirac system has 
$\sigma_{xx}\propto \sqrt{\frac{\Omega_{x}}{\varepsilon_{0}}}$,
$\sigma_{yy}\propto \sqrt{\frac{\varepsilon_{0}}{\Omega_{x}}}$\cite{Carbotte J P}.
  %{Optical properties of a semi-Dirac material}

\subsection{isotropic treatment: low carrier-density approximation}

For the case that the chemical potential lies within the conduction band but with low-enough carrier density\cite{Pyatkovskiy P K,Park S,Doh H},
  %{Semiclassical Boltzmann transport theory of few-layer black phosphorus in various phases}
  %{Dynamical polarization and plasmons in a two-dimensional system with merging Dirac points}
  %{Dirac-semimetal phase diagram of two-dimensional black phosphorus}
the dispersion of semi-Dirac quasiparticle can be treated isotropic but with different effective masses in different direction,
which leads to the Hamiltonian 
\begin{equation} 
\begin{aligned}
H=\frac{p_{x}^{2}}{2m_{x}}\sigma_{x}+\frac{p_{y}^{2}}{2m_{y}}\sigma_{y}+(D-\mu_{i})\sigma_{0},
  %{Semiclassical Boltzmann transport theory of few-layer black phosphorus in various phases}
  %{Collective modes in multi-Weyl semimetals support}
  %{Optical conductivity of multi-Weyl semimetals}
\end{aligned}
\end{equation}
and the spectrum reads 
\begin{equation} 
\begin{aligned}
\varepsilon=&D-\mu_{i}\pm\sqrt{\frac{p_{x}^{4}}{4m_{x}^{2}}+\frac{p_{y}^{2}}{4m_{y}^{2}}}\\
\approx& \frac{p_{x}^{2}}{2m_{x}}+\frac{p_{y}^{2}}{2m_{y}}+D-\mu_{i},
\end{aligned}
\end{equation}
where the approximation in second line is valid for small $p_{x}$ and $p_{y}$.
Note that due to the low carrier concentration,
the polaron-polaron interaction, which is short ranged, can be neglected.
Then it is useful for us to carry the following coordinate transformation
\begin{equation} 
\begin{aligned}
p_{x}=&\sqrt{\frac{m_{x}}{m}}\sqrt{r{\rm cos}\theta},\\
p_{y}=&\sqrt{\frac{m_{y}}{m}}\sqrt{r{\rm sin}\theta},\\
%p_{x}=&\sqrt{\frac{m_{x}}{m}}p{\rm cos}\theta \ 2^{1/4},\\
%p_{y}=&\sqrt{\frac{m_{y}}{m}}p{\rm sin}\theta \ 2^{1/4},\\
%p_{x}=&p_{0}\sqrt{(r{\rm cos}\theta-\frac{D}{\varepsilon_{0y}})\frac{\varepsilon_{0y}}{\varepsilon_{0x}}}\\
%=&p_{0}\sqrt{(r{\rm cos}\theta-\frac{D}{\varepsilon_{0y}})\frac{m_{x}}{m_{y}}},\\
%p_{y}=&p_{0}\sqrt{r{\rm sin}\theta},\\
%            %&p_{0}r{\rm sin}\theta,\\
\varepsilon_{0x}=&\frac{p_{0}^{2}}{2m_{x}},\\
%\approx v_{x}p_{0},\\
       %=\sqrt{\frac{2|D|}{m_{x}}}p_{0},\\
       %\varepsilon_{0y}=&v_{y}p_{0}\approx \frac{p_{0}^{2}}{2m_{y}}=\frac{p_{0}^{2}v_{y}^{2}}{2D}.
\varepsilon_{0y}=&\frac{p_{0}^{2}}{2m_{y}}=\frac{p_{0}^{2}v_{y}^{2}}{2D},
\end{aligned}
\end{equation}
where $m=\sqrt{m_{x}^{2}+m_{y}^{2}}$ and $\theta={\rm arctan}\frac{p_{y}^{2}}{m_{y}}/\frac{p_{x}^{2}}{m_{x}}$.
%and the factor $2^{1/4}$ is to make sure, when $m_{x}=m_{y}$, we have $p_{x}^{2}+p_{y}^{2}=p^{2}$.
Then similar to the above procedure, we can write the Hamiltonian as
\begin{equation} 
\begin{aligned}
H=&
\begin{pmatrix}
0 &  \frac{p_{x}^{2}}{2m_{x}}-i\frac{p_{y}^{2}}{2m_{y}} \\
\frac{p_{x}^{2}}{2m_{x}}+i\frac{p_{y}^{2}}{2m_{y}} & 0
\end{pmatrix}+D\sigma_{0}-\mu_{i}\sigma_{0},\\
=&\begin{pmatrix}
0 &  \frac{r}{2m}{\rm cos}\theta-i\frac{r}{2m}{\rm sin}\theta \\
 \frac{r}{2m}{\rm cos}\theta+i\frac{r}{2m}{\rm sin}\theta  & 0
\end{pmatrix}+D\sigma_{0}-\mu_{i}\sigma_{0},\\
=&
\frac{r}{2m}
\begin{pmatrix}
0 &  e^{-i\theta}\\
e^{i\theta} & 0
\end{pmatrix}+D-\mu_{i},
\end{aligned}
\end{equation}
thus we can obtain the eigenenergy $\varepsilon=\pm\frac{r}{2m}+D-\mu_{i}$.
The Jacobian transformation reads
\begin{equation} 
\begin{aligned}
dp_{x}dp_{y}
%\begin{vmatrix}
%\frac{\partial p_{x}}{\partial r} & \frac{\partial p_{x}}{\partial \theta}\\
%\frac{\partial p_{y}}{\partial r} & \frac{\partial p_{y}}{\partial \theta}
%\end{vmatrix}drd\theta\\
%=&\begin{vmatrix}
%p_{0}\frac{1}{2}\frac{{\rm cos}\theta    \frac{\varepsilon_{0y}}{\varepsilon_{0x}}       }
%{\sqrt{({\rm cos}\theta-\frac{D}{\varepsilon_{0y}})\frac{\varepsilon_{0y}}{\varepsilon_{0x}}   }}
%&
%p_{0}\frac{1}{2}\frac{-{\rm sin}\theta r \frac{\varepsilon_{0y}}{\varepsilon_{0x}}         }
%{\sqrt{({\rm cos}\theta-\frac{D}{\varepsilon_{0y}})\frac{\varepsilon_{0y}}{\varepsilon_{0x}}  }}
%\\
%p_{0}{\rm sin}\theta & p_{0}r{\rm cos}\theta
%\end{vmatrix}drd\theta\\
%=&\begin{vmatrix}
%\xi {\rm cos}\theta & \xi r(-{\rm sin}\theta)\\
%p_{0}{\rm sin}\theta & p_{0}r{\rm cos}\theta
%\end{vmatrix}drd\theta\\
%=2^{1/2}\frac{\sqrt{m_{x}m_{y}}}{m}pdpd\theta,
=\frac{\sqrt{m_{x}m_{y}}}{m}\frac{1}{4}\frac{1}{\sqrt{{\rm cos}\theta{\rm sin}\theta}}drd\theta,
\end{aligned}
\end{equation}
Then the DOS can be obtained as
\begin{equation} 
\begin{aligned}
\rho(\omega)
=&\frac{1}{(2\pi)^{2}}
\int^{\infty}_{0} dp\int^{\pi/2}_{0} d\theta 
\frac{\sqrt{m_{x}m_{y}}}{m}\frac{1}{4}\frac{1}{\sqrt{{\rm cos}\theta}\sqrt{{\rm sin}\theta}}
 \delta[\omega-(\pm\frac{r}{2m}+D-\mu_{i})]\\
=&\frac{1}{(2\pi)^{2}}\frac{\sqrt{m_{x}m_{y}}\sqrt{2\pi}\Gamma(\frac{5}{4})\Theta(m(-D+\mu_{i}+\omega))}{\Gamma(\frac{3}{4}){\rm sgn}[\omega]}\\
\approx &0.7397\sqrt{m_{x}m_{y}}\sqrt{2\pi}\Theta(m(-D+\mu_{i}+\omega))
{\rm ,for\ upper\ brance}\\
=&\frac{1}{(2\pi)^{2}}\frac{\sqrt{m_{x}m_{y}}\sqrt{2\pi}\Gamma(\frac{5}{4})\Theta(m(2D-2\mu_{i}-2\omega))}{\Gamma(\frac{3}{4}){\rm sgn}[\omega]}\\
=&\frac{1}{(2\pi)^{2}}0.7397\sqrt{m_{x}m_{y}}\sqrt{2\pi}\Theta(m(2D-2\mu_{i}-2\omega))
{\rm ,for\ lower\ brance}.
%=&\frac{\sqrt{m_{x}m_{y}}\pi(\pm D\mp \mu_{i}\mp \omega)\Theta(\pm\frac{-D+\mu_{i}+\omega}{\sqrt{2^{1/4}\frac{m_{x}m_{y}}{m}}})}
%{2m\sqrt{\frac{m_{x}m_{y}}{m}}\sqrt{\big|\frac{m_{x}m_{y}}{m}}\big|},
\end{aligned}
\end{equation}
It is easy to find that, unlike the result obtained base on anisotropic model as shown above,
 the DOS obtained in low-density approximation is independent of energy $\omega$ in some certain region:
for upper branch,
nonzero constant DOS requires $\omega+\mu_{i}>D$, while for
lower branch,
nonzero constant DOS requires $\omega+\mu_{i}<D$.
This is different to both the intrinsic Dirac systems 
(or some other graphene-related complex junction structures\cite{González J,Abdullah H M})
which has a DOS linear with energy in low-energy regime\cite{Hwang E H,Stauber T},
  %{Screening-induced temperature-dependent transport in two-dimensional graphene}
  %{Anomalous orbital magnetism in Dirac-electron systems: Role of pseudospin paramagnetism}
  %{Integer quantum Hall transition: An alternative approach and exact results}
and the semi-Dirac systems whose DOS is proportional to $\sqrt{\omega}$ as we stated above.
Such special phenomenon in DOS (constant, but not follow the power law behavior) is similar to the normal 2D electron gas, and
can also be found in the bilayer graphene\cite{Van Duppen B,Tabert C J}
 or other bilayer Dirac-like systems under magnetic field\cite{bx,Carbotte J P}
(if we ignore the largest peak in zero energy which is contributed by a doubly degenerated level).
  %{Propagating, evanescent, and localized states in carbon nanotube–graphene junctions}
That also implies in low carrier density approximation the semi-Dirac system can be approximately treated as 2D electron gas
although with different effective masses in different directions.
  %{Thermodynamic properties of the electron gas in multilayer graphene in the presence of a perpendicular magnetic field}
  %{Anomalous orbital magnetism in Dirac-electron systems: Role of pseudospin paramagnetism}
For constant DOS, the noninteracting fermi energy could also be a constant as $E_{F}=n_{\downarrow}/\rho$.
  %{Exchange intervalley scattering and magnetic phase diagram of transition metal dichalcogenide monolayers}
Note that since $D\gg (\mu_{\uparrow}-D)>0$ here,
the fermi momentum $k_{F}=\sqrt{\mu^{2}-D^{2}}$ could be very small.

%We note that the DOS of carbon nanotube-graphene junctions at very low energy regime\cite{González J} will shows the same behavior.
 
Base on above formulas,
we can make the substitute when integral over the momentu $k$
\begin{equation} 
\begin{aligned}
d{\bf k}=dk_{x}dk_{y}=\frac{\sqrt{m_{x}m_{y}}}{m}\frac{1}{4}\frac{1}{\sqrt{{\rm cos}\theta_{k}{\rm sin}\theta_{k}}}dr_{k}d\theta_{k}
=\frac{\sqrt{m_{x}m_{y}}}{m}\frac{1}{4}\frac{1}{\sqrt{{\rm cos}\theta_{k}{\rm sin}\theta_{k}}}dr_{k}d\theta,
\end{aligned}
\end{equation}
where for the electron within electron-hole pair in 2D electron gas, the isotropic momentum (with $m_{x}=m_{y}$) has
 $k=\sqrt{k_{x}^{2}+k_{y}^{2}}=\sqrt{\frac{1}{\sqrt{2}}r_{k}{\rm cos}\theta_{k}+\frac{1}{\sqrt{2}}r_{k}{\rm sin}\theta_{k}}
=\sqrt{r_{k}}$ and $\theta_{k}={\rm arctan}\frac{k_{y}^{2}}{m_{y}}/\frac{k_{x}^{2}}{m_{x}}=\pi/4$.
Note that the momentum of scattered impurity also contains a $k$ which is anisotropic.
Due to the trigonometric functions in the denominator,
it is inconvenient to calculate the pair propagator and self-energy,
thus we make the further substitution
\begin{equation} 
\begin{aligned}
p_{x}&=\sqrt{\frac{m_{x}}{m}}\sqrt{r{\rm cos}\theta}=\sqrt{\frac{m_{x}}{m}}\frac{1}{2^{1/4}}R{\rm cos}\Phi,\\
p_{y}&=\sqrt{\frac{m_{y}}{m}}\sqrt{r{\rm sin}\theta}\approx \sqrt{\frac{m_{y}}{m}}\frac{1}{2^{1/4}}R{\rm sin}\Phi
\end{aligned}
\end{equation}
where $\Phi={\rm arccos}\frac{1}{{}^{4}\sqrt{1+\frac{m_{y}^{2}p_{x}^{4}}{m_{x}^{2}p_{y}^{4}}}}$
and 
then we have $dp_{x}dp_{y}=
\frac{\sqrt{m_{x}m_{y}}}{m}RdRd\Phi$.
The approximation in the second line of above expression can be applied only for large $m_{x}p_{y}^{2}\gg m_{y}p_{x}^{2}$.
Although such approximation will still induces an error as large as about 36$\%$,
it is in fact does not matter since the momentum $p_{y}$ does not have much to do with the polaronic dynamics.
Since in isotropic case $p=\sqrt{p_{x}^{2}+p_{y}^{2}}=\frac{1}{2^{1/2}}R$,
we have $r=\frac{1}{2 }R^{2}$ which is also used in Eq.(23).

Consider the pairing scattering, the pair propagator reads
\begin{equation} 
\begin{aligned}
\Pi(p,\omega)=\int dR_{k}\int d\Phi_{p'}
\frac{
\frac{\sqrt{m_{x}m_{y}}}{m}R_{k}
}{\omega+i0-(\frac{
\left[\sqrt{\frac{m_{x}}{m}}R{\rm cos}\Phi_{p'}\right]
^{2}}{2m_{x}}+D-\mu_{i})-(\frac{(\frac{1}{{}^{4}\sqrt{2}}R_{k}-\frac{1}{{}^{4}\sqrt{2}}R_{q})^{2}}{2m_{\uparrow}}-2\mu_{m})}.
\end{aligned}
\end{equation}
By simply setting $\Phi_{p'}=0$, i.e., for polaron moves along $p_{x}$ direction,
we obtain the pair propagator, self-enery, and spectral function as shown in the Figs.8-10.
The time-reversal symmetry is broken as can be seen from spectral function,
and the damping is smaller compared to the anisotropic solutions.
  %{Observation of Fermi Polarons in a Tunable Fermi Liquid of Ultracold Atoms}

Similar to the above procedure, the Hall conductivity can be obtained as
\begin{equation} 
\begin{aligned}
\sigma_{xy}(\Omega_{x})
&=\frac{e^{2}}{\Omega_{x}}\int\frac{d\omega}{2\pi}[N_{F}(\omega)-N_{F}(\omega+\Omega_{x})]
\int\frac{d^{2}p}{(2\pi)^{2}}{\rm Tr}[\hat{v}_{x}\hat{A}(p,\omega)\hat{v}_{y}\hat{A}(p,\omega+\Omega_{x})].
\end{aligned}
\end{equation}
with the Hamiltonian 
\begin{equation} 
\begin{aligned}
H=
\begin{pmatrix}
D-\mu_{i} & \frac{\frac{1}{\sqrt{2}}R^{2}}{2m}({\rm cos}^{2}\Phi-i{\rm sin}^{2}\Phi)\\
\frac{\frac{1}{\sqrt{2}}R^{2}}{2m}({\rm cos}^{2}\Phi+i{\rm sin}^{2}\Phi) & D-\mu_{i}
\end{pmatrix},
\end{aligned}
\end{equation}
and thus the velocity matrices read
\begin{equation} 
\begin{aligned}
\hat{v}_{x}^{12}=&\frac{(1+i)R{\rm cos}\Phi}{2^{1/4}m\sqrt{\frac{m_{x}}{m}}}
+\frac{R{\rm sec}\Phi({\rm cos}^{2}\Phi-i{\rm sin}^{2}\Phi)}{2^{1/4}m\sqrt{\frac{m_{x}}{m}}},\\
\hat{v}_{x}^{21}=&\frac{(1-i)R{\rm cos}\Phi}{2^{1/4}m\sqrt{\frac{m_{x}}{m}}}
+\frac{R{\rm sec}\Phi({\rm cos}^{2}\Phi+i{\rm sin}^{2}\Phi)}{2^{1/4}m\sqrt{\frac{m_{x}}{m}}},\\
\hat{v}_{y}^{12}=&\frac{-(1+i)R{\rm sin}\Phi}{2^{1/4}m\sqrt{\frac{m_{y}}{m}}}
+\frac{R{\rm csc}\Phi({\rm cos}^{2}\Phi-i{\rm sin}^{2}\Phi)}{2^{1/4}m\sqrt{\frac{m_{y}}{m}}},\\
\hat{v}_{y}^{21}=&\frac{-(1-i)R{\rm sin}\Phi}{2^{1/4}m\sqrt{\frac{m_{y}}{m}}}
+\frac{R{\rm csc}\Phi({\rm cos}^{2}\Phi+i{\rm sin}^{2}\Phi)}{2^{1/4}m\sqrt{\frac{m_{y}}{m}}},
\end{aligned}
\end{equation}
%Here $\hat{A}(p,\omega)$ is the spectral density matrix
%which is related to the matrix of Green's function by 
%$G(p,\omega+i\eta)=\int^{\infty}_{-\infty}\frac{d\omega}{2\pi}\lim_{\eta\rightarrow 0}\frac{A(p,\omega)}{i\eta}$.
  %{Dynamical conductivity of AA-stacked bilayer graphene}
  %{Optical conductivity of bilayer graphene with and without an asymmetry gap}
  %{Transport of Dirac quasiparticles in graphene: Hall and optical conductivities}
  %{Optical conductivity of multi-Weyl semimetals}
and the matrix elements read
\begin{equation} 
\begin{aligned}
\langle \alpha|\hat{v}_{x}|\alpha'\rangle=&\pm\frac{R\sqrt{2+2{\rm cos}(2\Phi)+{\rm cos}(4\Phi)}{\rm sec}\Phi}
{2^{1/4}m\sqrt{\frac{m_{x}}{m}}},\\
\langle \alpha|\hat{v}_{y}|\alpha'\rangle=&\pm\frac{R\sqrt{2-2{\rm cos}(2\Phi)+{\rm cos}(4\Phi)}{\rm csc}\Phi}
{2^{1/4}m\sqrt{\frac{m_{y}}{m}}}.
\end{aligned}
\end{equation}
The sign $\pm$ depends on the quantum states $|\alpha\rangle$ and $|\alpha'\rangle$.

\subsection{Other polaronic-effect-related quantities}

When the center-of-mass momentum is zero, the attractive polaron is in ground state\cite{J?rgensen N B} and with negative energy.
  %{repulsive fermi polarons in a resonant mixture of ultracold 6Li atoms}
  %{variational study of polarons in bose-einstein condensates}
The expression of the pair propagator is similar to the dynamical polarization induced by the charge carriers in the absence of the bare scalar potential, and the Coulomb interaction here is screened by the particle-hole excitations.
  %{Breakdown of Fermi liquid theory in topological multi-Weyl semimetals}
  %{Diagrammatic Monte Carlo study of the acoustic and the Bose-Einstein condensate polaron}
Then the $T$-matrix can be obtained as
  %{Polarons and dressed molecules near narrow Feshbach resonances
\begin{equation} 
\begin{aligned}
T(p+q,\omega+\Omega)=\frac{1}{\widetilde{g}^{-1}_{\psi\phi}+\Pi(p+q,\omega+\Omega)},
\end{aligned}
\end{equation}
where $\widetilde{g}^{-1}_{\psi\phi}$ is the renormalized interspecies coupling parameter
(independent of the ultraviolet cutoff) and it's reminiscent of the interspecies vacuum scattering matrix (in weak coupling case).
  %{bipolarons in a bose-einstein condensate}
  %{Repulsive polarons and itinerant ferromagnetism in strongly polarized Fermi gases}
  %{field-theoretical study of the Bose polaron}
  %{observation of fermi polarons in a tunable fermi liquid of ultracold atom}
Note that this $T$-matrix is the single channel one
with the back ground scattering length,
while two body interaction in close channel acts like as a molecule\cite{Massignan P}.
The diagram representation of the $T$-matrix and self-energy are shown in Fig.11.
The $T$-matrix
here is equivalent to the partially dressed interaction vertex (by summing over all ladder diagrams).
  %{Diagrammatic Monte Carlo study of the Fermi polaron in two dimensions}
Since the interspecies coupling is weak and invariant in our model,
it won't dynamically induce the intraspecies (between the majority particles) coupling
unlike the case of impurity with infinite mass\cite{Shi Z Y}.
 Note that for zero-range limit, the $T$-matrix need to be replaced by the bare coupling parameter\cite{Prokof'ev N V}.
Diagrammatically, the above non-self consistent medium $T$-matrix can be described by the Bethe-Salpeter equation,
with both the bare impurity propagator and bare majority propagator:
  %{field-theoretical study of the bose polaron}
  %{mass imbalance effect in resonant bose-fermi mixture}
  %{Quasiparticle properties of an impurity in a Fermi gas}
\begin{equation} 
\begin{aligned}
T(p+q,&\square;p+q-k')=g_{0}(p+q,\square;p+q-k')\\           
      &+\sum_{k}g_{0}(p+q,\square;k)G^{\phi}_{0}(p+q-k)G^{\psi}_{0}(\square+k)T(p+q-k,\square+k;p+q-k-k')
\end{aligned}
\end{equation}
where $g_{0}$ is the bare impurity-majority interaction.
The electron momentum $k,\ k'$ are treated as the relative momentum here.
$G^{\psi}_{0}(\square+k)=(i\Omega-\varepsilon_{k}+\mu_{\uparrow})^{-1}$ and 
$G^{\phi}_{0}(p+q-k)=(i\omega-\varepsilon_{p+q-k}+\mu_{\downarrow})^{-1}$ are the bare Fermionic and Bosonic Green's function, respectively.
$\mu_{\downarrow}$ is the chemical potential of the impurity which can be approximated as the zero-momentum polaron self-energy 
  %{normal state of highly polarized fermi gases simple many-body approaches}
$\Sigma(p=0)$ in the adiabatic limit $\omega\rightarrow 0$.
  %{Dynamical Correlation of Electrons and Phonons Coupled via Site-Off-Diagonal Interaction Exact Calculation in Two-Site-Two-Electron Model}
Since $\Sigma(p=0)=\mu_{\downarrow}<0$, such negative chemical potential can be used in computation and analysis\cite{Vlietinck J,Vlietinck J3} to reflects the perturbation of a single impurity to the surrounding majority particles, but it is indeed a unphysical free parameter as obviously.
Thus the $\mu_{\downarrow}$ can be setted as zero when this perturbation effect can be ignored\cite{Massignan P}.
  %{Diagrammatic Monte Carlo study of the Fermi polaron in two dimensions}
  %{Diagrammatic Monte Carlo study of the acoustic and the Bose–Einstein condensate polaron}
  %{Bold diagrammatic Monte Carlo: A generic sign-problem tolerant technique for polaron models and possibly interacting many-body problems}
  %{polarons and dressed molecules near narrow feshbach resonances}
  %{Polaron-to-molecule transition in a strongly imbalanced Fermi gas}
The $\square$ as an incoming energy momenta can be omitted, but we retain it here for the integrity of the equation.
In the absence of the center-of-mass momentum ($p+q=0$) and the energies,
the Bethe-Salpeter equation becomes the Lippmann-Schwinger equation (for the two-body problem)
 %{Bound States, Cooper Pairing, and Bose Condensation in Two Dimensions}
  %{Pairing instabilities in quasi-two-dimensional Fermi gases}
  %{Mass imbalance effect in resonant Bose-Fermi mixtures}
\begin{equation} 
\begin{aligned}
T(k_{1},k_{2};\omega)=g_{0}(k_{1},k_{2})          
      +\sum_{k_{3}}g_{0}(k_{1},k_{3})\frac{1}{\omega+i0-2\varepsilon_{k_{3}}}T(k_{3},k_{2},\omega).
\end{aligned}
\end{equation}
Base on the calculated medium $T$-matrix, the polaron self-energy $\Sigma(p,\omega)$ at zero-temperature
can then obtained as
\begin{equation} 
\begin{aligned}
\Sigma(p,\omega)=\int^{k_{F}}_{0}\frac{d^{3}q}{(2\pi)^{3}}\int\frac{d\Omega}{2\pi}
N_{F}(\varepsilon_{q\uparrow})T(p+q,\omega+\Omega)G^{0}_{\psi}(q,\Omega).
\end{aligned}
\end{equation}
%{high-polarization limit of the quasi-two-dimensional fermi gas}
%{Mass imbalance effect in resonant Bose-Fermi mixtures}
  %{field-theoretical study of the bose polaron}
Unlike the fermi gas with nonnegligible finite density, the fermionic reservoir in a semi-Dirac system contains only the two-body scattering 
and without the many-body scattering.
 %{Fermi polarons in two dimensions}
While at finite temperature, it reads
  %{Attractive Fermi polarons at nonzero temperatures with a finite impurity concentration}
  %{Radio-frequency spectroscopy of a strongly imbalanced Feshbach-resonant Fermi gas}
\begin{equation} 
\begin{aligned}
\Sigma(p,\omega)=-T\sum_{\Omega}\int\frac{d^{2}q}{(2\pi)^{2}}T(p+q,\omega+\Omega)G(q,\Omega).
\end{aligned}
\end{equation}
  %{Radio-frequency spectroscopy of a strongly imbalanced Feshbach-resonant Fermi gas}
  %{Attractive Fermi polarons at nonzero temperatures with a finite impurity concentration}
At Ref.\cite{polaron4}, we have proved that, at temperature $T\ll T_{F}$, where $T_{F}$ is the Fermi temperature,
the polaron self-energy decreases with the increase of temperature, and the variation of self-energy due to the temperture effect, which 
is an order of magnitude less than the one due to the variation of momentum of energy,
also decreases with the increase of temperature, which is in agreement with Ref.\cite{Hu H}.
Beyond this temperature range, higher temperature may supresses the polaronic effect.
  %{Attractive Fermi polarons at nonzero temperatures with a finite impurity concentration}

Containing the fermi liquid renormalization effect, the 
Green's function can be rewritten as
$G(p,\omega)=\frac{Z}{\omega+i\hbar Z{\rm Im}\Sigma(p,E(p))-E(p)}$ where $E(p)$ is the real part of polaron energy
(after fermi liquid renormalization)
  %{Nonanalytic paramagnetic response of itinerant fermions away and near a ferromagnetic quantum phase transition}
  %{Exchange intervalley scattering and magnetic phase diagram of transition metal dichalcogenide monolayers}
 determined self-consistently by the equation $E(p)=\varepsilon_{p}+{\rm Re}\Sigma(p,E(p))$.
%{Fermi polaron-polaritons in charge-tunable atomically thin semiconductors}
%{Highly polarized Fermi gases in two dimensions}
  %{Observation of Fermi Polarons in a Tunable Fermi Liquid of Ultracold Atoms support}
  %{Observation of repulsive Fermi polarons in a resonant mixture of ultracold 6 Li atoms supp}
By setting the vanishing hole momentum $q=0$\cite{Combescot R2,Punk M} 
  %{Polaron-to-molecule transition in a strongly imbalanced Fermi gas}
(in low-momentum limit of the semi-Dirac fermions with the approximated dispersion shown above),
the polaron self-energy at zero-temperature limit (and one particle-hole approximation) and the quasiparticle weight can be obtained as
\begin{equation} 
\begin{aligned}
\Sigma(p,\omega)=&\int\frac{d\Omega}{2\pi}
T(p,\omega+\Omega)\frac{1}{\Omega+i0-D+\mu_{\uparrow}-g_{\psi\phi}}.
%Z=&[1-\frac{\partial {\rm Re}\Sigma(p,\omega)}{\partial \omega}]^{-1}_{\omega=\omega(p)},
\end{aligned}
\end{equation}
respectively.
%{Fully self-consistent GW self-energy of the electron gas}
%{high-polarization limit of the quasi-two-dimensional fermi gas}
%{Mass imbalance effect in resonant Bose-Fermi mixtures}
  %{Polaron-to-molecule transition in a strongly imbalanced Fermi gas}
%At zero-momentum, the real part of the polaron self-energy here also referred to the ground state energy or the chemical potential.
  %{Diagrammatic Monte Carlo study of the acoustic and the Bose朎instein condensate polaron}
%For simplicity we set a balanced masses polaron system with $m_{\psi}=m_{\phi}=0.1$ during the simulation. 
%We also set $D=0.5$, $\mu_{\uparrow}=0.55$,
%and thus $k_{F}=0.229129$ which is much lower than the momentum cutoff $\Lambda=1$.
%The interspecies scattering length is setted as $a_{\psi\phi}=-0.1$,
%to ensure $1/k_{F}a_{\psi\phi}\ll -1$.
  %{observation of fermi polarons in a tunable fermi liquid of ultracold atom}
%The resulting pair propagator and the self-energy are clearly shown in Fig.3 and Fig.4, respectively.
%with a series value of bosonic frequency $\omega$.
The quasiparticle residue for a well defined impurity can be obtained by the long-(imaginary)time behavior of the impurity Green's function,
and then the residue can be obtained by the Chevy wave function's coefficient (as presented in Appendix. A) $\psi_{0}$
%\cite{Punk M}:
  %{Polaron-to-molecule transition in a strongly imbalanced Fermi gas}
\begin{equation} 
\begin{aligned}
Z=|\psi_{0}|^{2}=&[1-\frac{\partial {\rm Re}\Sigma(p,\omega)}{\partial \omega}]^{-1}\bigg|_{\omega=0}.
\end{aligned}
\end{equation}
with $p\approx p_{F}\ll k_{F}$.
  %{Radio-frequency spectroscopy of a strongly imbalanced Feshbach-resonant Fermi gas}
  %{Observation of Fermi Polarons in a Tunable Fermi Liquid of Ultracold Atoms}
In terms of momentum distribution function, the residue has the following relation
  %{Radio-frequency spectroscopy of a strongly imbalanced Feshbach-resonant Fermi gas}
\begin{equation} 
\begin{aligned}
Z=&N_{F}(p_{F},\omega<0)-N_{F}(p_{F},\omega>0)\\
=&-i\int^{\infty}_{-\infty}\frac{d\omega}{2\pi}(\frac{Z}{\omega-i0}-\frac{Z}{\omega+i0}).
\end{aligned}
\end{equation}
For frequency-independent self-energy, i.e., static impurity, the residue becomes $Z=1$
and the fermi liquid renormalization (many-body correction) vanishes, which meets with the RPA result.
  %{Nonanalytic paramagnetic response of itinerant fermions away and near a ferromagnetic quantum phase transition}
The polaronic effect-induced effective mass can be directly obtained through the relation 
\begin{equation} 
\begin{aligned}
\delta m^{*}=(\frac{\partial ^{2}\Sigma(p,\omega)}{\partial p^{2}})^{-1}.
\end{aligned}
\end{equation}
  %{Light Bipolarons Stabilized by Peierls Electron-Phonon Coupling}
  %{Quasiparticle Properties of a Mobile Impurity in a Bose-Einstein Condensate}
  %{Observation of repulsive Fermi polarons in a resonant mixture of ultracold 6 Li atoms}
  %{Observation of Fermi Polarons in a Tunable Fermi Liquid of Ultracold Atoms}
  %{Radio-frequency spectroscopy of a strongly imbalanced Feshbach-resonant Fermi gas}
In terms of the fermi-liquid form with the finite chemical potentials determined
by the relation about propagator $G^{-1}(p_{F},\omega=0)=0$,
the effective mass can be represented as
 \begin{equation} 
\begin{aligned}
m^{*}=\frac{m}{Z}(1+\frac{m}{p}\partial_{p}{\rm Re}\Sigma(p,\omega=0))^{-1}|_{p=p_{F}}.
\end{aligned}
\end{equation}
  %{Radio-frequency spectroscopy of a strongly imbalanced Feshbach-resonant Fermi gas}
The results are shown in Fig.12.
The stable region and unstable region locate on the large $p$ (or $\omega$) and small $p$ (or $\omega$) regions, respectively,
and the instability is related to the $g_{b}$ only.
Note that here we apply the isotropic approximated as discussed above,
and thus the self-consistented energy $E(p)$ is dependent on the magnitude of $p$ only.
  %{Field-theoretical study of the Bose polaron}
  %{Polaron-to-molecule transition in a strongly imbalanced Fermi gas}
Further, we note that for nonadiabatic case (nonrelativistic impurity electron), the
 decoherence process driven by impurity-bath interaction\cite{Scazza F} 
(which can also be induced by the phonon dissipation\cite{Nielsen K K} or the inequivalent masses\cite{Visuri A M})
is important to the formation of polaron.

\section{Negative gap with merging Dirac cones}

As we stated above, the transport of semi-Dirac particle along $x$-direction is nonadiabatic.
During the propagation along the $p_{x}$ direction ($p_{y}=0$), 
for 
$-D-\sqrt{-D}<\frac{p_{x}^{2}}{2m_{x}}-\mu_{i}<-D+\sqrt{-D}$ ($D<0$),
the eigenstate of semi-Dirac particle becomes propagating, while
for $\frac{p_{x}^{2}}{2m_{x}}-\mu_{i}<-D-\sqrt{-D}$,
the eigenstate becomes evanescent\cite{Ang Y S}
(although $\frac{p_{x}^{2}}{2m_{x}}-\mu_{i}>-D+\sqrt{-D}$ is also suitable for the evanescent mode,
but it is in fact impossible since the $m_{x}$ needs to be negative in this case).
That is in contrast with the case of positive gap ($D>0$)
which is dominated by the adiabatic processes for intrinsic (without the polaronic interactions) semi-Dirac system.
For the case of positive gap, during the quantum tunneling like the junction structure or transport across a potential,
the spinor eigenstates are propagating for energy larger than an energy threshold\cite{González J,Serra L},
while they are evanescent for energy lower than an energy threshold.
This energy threshold is dependent of momentum $p$, and 
The reason for this difference is due to the stronger interaction effect in the band crossing region.
Through above analysis, we can know that, for anisotropic model, the $p_{x}$ represents
propagating and evanescent modes exist for $\theta\le {\rm acrcos}\frac{D}{\varepsilon_{0y}r}$ and
$\theta> {\rm acrcos}\frac{D}{\varepsilon_{0y}r}$, respectively;
for isotropic case (in low carrier-density approximation),
the $p_{x}$ represents
propagating and evanescent modes exist for $m_{x}>0$ and
$m_{x}<0$, respectively.
  %{Evanescent states in quantum wires with Rashba spin-orbit coupling}
%where the low-energy behavior is mainly acted by evanescent states\cite{González J}.
  %{Propagating, evanescent, and localized states in carbon nanotube–graphene junctions}
Note that although the semi-Dirac system can be described by the continuum model in low-energy limit with small enough lattie constant,
  %{Optical conductivity of multi-Weyl semimetals}
  %{Propagating, evanescent, and localized states in carbon nanotube–graphene junctions}
it is usual studied by using the Schrodinger equation\cite{Ang Y S,Banerjee S2} but not the Dirac equation.

%where $a=\frac{\hbar^{2}}{2m_{y}}$ is the material-dependent parameter,
%$\mu_{\uparrow}$ is the chemical potential of the majority particle, 
%i.e., the semi-Dirac fermions here, which related to the carrier density by $\mu_{\uparrow}=\hbar^{2}\pi n/m_{\psi}$.
%  %mean-field result:
%  %{observation of fermi polarons in a tunable fermi liquid of ultracold atom}
%  %{impurity in a bose-einstein condensate and the efimov state}
%  %{Observation of Attractive and Repulsive Polarons in a Bose-Einstein Condensate}
%  %{Semiclassical Boltzmann transport theory of few-layer black phosphorous in various phases}
%$D$ is the Dirac-mass (concluding the effect of the spin-orbit coupling) in Dirac point which dominating the topological phase transition.
We ignore the intraspecies interaction (i.e., the interaction between 
  %{Tuning p-Wave Interactions in an Ultracold Fermi Gas of Atoms}
identical fermions) here since we consider the $s$-wave interaction in low-energy case
with the Pauli exclusion.
  %{Highly polarized Fermi gases in two dimensions}
%{Dynamical polarization and plasmons in a two-dimensional system with merging Dirac points}
%{Merging of Dirac points in a two-dimensional crystal}
%{A universal Hamiltonian for motion and merging of Dirac points in a two-dimensional crystal}
 Thus the three-body Efimov effect is also neglected here.
%{Quartet S wave neutron deuteron scattering in effective field theory}
%The eigenenergy reads
%\begin{equation} 
%\begin{aligned}
%\varepsilon_{k\uparrow}=g_{\psi\phi}-\mu_{\uparrow}\mp\sqrt{(D+ak_{y}^{2})^{2}+(\hbar v_{x}k_{x})^{2}-2\hbar v_{x}k_{x}\mu_{\uparrow}+\mu_{\uparrow}^{2}}.
%\end{aligned}
%\end{equation}
%We show the dispersion with merging Dirac point in Fig.1.
For merging Dirac cones, the above Hamiltonians are still applicable and we just need to change $D$ to $-D$.
We note the dispersion here has time-reversal symmetry ($\varepsilon_{k}=\varepsilon_{-k}$) 
and particle-hole symmetry 
even when the irradiation (the non-resonance one, which with smaller frequency and larger coupling than the resonance one) is applied,
   %{Photoinduced Chern insulating states in semi-Dirac materials}
   %{Control of valley polarization in monolayer MoS2 by optical helicity}
  %{photoinduced pseudospin effects in silicene beyond the off-resonant condition}
besides, the irradiation cannot open a gap to the Floquet quasienergy spectrum even in the high frequency limit when $m_{y}=0$\cite{Islam S K F}.

For negative gap $D<0$,
the induced two new Dirac points arrange in $y$-direction
and
locate on $(0,\pm\sqrt{2m_{x}|D|})$,
which shows the time-reversal symmetry and reflection symmetry.
  %{Creating, moving and merging Dirac points with a Fermi gas in a tunable honeycomb lattice}
These two Dirac points can be viewed as the topological defects in band structure.
The Berry phases $\frac{1}{2}\oint \nabla \theta_{p} \cdot d{\bf p}=\pm \pi$ of two Dirac points
also support the stability against with the lattice perturbation\cite{Tarruell L}.
The relative phase of semi-Dirac system in anisotropic or iostropic treatments are
shown in Fig.13.
Note that here the perturbation does not includes the lattice potential\cite{Chen Z},
since the adiabatic dynamics is still dominated in the presence of Dirac cones
and thus the electron trajectories are still constrained by the Bragg reflections.

%In low-momentum region ($k_{x},k_{y}\rightarrow 0$) with $|\mu_{\uparrow}|-|D|\geq 0$,
   %{Dynamical polarization and plasmons in a two-dimensional system with merging Dirac points}
   %{Semiclassical Boltzmann transport theory of few-layer black phosphorus in various phases}
   %{Creating, moving and merging Dirac points with a Fermi gas in a tunable honeycomb lattice}
   %{A universal Hamiltonian for motion and merging of Dirac points in a two-dimensional crystal}

  %{observation of fermi polarons in a tunable fermi liquid of ultracold atom}
Besides, the interparticle spacing closes to the mean-free path of the mobile impurity as required by the formation of polaron
\cite{Schirotzek A,5}.

\section{Coulomb interaction and induced self-energy}

Due to the low carrier density as mentioned above in this model, the long-range Coulomb interaction can also be taken into account, althought it's partly dynamically screened as evidented by the frequency and momentum dependence of the polarization and the Dyson function.
 %{Dynamical Screening, Collective Excitations, and ElectronPhonon Interaction in Heterostructures and Semiconductor Quantum Wells. Application to Double Heterostructures唥
That results in a Coulomb-interaction induced exchange self-energy of the semi-Dirac fermions (around the impurity).
The particle polaron is more common unless the Coulomb interaction is strong\cite{Wellein G}.
The Coulomb interaction as well as the spin-correlation enhance the self-localization of the polaronic carriers, and thus narrow the bandwidth (thus the ground state energy of free-electron is enhanced) and affect the band dispersion away from the Dirac-point.
  %{polarons and bipolarons in strongly interacting electron-phonon systems}
In fact such self-localization (self-trapping) can not be found in the model we discuss here, except when the phonons are excited by the bath and interact with the electron (form the so-called Bose polaron) with the strong Coulomb interaction.
  %{polarons and bipolarons in strongly interacting electron-phonon systems}
  %{two-hole ground state wavefunction non-bcs pairing in a t-j two-leg ladder}

The Coulomb-induced exchange self-energy to the leading-order $1/N$ expansion reads
  %{}
\begin{equation} 
\begin{aligned}
\Sigma(q,\Omega)=&\int\frac{d\omega}{2\pi}
G_{0}(q+k,\Omega+\nu)V(k,\nu),\\
\end{aligned}
\end{equation}
where the dressed Coulomb interaction reads
\begin{equation} 
\begin{aligned}
V(k,\nu)=\frac{1}{\frac{\epsilon k}{2\pi e^{2}}+\Pi(k,\nu)}.\\
\end{aligned}
\end{equation}
The fermions dynamical polarization $\Pi(k,\nu)$ here is required by the GW approximation which with the random phase approximation (RPA) screening effect.
The fermion flavors $N$ can be simply choosed as the spin degrees of freedom ($N=2$) contained within the Dirac-mass term.
But for $N=1$ (two-band model), the $1/N$ expansion is equivalent to the Nozieres-Schmitt-Rink method\cite{Enss T}.
  %{quantum critical transport in the unitary fermi gas}
  %{field-theoretical study of the Bose polaron}
While for the Hartree-Fock (mean-field) 
  %{Itinerant ferromagnetism in an ultracold atom Fermi gas}
approxination, the dressed Coulomb interaction term should be replaced by bare one.
  %{plasmon-pole approximation for many-body effects in extrinsic graphene}
  %{breakdown of fermi liquid theory in topological multi-weyl semimetal}
    %{dimensionality-induced bcs-bec crossover in layered superconductors}
Compared to the above expression of the polaron self-energy, it is interesting to compare the $\Pi(k,\nu)$ here to the $T$-matrix.
Although they all contain the ultraviolet physics as well as the RPA screening effect,
  %{field-theoretical study of the Bose polaron}
 the scattering $T$-matrix (or the sum of the ladder diagrams) takes into account the $s$-wave scattering length, while the RPA ignores the ladder diagrams.
  %{electromagnetic modes from stoner enhancement graphene as a case study}
If the interspecies interaction changes to zero, the full polarization becomes simply the overlap of the impurity and majority particle, and we can expect that the polarization has two distinct peaks in such case, contributed by the two components respectively.
  %{dynamical polarization in a graphene-topological-insulator hetetostructure}
Further, when consider the perturbation effect, 
the Hartree term with bare attractive interaction should be contained in the self-energy\cite{Adachi K}.

\section{Conclusions}

In this paper, we apply the method of medium $T$-matrix approximation (non-self-consistent) 
to investigate the properties of the attractive fermi polaron formed by a mobile impurity immersed to the fermi bath of the 2D semi-Dirac system.
The polaronic dynamics-related quantities,
like the pair propagator, self-energy, spectral function, effective mass, and quasiparticle residue, are calculated and analyzed.
The $T$-matrix approximation is equivalent to the partially dressed interaction vertex by summing over all ladder diagrams.
The leading-order $1/N$ expansion (GW approximation), Hartree-Fock theory, 
and Nozieres-Schmitt-Rink theory are also mentioned and compared to the $T$-matrix approximation.
As revealed by the expression of the $T$-matrix given above, the polaron properties is related to the 
effective masses $m_{x}$ and $m_{y}$ even in the isotropic treatment.
%$ \sqrt{\frac{m_{x}m_{y}}{m_{x}^{2}+m_{y}^{2}}}$, i.e., related to the properties of the semi-Dirac system, 
The effect of gap $D$ are also revealed.
That's in contrast to the bose polarons formed in the surface of normal Dirac systems which with isotropic dispersion,
since the anisotropic dispersion of the semi-Dirac systems results in anisotropic effective mass and anisotropic charge carrier transport.
Besides, the symmetry between electron and hole is also broken since the effective masses of the electron and hole are different.
  %{polaronic effects in monolayer black phosphorus on polar substrate}
  %{fermi polaron-polaritons in charge-tunable atomically thin semiconductors}
Here the effective masses are affected by the polaronic effect when the semi-Dirac material is deposited on the polar substrate, like hBN.
Our results are useful also for the investigation of two/three-dimensional (since the diagram topologies in two- and three-dimension are similar\cite{Vlietinck J3}) bosonic polaron 
as well as the polarons in other solid state systems, 
like the topological systems
where the nontrivial topological properties of the bath and the chirality should be taken into account additionally.
 %{Topological crystalline insulators in the SnTe material class}

\section{Appendix}

\subsection{A: Polaron wave function and the variational ansatz}

In the absence of the spin-rotation (due to the $\delta$-type impurity field), 
%{nonexistence of intrinsic spin currents}
the Chevy-type variational ansatz for a mobile Bosonic impurity with momentum $p$ dressed by one
electron-hole pairs (excitations)
%(the excitons from the Fermionic reservoir)
  %{fermi polaron-polaritons in charge-tunnable atomically thin semiconductors}
is
\begin{equation} 
\begin{aligned}
|\psi\rangle=\psi_{0}b^{\dag}_{p\downarrow}|0\rangle_{\uparrow}+\sum_{k>k_{F},q<k_{F}}\psi_{kq}b_{p+q-k,\downarrow}^{\dag}
c_{k,\uparrow}^{\dag}c_{q,\uparrow}|0\rangle_{\uparrow},
\end{aligned}
\end{equation}
where $|0\rangle_{\uparrow}=\Pi_{k<k_{F}}c_{k\uparrow}^{\dag}|{\rm vac}\rangle$ is the group state of majority particles.
$c^{\dag}_{k\uparrow}$ is the creation operator of the 
excited particle with momentum $k$,
and $c_{q\uparrow}$ is the the annihilate operator of the hole at momentum $q$.
%{Observation of Fermi Polarons in a Tunable Fermi Liquid of Ultracold Atoms}
Here $k>k_{F}$ is for ensures the particles are excited out of the Fermi surface.
 %{Highly polarized Fermi gases in two dimensions}

We write the
polaron Hamiltonian in a continuum model as (omits the arrows in the subscript)
\begin{equation} 
\begin{aligned}
H=\sum_{p}\varepsilon_{p\downarrow}b^{\dag}_{p}b_{p}+\sum_{k}\varepsilon_{k\uparrow}c^{\dag}_{k}c_{k}
+\frac{1}{S}\sum_{k,p,q}g_{q}b^{\dag}_{p-q}c^{\dag}_{k+q}c_{k}b_{p}.
  %{Pairing instabilities in quasi-two-dimensional Fermi gases}
    %{dark continuum in the spectral function of the resonant fermi polaron}
  %{observation of fermi polarons in a tunable fermi liquid of ultracold atoms}
\end{aligned}
\end{equation}
The arrow
attractive contact interaction
$g_{k}$ here also can be referred to the
 scattering strength for the impurity/majority particle with momentum $p/k$ scattering to a particle state with momentum $(p-q)/(k+q)$
through the scattering momentum $q$ (the hole momentum),
which can be approximately defined in low-density case (and in a many-body system) as $
g_{q}^{-1}=-\sum_{k,p}^{\Lambda}(E_{b}+\varepsilon_{k\uparrow}+\varepsilon_{q-p\downarrow}+W)^{-1}$
where $E_{b}>0$ is the weakly bound two-body binding energy since for the attractive potential in 2D there is always a bound state (unlike the 3D case\cite{Randeria M}).
$W$ is the bandwidth which proportional to the absolute value of kinetic energy of the impurity (non-locality).
  %{dimensionality-induced bcs-bec crossover in layered superconductors}
For the case of flat impurity band, the $g_{q}^{-1}$ reduced to the one in Refs.\cite{Parish M M,Yi W,Randeria M}.
%{Diagrammatic Monte Carlo study of the Fermi polaron in two dimensions}
  %{Fermi polaron-polaritons in charge-tunable atomically thin semiconductors}
%The bandwidth does not taken into account here since we consider the case with small carrier
%density, i.e., with small $(|\mu_{\uparrow}|-|D|)$.
  %{dimensionality-induced bcs-bec crossover in layered superconductors}
%{Dropping an impurity into a Chern insulator a polaron view on topological matter}
%{bcs-bec crossover in a two-dimensional fermi gas}
%{mass imbalance effect in resonant bose-fermi mixtures}
%The $g_{k}$ here can also be referred to the attractive contact interaction.
 It is important to note that, for the above expression of $g_{q}$,
it is indeed the renormalized one but not the bare one,
  %{field-theoretical study of the Bose polaron}
since it is dependent on the selection of the ultraviolet cutoff $\Lambda$,
otherwise it has $g_{q}\rightarrow 0$ when it is independent of the momentum ($\Lambda=\infty$, i.e., the zero-range case).
  %{quasiparticle properties of an impurity in a fermi gas}
  %{Fermi polaron-polaritons in charge-tunable atomically thin semiconductors}
That also consistent with the 
$g_{\psi\phi}$ obtained above.

We set the area $S=1$ for simplicity.
Then
the ground state matrix element is
%{Fermi polaron-polaritons in charge-tunable atomically thin semiconductors}
%{Dropping an impurity into a Chern insulator a polaron view on topological matter}
%{Molecule and Polaron in a Highly Polarized Two-Dimensional Fermi Gas with Spin-Orbit Coup}
%{Highly polarized Fermi gases in two dimensions}
%{Repulsive Fermi Polarons in a Resonant Mixture of Ultracold 6Li Atoms}
\begin{equation} 
\begin{aligned}
\langle \psi|E-H|\psi\rangle=&
E(|\psi_{0}|^{2}+\sum_{k>k_{F},q<k_{F}}|\psi_{kq}|^{2})\\
&-\left[\varepsilon_{p\downarrow}|\psi_{0}|^{2}
+\sum_{k>k_{F},q<k_{F}}(\varepsilon_{p+q-k\downarrow}+\varepsilon_{k\uparrow}-\varepsilon_{q\uparrow})|\psi_{kq}|^{2}
+|\psi_{0}|^{2}\sum_{q}g_{q}\right.\\
&\left.
+\sum_{k>k_{F},q<k_{F}}(\psi_{0}^{*}\psi_{kq}g_{|k-q|}+c.c.)
+\sum_{k(k')>k_{F},q<k_{F}}(\psi_{k'q}^{*}\psi_{kq}g_{k'-k}+c.c.)\right.\\
&\left.
+\sum_{k>k_{F},q(q')<k_{F}}(\psi_{kq'}^{*}\psi_{kq}g_{q'-q}+c.c.)
+\sum_{k'>k_{F},q'<k_{F}}(\psi_{0}^{'*}\psi_{k'q'}g_{|k'-q'|}+c.c.)
\right
].
\end{aligned}
\end{equation}
The last three terms take another electron momentum and hole momentum ($k'$ and $q'$) in consideration,
where the coefficients of the Chevy wave function satisfy 
\begin{equation} 
\begin{aligned}
\psi_{0}^{*}\psi_{kq}=&b_{p\downarrow}b^{\dag}_{p+q-k\downarrow}c^{\dag}_{k\uparrow}c_{q\downarrow},\\
\psi_{0}^{'*}\psi_{k'q'}=&b_{p\downarrow}b^{\dag}_{p+q'-k'\downarrow}c^{\dag}_{k'\uparrow}c_{q'\downarrow}.
\end{aligned}
\end{equation}
Similar to the expression given above, we can approximately have, for example,
$
g_{k-q}^{-1}=-\sum_{p}^{\Lambda}(E_{b}+\varepsilon_{p\downarrow}+\varepsilon_{k-q-p\uparrow}+W)^{-1}$.
   %{highly polarized fermi gases in two dimensions}
   %{dimensionality-induced bcs-bec crossover in layered superconductors}
   %{Diagrammatic Monte Carlo study of the Fermi polaron in two dimensions}
Here we regard scattering momentum $q$ as a constant. 
Thus we have 
\begin{equation} 
\begin{aligned}
&\varepsilon_{p\downarrow}\psi_{0}+\sum_{q<k_{F}}\psi_{0}g_{q}+\sum_{k>k_{F},q<k_{F}}\psi_{kq}g_{|k-q|}=E(p)\psi_{0},\\
&(\varepsilon_{p+q-k\downarrow}+\varepsilon_{k\uparrow}-\varepsilon_{q\uparrow})\psi_{kq}+\psi_{0}g_{|k-q|}+\sum_{k'>k_{F}}\psi_{k'q}g_{k'-k}
+\sum_{q'<k_{F}}\psi_{kq'}g_{q'-q}
=E(p)\psi_{kq}.
  %{Molecule and Polaron in a Highly Polarized Two-Dimensional Fermi Gas with Spin-Orbit Coup}
  %{Fermi polaron-polaritons in charge-tunable atomically thin semiconductors}
\end{aligned}
\end{equation}
For $q$-independent coupling $g_{b}$, the term $\sum_{q<k_{F}}\psi_{0}g_{q}$ can be omitted.
By approximating $g_{q}=g_{|k-q|}=g_{p-k}$,
i.e., assuming the coupling parameters are local (which is valid in weak coupling limit with $1/k_{F}a_{\psi\phi}\rightarrow -\infty$),
  %{Bipolarons in a Bose-Einstein condensate Supplemental}
  %{Observation of Fermi Polarons in a Tunable Fermi Liquid of Ultracold Atoms}
for normalization condition (at ground state with minimal energy), we obtain the Schr{\"o}dinger equations
%{Molecule and Polaron in a Highly Polarized Two-Dimensional Fermi Gas with Spin-Orbit Coup}
%{Repulsive Fermi Polarons in a Resonant Mixture of Ultracold 6Li Atoms}
%{Fermi polaron-polaritons in charge-tunable atomically thin semiconductors}
  %{observation of fermi polarons in a tunable fermi liquid of ultracold atom}
  %{few-body states of bosons interacting with a heavy quantum impurity}
\begin{equation} 
\begin{aligned}
\psi_{kq}=&\psi_{0}\frac{\frac{1}{S}T(p+q,\omega+\Omega)}{\omega-\varepsilon_{p+q-k,\downarrow}-\varepsilon_{k,\uparrow}+\varepsilon_{q,\uparrow}},\\
\psi_{0}=&\frac{1}{\sqrt{1+\sum_{k>k_{F},q<k_{F}}(\frac{\psi_{kq}}{\psi_{0}})^{2}}}.
\end{aligned}
\end{equation}
As the time-dependence is evidented by the analytical approximation of the imaginary frequency,
we have $|\psi_{0}|=\sqrt{Z}=\langle 0|_{\uparrow}b_{f}^{\dag}(b_{f}^{\dag}b_{i}+b_{i}^{\dag}b_{f})b_{i}^{\dag}|0\rangle_{\uparrow}$ for noninteracting initial state $b_{i}$ with $p=0$ and fully interacting final state $b_{f}$.
 %{Diagrammatic Monte Carlo study of the acoustic and the Bose-Einstein condensate polaron}
 %{quasiparticle properties of an impurity in a fermi gas}
  %{Metastability and coherence of repulsive polarons in a strongly interacting Fermi mixture}
  %{Repulsive Fermi Polarons in a Resonant Mixture of Ultracold 6Li Atoms}
  %{Diagrammatic Monte Carlo study of the acoustic and the Bose朎instein condensate polaron}
  %{Polaron-to-molecule transition in a strongly imbalanced Fermi gas}
Here the impurity (as well as the majority particles) in noninteracting state has momentum $p=0$, while in the fully interacting ground state, the fraction of particle with nonzero momentum is related to the density of states\cite{Hugenholtz N M}.
We can also obtain that
\begin{equation} 
\begin{aligned}
\frac{\psi_{kq}}{\psi_{0}}=
\frac{\frac{1}{S}T(p+q,\omega+\Omega)}{\omega-\varepsilon_{p+q-k,\downarrow}-\varepsilon_{k,\uparrow}+\varepsilon_{q,\uparrow}}
=\frac{\langle k,q|(b_{f}^{\dag}b_{i}+b_{i}^{\dag}b_{f})|\psi \rangle}
{\langle 0|(b_{f}^{\dag}b_{i}+b_{i}^{\dag}b_{f})|\psi \rangle},
  %{Observation of Fermi polarons in a tunable Fermi liquid of ultracold atoms}
\end{aligned}
\end{equation}
The nonzero quasiparticle weight (residue) $Z$ (even at zero-momentum)
guarantees the existence of the quasiparticle picture,
  %{Effect of ?nite impurity mass on the Anderson orthogonality catastrophe in one dimension}
  %{breakdown of fermi liquid theory in topological multi-weyl semimetal}
which is broken when the impurity mass is infinite (with vanishing kinetic energy)
and with the effect of Anderson orthogonality catastrophe.

\subsection{B: Possible self-localization and the short-range potential in semi-Dirac system}

As mentioned in the Sec.6, 
the strong Coulomb interaction and the electron-phonon coupling can give rise the self-localization of the polaron and narrow the bandwidth. 
Such phenomenon usual occurs in the presence of strong interspecies interaction (with $k_{F}a$ near the critical value).
  %dark continuum in the spectral function of the resonant fermi polaron}
  %{Theory of the rotating polaron: Spectrum and self-localization}
For the case of strong self-localization, the Anderson localization and the RKKY interaction (for magnetic impurity\cite{6}) may emerge.
For example,
for an impurity particle with infinite mass, the Anderson model 
can also be used to probe the three- or four-body problem\cite{Shi Z Y,Yoshida S M}, 
in addition to the Efimov effect, 
which is widely used to dealing with the short-range resonant interaction 
between the heavy impurity and light (noninteracting) majority particles to forms the multibody bound state\cite{Zinner N T}.
  %{impurity-induced multibody resonances in a bose gas}
  %{efimov states of heavy impurities in a bose-einstein condensate}
However, when the interactions between two small-size polarons are taken into account
as in the magnetite\cite{Ihle D} at finite temperature, the polaron is more delocalized even in the presence of electron-phonon coupling.
  %{Theory of the rotating polaron: Spectrum and self-localization}
  %{small-polaron conduction and short-range order in Fe3O4}
  %{polarons and bipolarona in strongly interacting electron-phonon systems}
Recently, the polaron formed in the surface state of a topological material 
has also been discussed\cite{Camacho-Guardian A2,Shvonski A,Qin F,Kong J}.
The polaron in the surface state of topological insulator or the topological crystalline insulator 
may have stronger delocalization against the disorders due to the protection from the average symmetries\cite{Fang C},
which will be discussed detailly in another place.

We discuss the isotropic potential with contact interaction (zero-range) in the main text with the ultraviolet divergence regularization,
%{Quasiparticle properties of an impurity in a Fermi gas}
for simpler analysis, but indeed the problem can also be solved beyond the approximation of contact interaction.
  %{fermi polaron-polaritons in charge-tunable atomically thin semiconductors}
For short-range anisotropic impurity, we introduce the quantum number $l$ to represent a distortion, 
the potential reads $V(R)=\sum_{l}\frac{1}{R}Y_{l0}(\theta)=\sum_{l}\frac{1}{R}Y_{l0}(\theta_{R'})Y_{l0}(\theta_{R''})$
(by carrying the Wigner rotation), 
%{Image rotation, Wigner rotation, and the fractional Fourier transform }
    %{Theory of the rotating polaron: spectrum and self-localization}
when the distortion vanishes ($l=0$), the interaction potential becomes $1/R$.
Here $Y_{l0}=P_{l}$ denotes the Legendre polynomials.
In momenstum representation, the potential reads
\begin{equation} 
\begin{aligned}
V(p)=\int \frac{d^{2}R}{(2\pi)^{2}}e^{-i{\bf p}\cdot{\bf R}}V(R),
\end{aligned}
\end{equation}
where the Rayleigh equation is used,
\begin{equation} 
\begin{aligned}
e^{-i{\bf p}\cdot{\bf R}}=&4\pi i^{-l}j_{l}(pR)Y_{l0}^{*}(\theta_{R'})Y_{l0}(\theta_{p}).
\end{aligned}
\end{equation}
Here ${\bf p}={\bf p}_{\parallel}+{\bf p}_{\bot}$ but only the part parallel to $R$ needed to be taken into account in the plane-wave expansion.
  %{Theory of the rotating polaron: spectrum and self-localization}
  %{rkky interaction of magnetic impurities in Dirac amd weyl semimetals}
  %{Ruderman-Kittel-Kasuya-Yosida interaction in Weyl semimetals}
 %{RKKY interaction in carbon nanotubes and graphene nanoribbons}
Then we have
\begin{equation} 
\begin{aligned}
V(p)=4\pi 
\sum_{l}i^{-l}
\int d\theta_{R''} \int^{\infty}_{0} dR 
j_{l}(pR)Y_{l0}(\theta_{R''})Y_{l0}(\theta_{p}).
\end{aligned}
\end{equation}

\subsection{C. Possible Experimental realization}

As we mentioned above, the resonantly bound multibody states can be investigated by consider
the Efimov correlations, for a configuration with several impurities (like the trimer or even the tetramer\cite{von Stecher J} bound states) 
short-range resonantly interact with the identical (noninteracting) bath particles (usually the bosons) in a 
quantum condensed phase.
For example, two heavy electrons interact with the excited phonons through the $s$-wave interaction, 
which can be realized also in the semi-Dirac system described in this article, in addition to the Bose gas.
  %{impurity-induced multibody resonances in a bose gas}
  %{impurity in a bose-einstein condensate and the efimov effect}
  %{inpurity in a bose-einstein condensate: study of the attractive and repulsive branch using quantum monte carlo methods}
Here the impurity-boson scattering length is very large (compared to the effective range), and
the boson-boson scattering length is also much shorter than the boson-impurity scattering length.
As it is well known, the Efimov trimer requires the university limit $a\rightarrow \infty$.
  %{impurity in a bose-einstein condensate and the efimov effect}
Apart from the BEC which requires the light enough bosons and low temperature, the condensed phase like the superconductivity or the supercurrent (e.g., in a Josephson junction setup),
also provide us opportunities to study, e.g., the 
Fulde-Ferrell-Larkin-Ovchinikov (FFLO) pairing mechanism, 
%pairing is a typical pairing mechanism where only the odd-frequency (triplet) Cooper pair can be 
%penetrated into the middle part\cite{Keizer R S,Asano Y}.
  %{Josephson effect due to odd-frequency pairs in diffusive half metals}
  %{attractive and repulsive fermi polarons in two dimensions}
  %{fermi polaron-polaritons in charge-tunable atomically thin semiconductors}
expecially for a spin-imbalanced system as also revealed in the mesoscopic Josephson junction\cite{Asano Y,Kreula J M}.
%{spin-asymmetry josephson plasma oscillations}
%{geometric phase in a mesoscopic josephson junction with classical driving source}
The creation of exciton polaritons is reported\cite{Kasprzak J,Sidler M} in the experiments base on the semiconductor microcavity
where the photons are confined within a cavity consist of the non-magnetic dielectric material,
and strongly coupled to excitations.
Then the cavity spectroscopy can used to probes the formation of polaron or polariton,
just like the radiofrequency (rf) spectroscopy (of the impurity) for a fermi gas,
or the Bragg spectroscopy for a BEC\cite{Bruun G M},
which can be applied to the semi-Dirac system,
like the VO$_{2}-$TiO$_{2}$ heterostructure, SnPSe$_{3}$\cite{Damljanovi? V} or the organic conductor\cite{Kobayashi A},
  %{Photoinduced Chern insulating states in semi-Dirac materials}
as its low-energy (the region we focus on) characteristics lies somewhere between the gapped semiconductors and the Dirac semimetal\cite{Banerjee S},
and it behaves more like the semiconductor when the gap is possitive ($D>0$).
Bse on the cavity spectroscopy,
ont only the cavity resonance can be probed, the disperion of the semi-Dirac system (i.e., the real part of susceptibility) can also be 
obtained, which is similar to the technique of angle-resolved photoemission spectroscopy (ARPES).
  %{quasiparticle properties of an impurity in a fermi gas}
The obtained dispersion certainly should be continuous although the existence of the Fermion polaron, 
unlike the atomic system which with discrete levels.
More importantly, since we consider the weak coupling regime in this paper, the quasiparticle weight (or spectral weigth)
should be very large (closes one), which can be proved by the large overlap between the upper and lower levels (resonances).
That's due to the attractive Fermi polaron which 
prevents the electron-hole pair excitation from across the Fermi surface\cite{Sidler M}.
While when the two-body bound state formed (for the scattering length larger than a critical value),
the spectral weigth vanishes and the metastability results in weak polaron peaks,
as displayed during the polaron-to-molecule transition,
  %{Quasiparticle properties of an impurity in a Fermi gas}
which can be revealed in experiment by the vanishing peak of the rf spectrum for fermi gas.
  %{Quasiparticle properties of an impurity in a Fermi gas}
In such case, the problem is much easier to deal with simply by multiples the dimers,
while for the multibody resonance where the immobile impurity (with infinite mass) induce the repulsions between the light majority particles\cite{Shi Z Y}.

As we discussed above, for
the exciton-electron interaction-induced fermi polaron, the direct Coulomb interaction need not to be taken into account since the exciton is a neutral particle except when there exits the Coulomb-type impurity.
%{Sidler M}
This mechanism
is also considered experimently in Ref.\cite{Sidler M} base on a MoSe$_{2}$ heterostructure
upon the hexagonal boron nitride (hBN) substrate,
however, in fact, 
the bose polaron should also exists in such system
due to the electron-phonon coupling which is nonnegligible since the hBN is a 
polar substrate with large surface-optical phonon mode energy (lattice vibrations) as experimently measured\cite{Wang Z W,Scharf B}.
Thus for semi-Dirac system, the bose polaron can be obtained by controlling the electron-phonon coupling 
(through tunning the strain and change the distance between two-dimensional semi-Dirac sample and the substrate).
Such tunning way is evidented by the distance (from electron to phonon and from substrate to 
semi-Dirac sample)-dependent factors within the expression of the electron-phonon coupling Hamiltonian,
  %{polaronic effects in monolayer black phosphorus on substrates}
  %{study of non-extensive entropy of bound polaron in monolayer garphene}
  %{spin-dependent polaron formation in pristine graphene}
  %{possible formation of chiral polarons in graphene}
which reads (consider the longitudinal branch only)
\begin{equation} 
\begin{aligned}
H_{e-ph}=\sum_{Q}M_{Q}b_{Q}e^{i{\bf Q}\cdot{\bf r}}+\sum_{Q}M^{*}_{Q}b^{\dag}_{Q}e^{-i{\bf Q}\cdot{\bf r}},
\end{aligned}
\end{equation}
where $M_{Q}$ is the complex amplitude of the electron-phonon coupling,
$b_{Q}$ is the anihilation operator of phonon which obeys $[b_{Q},b_{Q}^{\dag}]={\bf I}$.
  %{http://www.physics.drexel.edu/~bob/LieGroups/LG_06.pdf}
A complete Hamiltonian can be obtained by adding the above term into the Eq.(23).
The interaction between electron from the semi-dirac material surface and the longitudinal optical phonons from the surface of polar substrate
  %{study of non-extensive entropy of bound polaron in monolayer garphene}
  %{polaronic effects in monolayer black phosphorus on polar substrates}
  %{The Motion of Slow Electrons in a Polar Crystal}
  %{spin-dependent polaron formation in pristine graphene}
  %{possible formation of chiral polarons in graphene}
can be explored by the variational method where the ground state energy is obtained by minimizing the system expectation value, which should also be matched with the diagrammatic Monte Carlo method.
  %{Polaron-to-molecule transition in a strongly imbalanced Fermi gas}
  %DOI:https://doi.org/10.1103/PhysRevA.80.053605
According to the Lee-Low-Pines (LLP) unitary transformations\cite{Lee T D}, the ground state energy can
be solved through a routine process:
\begin{equation} 
\begin{aligned}
U_{1}=&e^{i({\bf k}-\sum_{Q}{\bf Q}b_{Q}^{\dag}b_{\dag})\cdot{\bf r}},\\
U_{2}=&e^{\sum_{Q}(f_{Q}b^{\dag}_{Q}-f^{*}_{Q}b_{Q})},\\
\end{aligned}
\end{equation}
where $f_{Q}$ is the variational function 
and $\varepsilon'$ is the eigenvalue of the transformed Hamiltonian.
The relation 
$
e^{-A}Be^{A}\approx B+[B,A]$ is used during this process.
That leads to
\begin{equation} 
\begin{aligned}
b_{Q}\rightarrow b_{Q}+f_{Q},\ e^{-i{\bf q}\cdot{\bf r}}b_{Q}^{\dag}\rightarrow f_{Q}^{*},\ b_{Q}e^{i{\bf q}\cdot{\bf r}}\rightarrow f_{Q},
k\rightarrow k-\sum_{Q}Qb_{Q}^{\dag}b_{Q}.
\end{aligned}
\end{equation}
Then the ground state energy can be obtained by solving 
\begin{equation} 
\begin{aligned}
\frac{\delta \varepsilon'}{\delta f_{Q}}=&\frac{\delta \varepsilon'}{\delta f^{*}_{Q}}=0.
\end{aligned}
\end{equation}
%{Fermi polaron-polaritons in charge-tunable atomically thin semiconductors}
%{Polaronic effects in monolayer black phosphorus on polar substrates}
%{Study of non-extensive entropy of bound polaron in monolayer graphene}
  %{The Motion of Slow Electrons in a Polar Crystal}

\end{large}
\renewcommand\refname{References}

\clearpage

Fig.1
\begin{figure}[!ht]
   \centering
 \centering
   \begin{center}
     \includegraphics*[width=0.9\linewidth]{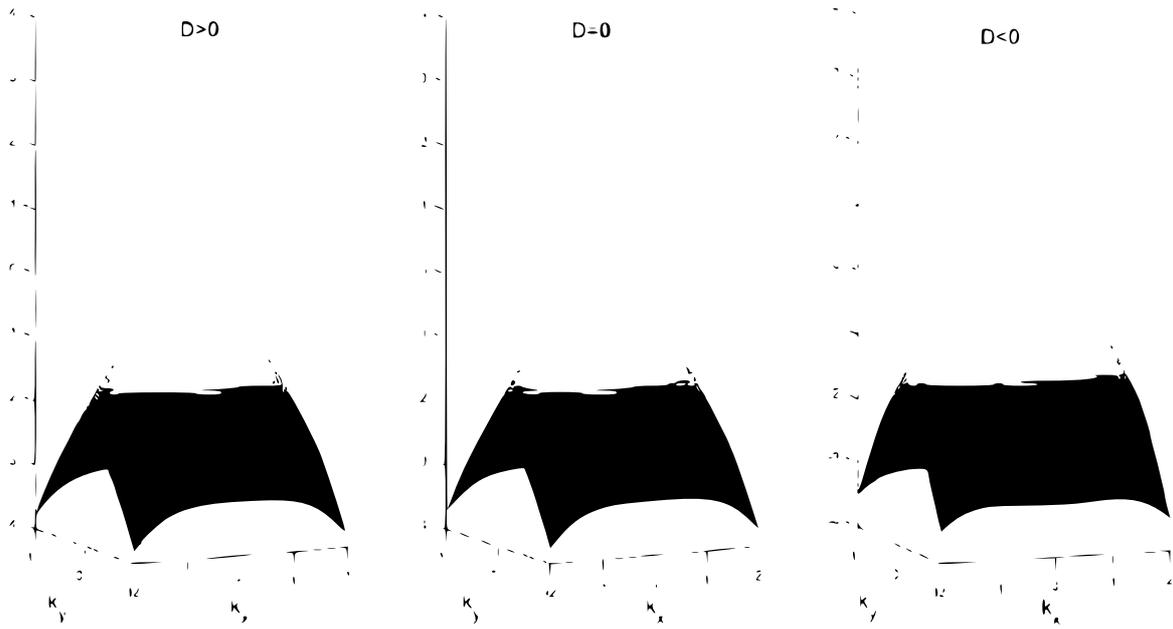}
\caption{
%Real part (left) and imaginary part (right) of the pair propagator at non-chiral case
%as a function of the impurity momentum $p$ and majority momentum $q$.
%The rows from top to bottom correspond to the Bosonic frequency (impurity) $\omega=-1,\ 0,\ %1,\ 2$, respectively.
%The momentum cutoff $\Lambda$ is setted as 1 and the chemical potential is zero.
%The vertical axis is in unit of $\frac{1}{2\pi}$.
Evolution of the low-energy dispersion of semi-Dirac system.
}
   \end{center}
\end{figure}

Fig.2
\begin{figure}[!ht]
   \centering
 \centering
   \begin{center}
     \includegraphics*[width=1\linewidth]{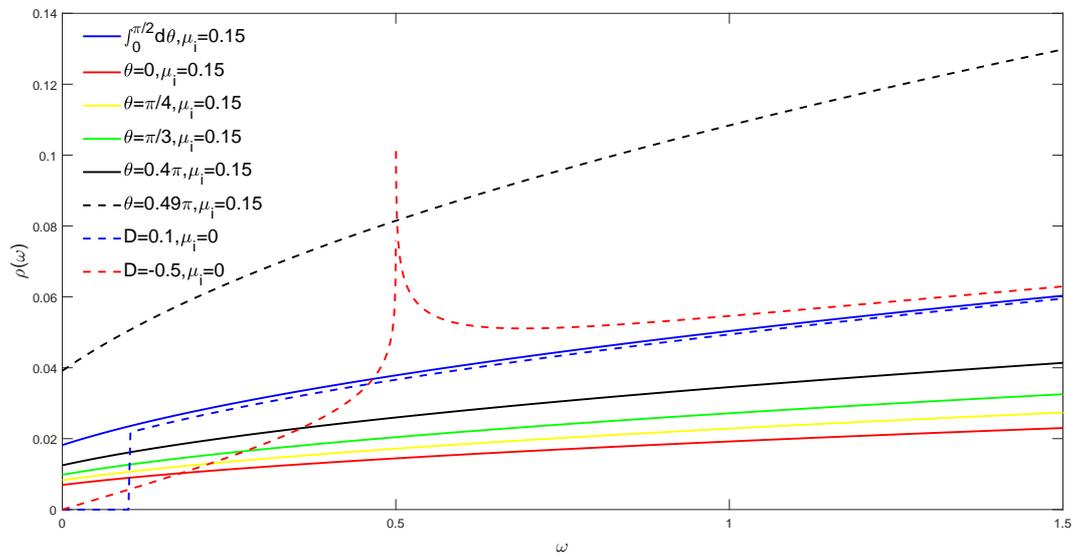}
\caption{
DOS of the semi-Dirac quasiparticle as a dunction of energy.
}
   \end{center}
\end{figure}
\clearpage

Fig.3
\begin{figure}[!ht]
   \centering
 \centering
   \begin{center}
     \includegraphics*[width=0.8\linewidth]{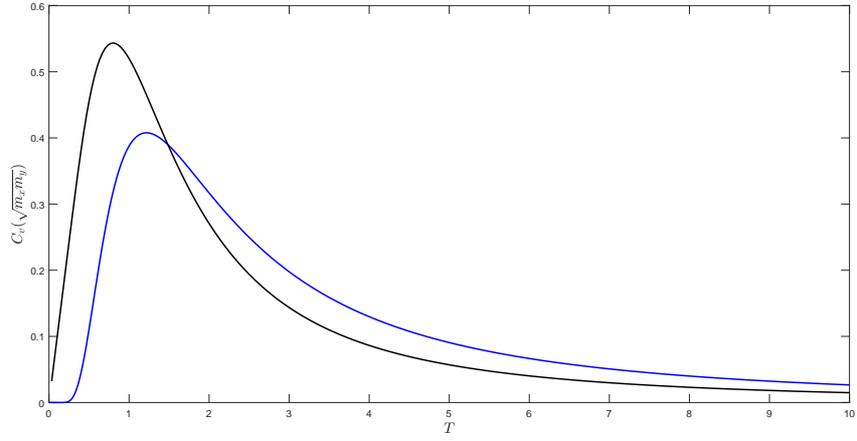}
\caption{
specific heat in unit of $\sqrt{m_{x}m_{y}}$.
The blue line corresponds to result obtained by Eq.(79) in perspective of $T$-dependent free energy.
The black line corresponds to result obtained by Eq.(80) within low carrier density approximation.
}
   \end{center}
\end{figure}
Fig.4
\begin{figure}[!ht]
   \centering
 \centering
   \begin{center}
     \includegraphics*[width=0.9\linewidth]{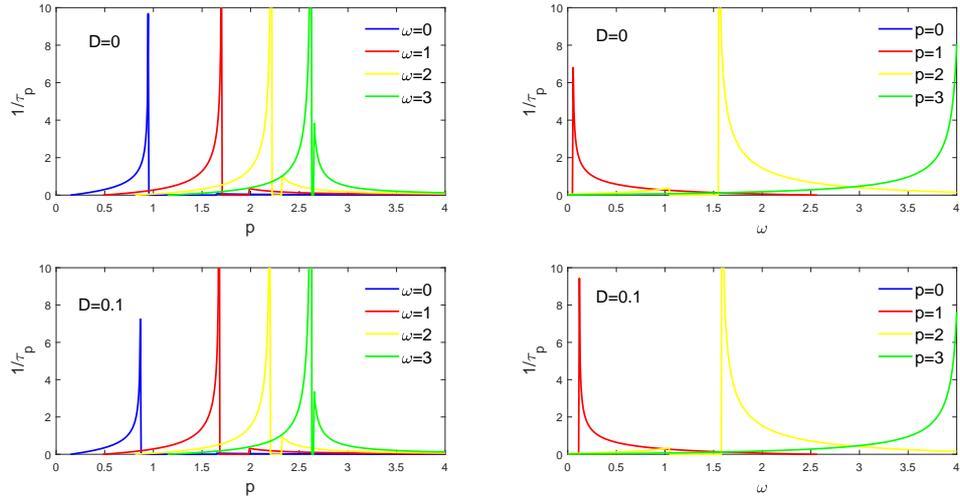}
\caption{
Relaxation time of the semi-Dirac system as a function of $p$ and $\omega$.
}
   \end{center}
\end{figure}

\clearpage
Fig.5
\begin{figure}[!ht]
   \centering
 \centering
   \begin{center}
     \includegraphics*[width=0.9\linewidth]{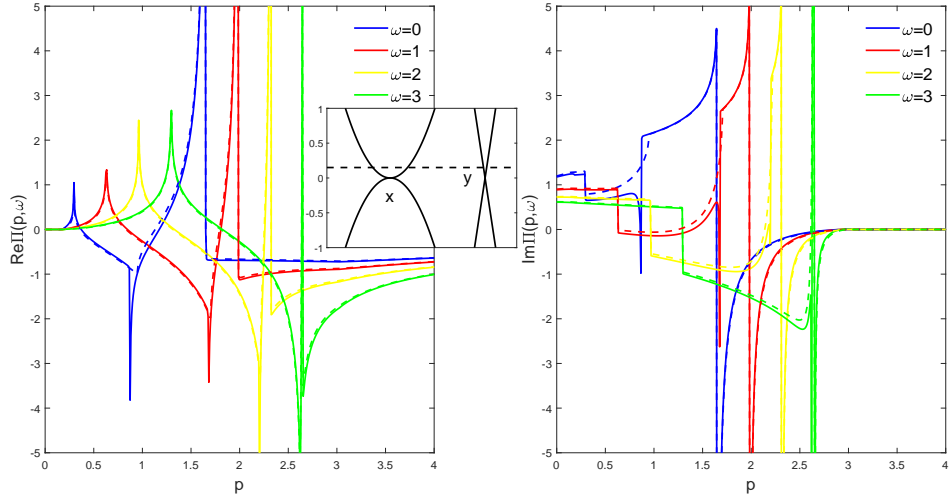}
\caption{
%Real part (left) and imaginary part (right) of the pair propagator at non-chiral case
%as a function of the impurity momentum $p$ and majority momentum $q$.
%The rows from top to bottom correspond to the Bosonic frequency (impurity) $\omega=-1,\ 0,\ %1,\ 2$, respectively.
%The momentum cutoff $\Lambda$ is setted as 1 and the chemical potential is zero.
%The vertical axis is in unit of $\frac{1}{2\pi}$.
Pair-propagator of the polaron formed in semi-Dirac system in anisotropic teatment.
The inset shoes the band structure.
%The horizontal axis corresponds to the $p=2^{1/2}p_{x}$ since we consider the $\Phi=0$ case.
}
   \end{center}
\end{figure}
Fig.6
\begin{figure}[!ht]
   \centering
 \centering
   \begin{center}
     \includegraphics*[width=0.9\linewidth]{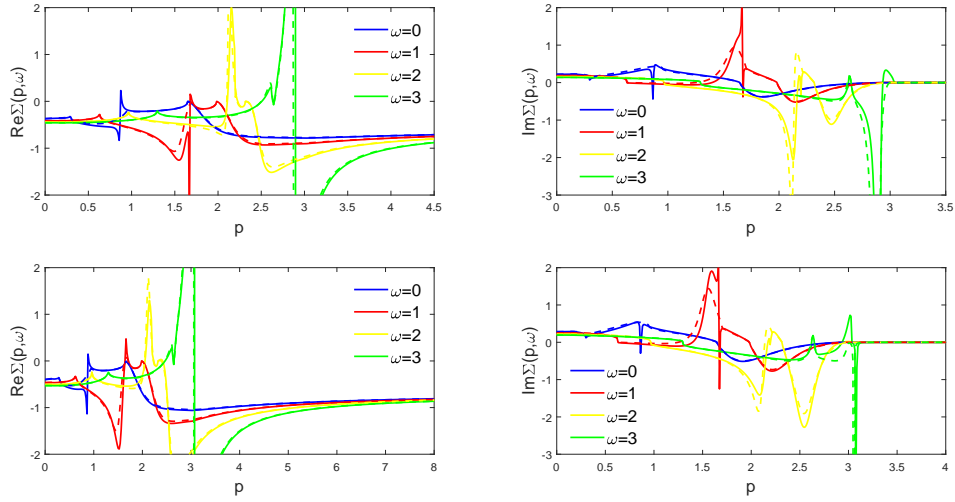}
\caption{
%Real part (left) and imaginary part (right) of the pair propagator at non-chiral case
%as a function of the impurity momentum $p$ and majority momentum $q$.
%The rows from top to bottom correspond to the Bosonic frequency (impurity) $\omega=-1,\ 0,\ %1,\ 2$, respectively.
%The momentum cutoff $\Lambda$ is setted as 1 and the chemical potential is zero.
%The vertical axis is in unit of $\frac{1}{2\pi}$.
Self-energy of the polaron in anisotropic teatment as
a function of initial impurity momentum $p$.
}
   \end{center}
\end{figure}

\clearpage
Fig.7
\begin{figure}[!ht]
   \centering
 \centering
   \begin{center}
     \includegraphics*[width=0.9\linewidth]{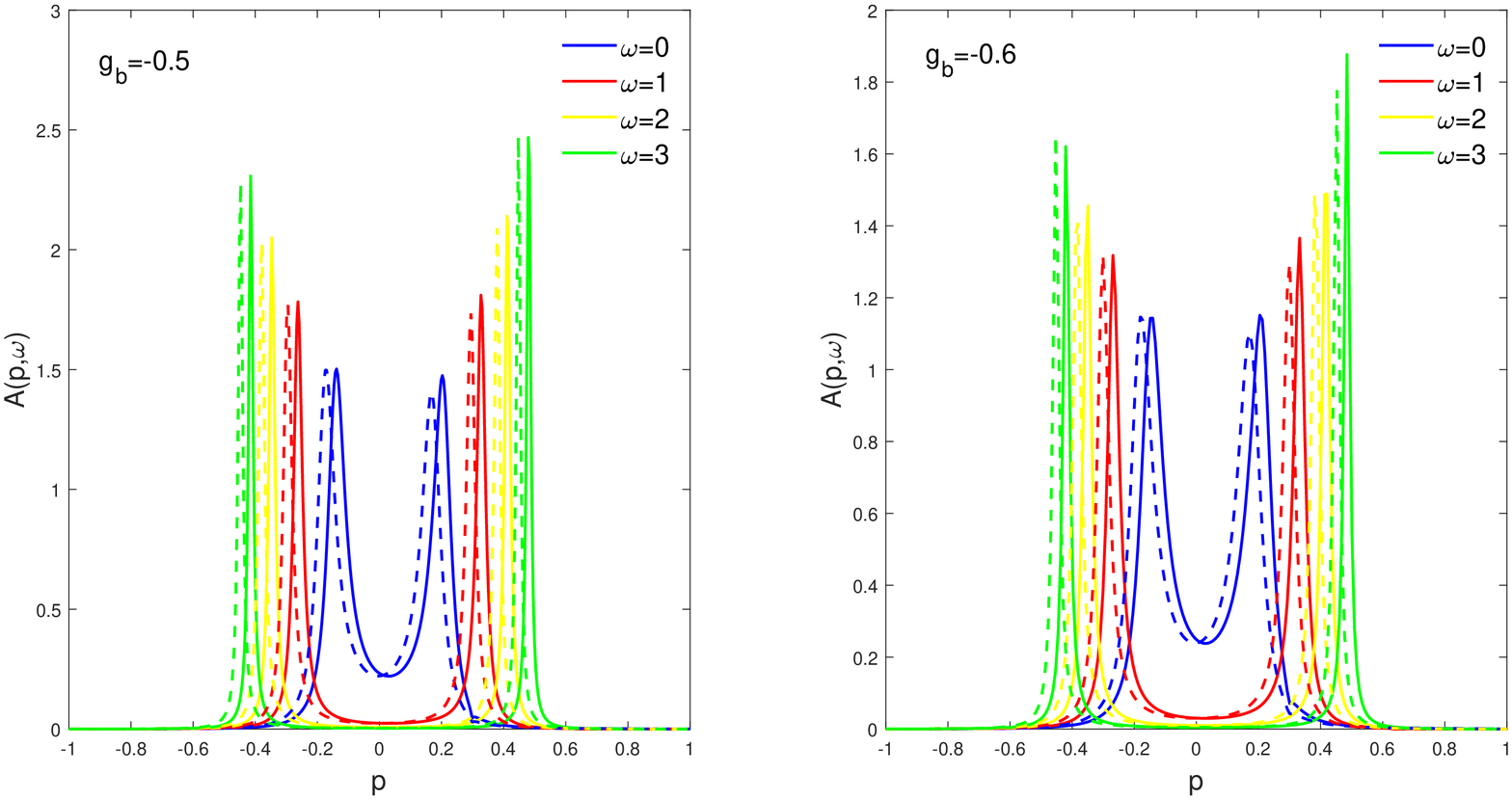}
\caption{
%Real part (left) and imaginary part (right) of the pair propagator at non-chiral case
%as a function of the impurity momentum $p$ and majority momentum $q$.
%The rows from top to bottom correspond to the Bosonic frequency (impurity) $\omega=-1,\ 0,\ %1,\ 2$, respectively.
%The momentum cutoff $\Lambda$ is setted as 1 and the chemical potential is zero.
%The vertical axis is in unit of $\frac{1}{2\pi}$.
Spectral function of the polaron in anisotropic teatment.
}
   \end{center}
\end{figure}
\clearpage
Fig.8
\begin{figure}[!ht]
   \centering
 \centering
   \begin{center}
     \includegraphics*[width=0.9\linewidth]{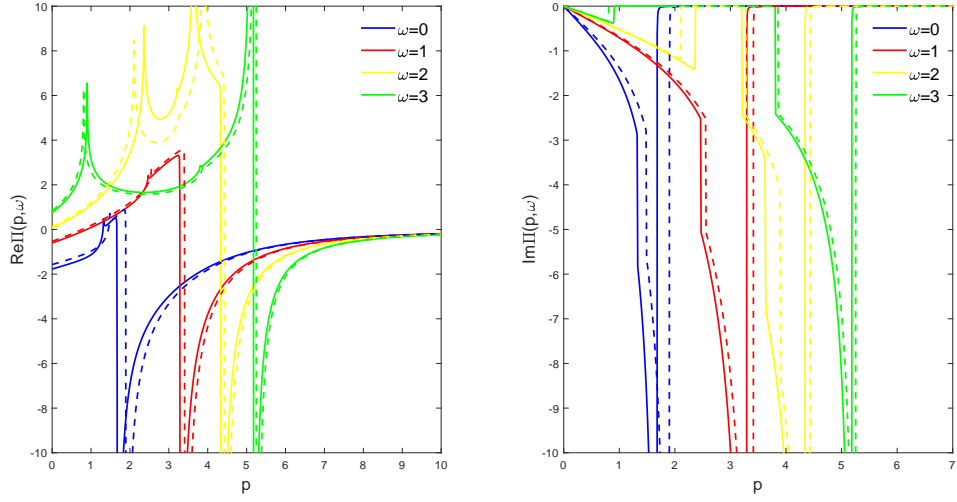}
\caption{
%Real part (left) and imaginary part (right) of the pair propagator at non-chiral case
%as a function of the impurity momentum $p$ and majority momentum $q$.
%The rows from top to bottom correspond to the Bosonic frequency (impurity) $\omega=-1,\ 0,\ %1,\ 2$, respectively.
%The momentum cutoff $\Lambda$ is setted as 1 and the chemical potential is zero.
%The vertical axis is in unit of $\frac{1}{2\pi}$.
Pair propagator in low carrier-density approximation.
The horizontal axis corresponds to the $p=2^{1/2}p_{x}$ since we consider the $\Phi=0$ case
i.e., the impurity as well as the polaron moves along the nonadiabatic ($p_{x}$) direction.
}
   \end{center}
\end{figure}
Fig.9
\begin{figure}[!ht]
   \centering
 \centering
   \begin{center}
     \includegraphics*[width=0.9\linewidth]{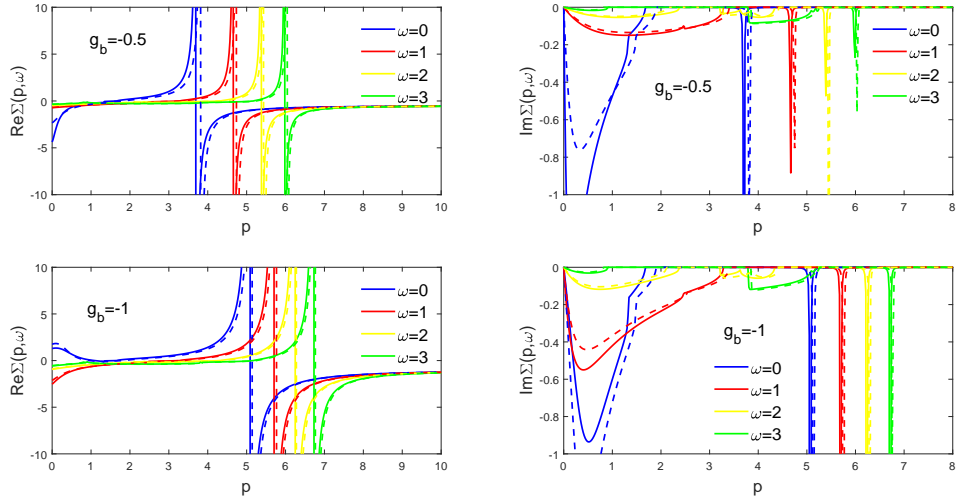}
\caption{
%Real part (left) and imaginary part (right) of the pair propagator at non-chiral case
%as a function of the impurity momentum $p$ and majority momentum $q$.
%The rows from top to bottom correspond to the Bosonic frequency (impurity) $\omega=-1,\ 0,\ %1,\ 2$, respectively.
%The momentum cutoff $\Lambda$ is setted as 1 and the chemical potential is zero.
%The vertical axis is in unit of $\frac{1}{2\pi}$.
Self-energy of the polaron in isotropic teatment.
}
   \end{center}
\end{figure}
\clearpage
Fig.10
\begin{figure}[!ht]
   \centering
 \centering
   \begin{center}
     \includegraphics*[width=0.9\linewidth]{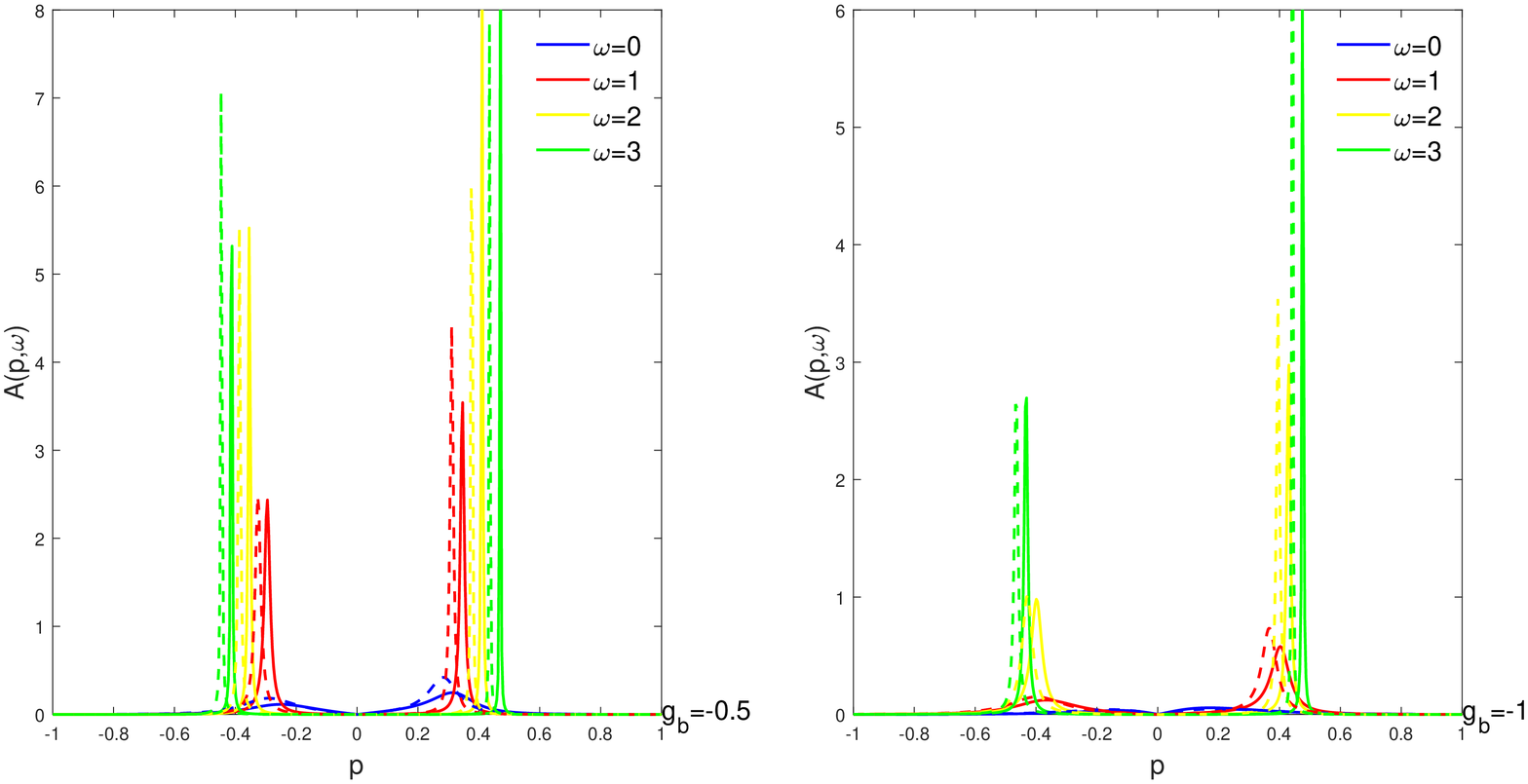}
\caption{
Spectral function of the polaron in isotropic teatment.
}
   \end{center}
\end{figure}

\clearpage
Fig.11
\begin{figure}[!ht]
   \centering
 \centering
   \begin{center}
     \includegraphics*[width=0.9\linewidth]{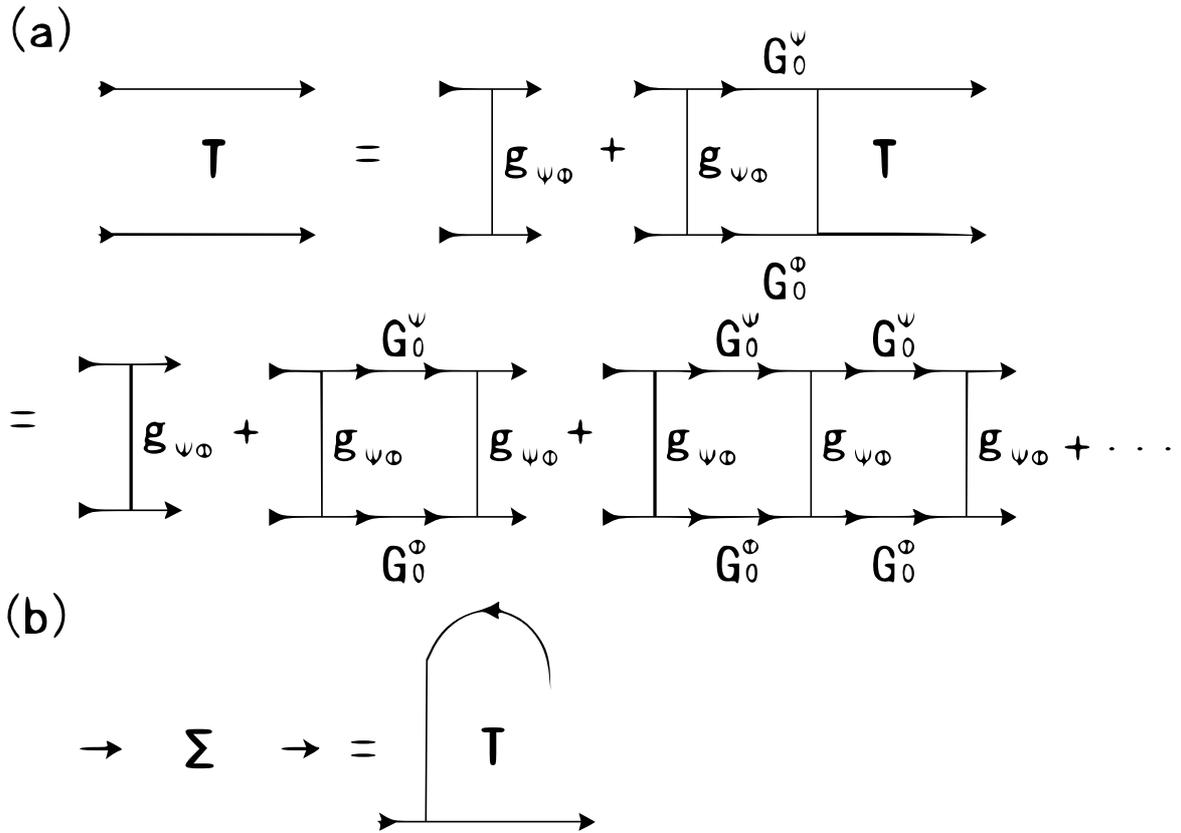}
 %{Repulsive polarons and itinerant ferromagnetism in strongly polarized Fermi gases}
 %{Bipolarons in a Bose-Einstein condensate}
 %{Mass imbalance effect in resonant Bose-Fermi mixtures}
 %{Twin peaks in rf spectra of Fermi gases at unitarity}
  %{dimensionality-induced bcs-bec crossover in layered superconductors}
  %{Diagrammatic Monte Carlo study of the Fermi polaron in two dimensions}
  %{electromagnetic modes from stoner enhancement graphene as a case study}
\caption{
(a) Diagrammatic representation of the medium $T$-matrix (the Bethe-Salpeter equation).
(b) The impurity self-energy.
All the black lines along the bottom edge of the $T$-matrix denote the bare impurity propagator while the ones along (or above the) the upper edge of
the $T$-matrix
denote the bare majority propagator.
The vertical line denotes the bare interaction vertex as labeled in the plot.
The impurity and the majority Green's function as well as the bare impurity-majority coupling are also labeled.
}
   \end{center}
\end{figure}
\clearpage

Fig.12
\begin{figure}[!ht]
   \centering
 \centering
   \begin{center}
     \includegraphics*[width=0.9\linewidth]{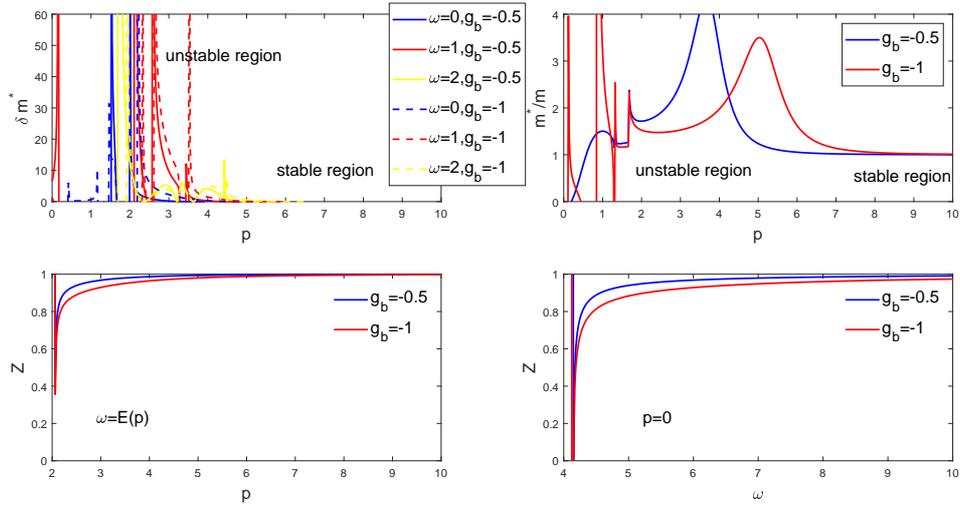}
\caption{General induced effective mass $\delta m^{*}$ (calculated by Eq.()), effective mass $m^{*}/m=(\delta m^{*}+m)/m$ in fermi liquid form 
(consider the residue; Eq.()), and the quasiparticle residue as a function of initial momentum $p$ and $\omega$. 
}
   \end{center}
\end{figure}

Fig.13
\begin{figure}[!ht]
   \centering
 \centering
   \begin{center}
     \includegraphics*[width=0.9\linewidth]{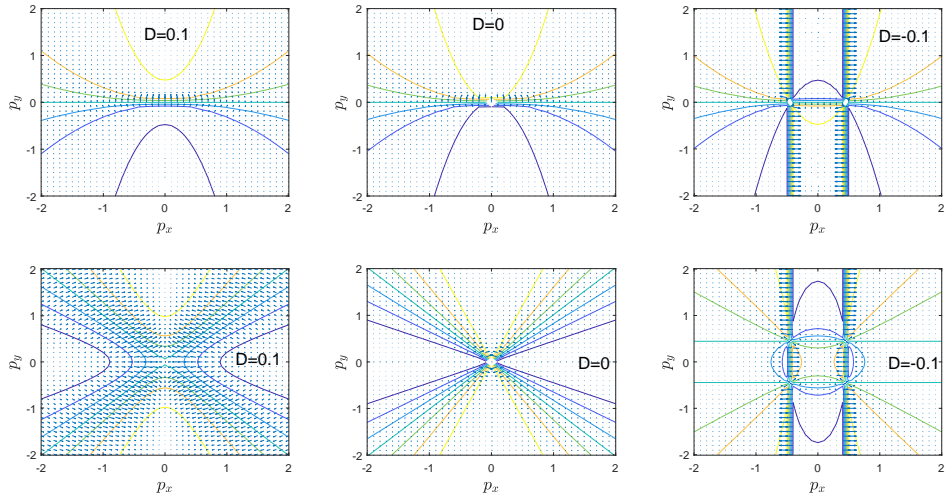}
\caption{
Relative phase $\theta$ of semi-Dirac system with anisotropic dispersion (first row) and
in low carrier density approximation (second row).
}
   \end{center}
\end{figure}

\end{document}